\PassOptionsToPackage{dvipsnames,svgnames*,table}{xcolor} 

\documentclass[10pt,
               letterpaper,
]{article}
\usepackage[top=0.85in,left=2.75in,footskip=0.75in]{geometry}

\usepackage{amsmath,amssymb}

\usepackage{changepage}

\usepackage{textcomp,marvosym}

\usepackage{cite}

\usepackage{nameref}

\usepackage[right]{lineno}

\usepackage{microtype}
\DisableLigatures[f]{encoding = *, family = * }

\usepackage[table]{xcolor}

\usepackage{array}

\newcolumntype{+}{!{\vrule width 2pt}}

\newlength\savedwidth

\raggedright
\usepackage[aboveskip=1pt,labelfont=bf,labelsep=period,justification=raggedright,singlelinecheck=off]{caption}

\renewcommand{\tablename}{Table}

\renewcommand{\thesubsection}{}

\makeatletter
\renewcommand{\@biblabel}[1]{\quad#1.}
\makeatother

\usepackage{lastpage,fancyhdr,graphicx}
\usepackage{epstopdf}
\fancyheadoffset[L]{2.25in}
\fancyfootoffset[L]{2.25in}
\usepackage[T1]{fontenc}
\usepackage[UKenglish]{babel}

\addto\captionsUKenglish{}
\addto\captionsUKenglish{\renewcommand{\tablename}{Table}}

\usepackage{microtype}

\usepackage{lmodern}
\usepackage{amssymb}
\usepackage{amsmath}
\usepackage{mathtools}
\usepackage[version=4]{mhchem}

\PassOptionsToPackage{hyphens}{url}
\usepackage[unicode=true,
            hidelinks,
]{hyperref}
\addto\extrasUKenglish{}
\addto\extrasUKenglish{}
\addto\extrasUKenglish{}

\usepackage[inline]{enumitem}

\usepackage{todonotes}
\usepackage{booktabs}
\usepackage{threeparttable}
\usepackage{longtable}
\usepackage{multirow}

\usepackage{lscape}

\usepackage{graphicx}

\usepackage{subcaption}

\newcommand{\captionwithlegend}[2]{\sisetup{detect-weight=true, detect-family=true} \caption{\textbf{#1} #2}}
\usepackage{csvsimple} \usepackage{csquotes} 
\usepackage{siunitx} 

\DeclareSIQualifier{\el}{el}
\DeclareSIQualifier{\th}{th}
\DeclareSIQualifier{\lhv}{LHV}
\DeclareSIQualifier{\hhv}{HHV}
\DeclareSIQualifier{\ammonia}{NH3}
\DeclareSIQualifier{\methane}{CH4}
\DeclareSIQualifier{\hydrogen}{H2}
\DeclareSIQualifier{\lohc}{LOHC}
\DeclareSIQualifier{\methanol}{MeOH}
\DeclareSIQualifier{\diesel}{Diesel}
\DeclareSIQualifier{\cd}{\ce{CO2}}
\DeclareSIUnit{\EUR}{EUR}
\DeclareSIUnit{\USD}{USD}
\DeclareSIUnit{\kW}{{\kilo\watt}}
\DeclareSIUnit{\MW}{{\mega\watt}}
\DeclareSIUnit{\MWh}{{\mega\watt\hour}}
\DeclareSIUnit{\GWh}{{\giga\watt\hour}}
\DeclareSIUnit{\TWh}{{\tera\watt\hour}}
\DeclareSIUnit{\PWh}{{\peta\watt\hour}}
\DeclareSIUnit{\percentpa}{{\percent}~p.a.}
\DeclareSIUnit{\wtpercent}{{wt.~\percent}}
\DeclareSIUnit{\km}{\kilo\meter}
\DeclareSIUnit{\day}{d}
\DeclareSIUnit{\EURpMWh}{\EUR\per\MWh\lhv}
\DeclareSIUnit{\EURpkgHydrogen}{\EUR\per\kg\hydrogen}
\DeclareSIUnit{\EURptCD}{\EUR\per\tonne\cd}

\usepackage[toc=false,section=section,
    numberedsection=false,]{glossaries}
\usepackage{glossaries-extra}
\AtBeginDocument{\colorlet{defaultcolor}{.}} 

\setabbreviationstyle[acronym]{long-short}

\newacronym[longplural={Levelised Cost of Energy}, shortplural=LCoE]{lcoe}{LCoE}{\text{Levelised Cost of Energy}}
\newacronym[longplural={Levelised Cost of Hydrogen}, shortplural=LCoH]{lcoh}{LCoH}{\text{Levelised Cost of Hydrogen}}
\newacronym[longplural={Marginal Costs of Energy}]{mcoe}{MCoE}{\text{Marginal Cost of Energy}}
\newacronym{sf}{SF}{\text{Surplus Factor}}
\newacronym{opex}{OPEX}{\text{Operational Expenditures}}
\newacronym{capex}{CAPEX}{\text{Capital Expenditures}}
\newacronym{fom}{FOM}{\text{Fixed Operation \& Maintenance}}
\newacronym{gegis}{GEGIS}{\text{GlobalEnergyGIS}}
\newacronym{lohc}{LOHC}{\text{Liquid Organic Hydrogen Carrier}}
\newacronym{lng}{LNG}{liquefied natural gas}
\newacronym{dbt}{DBT}{dibenzyltoluene}
\newacronym{wacc}{WACC}{\text{Weighted Average Cost of Capital}}
\newacronym{gwa}{GWA}{\text{Global Wind Atlas}}
\newacronym{eac}{EAC}{\text{Equivalent Annual Cost}}
\newacronym{esc}{ESC}{\text{Energy Supply Chain}}
\newacronym{esf}{ESF}{energy surplus factor}
\newacronym{lhv}{LHV}{\text{Lower Heating Value}}
\newacronym{hhv}{HHV}{\text{Higher Heating Value}}
\newacronym{dac}{DAC}{\text{Direct Air Capture}}
\newacronym{asu}{ASU}{air separation unit}
\newacronym{ae}{AE}{alkaline electrolysis}
\newacronym{pem}{PEM}{proton exchange membrane}
\newacronym[longplural={Renewable Energy Sources},shortplural={RES}]{res}{RES}{Renewable Energy Source}
\newacronym{cf}{CF}{Capacity Factor}
\newacronym{pt}{PT}{Pipeline Transport}
\newacronym{st}{ST}{Ship Transport}
\newacronym{hvdc}{HVDC}{High-Voltage Direct Current}
\newacronym{pv}{PV}{photovoltaics}
\newacronym{gdp}{GDP}{gross domestic product}
\newacronym{irena}{IRENA}{International Renewable Energy Agency}
\newacronym{oecd}{OECD}{Organisation for Economic Co-operation and Development}
\newacronym{uf}{UF}{utilisation factor}
\newacronym{fcev}{FCEV}{fuell-cell electric vehicle}
\newacronym{gis}{GIS}{geographic information system}
\newacronym{ftd}{FTD}{Fischer-Tropsch-Diesel}
\newacronym{ftf}{FT fuel}{Fischer-Tropsch fuel}
\newacronym{sng}{SNG}{synthetic natural gas}
\newacronym{smr}{SMR}{steam methane reforming}
\newacronym{msr}{MSR}{methanol steam reforming}
\newacronym{gwp}{GWP}{global warming potential}
\newacronym{cc}{CC}{carbon capture}
\newacronym{csr}{CSR}{corporate social responsibility}
\newacronym{esg}{ESG}{environmental social and corporate governance}
\newacronym{epc}{EPC}{engineering, procurement and construction}
\newacronym{au}{AU}{Australia}
\newacronym{ar}{AR}{Argentina}
\newacronym{de}{DE}{Germany}
\newacronym{dk}{DK}{Denmark}
\newacronym{eg}{EG}{Egypt}
\newacronym{es}{ES}{Spain}
\newacronym{ma}{MA}{Morocco}
\newacronym{sa}{SA}{Saudi Arabia}
\newacronym{eu}{EU}{European Union}
\newacronym{wana}{WANA}{Western Asia and Northern Africa}
\newacronym{usa}{USA}{United States of America}

\makeglossaries

\begin{document}
\vspace*{0.2in}

\begin{flushleft}
    {\Large
        \textbf\newline{Import options for chemical energy carriers from renewable sources to Germany} }
    \newline
\\
    Johannes Hampp\textsuperscript{1*},
    Michael Düren\textsuperscript{1},
    Tom Brown\textsuperscript{2,3}
\\
    \bigskip
    \textbf{1} Center for International Development and Environmental Research, Justus Liebig University Gießen, Hesse, Germany
    \\
    \textbf{2} Department of Digital Transformation in Energy Systems, Technische Universität Berlin, Berlin, Germany
    \\
    \textbf{3} Institute for Automation and Applied Informatics, Karlsruhe Institute of Technology, Baden-Württemberg, Germany
    \bigskip

* johannes.hampp@zeu.uni-giessen.de (JH)
    
\end{flushleft}
\section{Abstract}
    
Import and export of fossil energy carriers are cornerstones of energy systems world-wide.
If energy systems are to become climate neutral and sustainable, fossil carriers need to be
substituted with carbon neutral alternatives or electrified if possible.
We investigate synthetic chemical energy carriers, hydrogen, methane, methanol, ammonia and \glsxtrlongpl{ftf}, produced using 
electricity from \gls{res} as fossil substitutes.
\gls{res} potentials are obtained from \glsxtrshort{gis}-analysis and hourly resolved time-series are derived using reanalysis weather data.
We model the sourcing of feedstock chemicals, synthesis and transport along nine different \glsxtrlongpl{esc} to \glsxtrlong{de}
and compare import options for seven locations around the world against each other and with domestically sourced alternatives
on the basis of their respective cost per unit of hydrogen and energy delivered.
We find that for each type of chemical energy carrier, there is an import option with lower costs compared to domestic production in \glsxtrlong{de}.
No single exporting country or energy carrier has a unique cost advantage, since for each energy carrier and country there are cost-competitive alternatives.
This allows exporter and infrastructure decisions to be made based on other criteria than energy and cost.
The lowest cost means for importing of energy and hydrogen are by hydrogen pipeline from \glsxtrlong{dk}, \glsxtrlong{es} and 
\glsxtrlong{wana} starting at \SIrange{36}{42}{\EURpMWh\lhv} or \SIrange{1.0}{1.3}{\EURpkgHydrogen} (in 2050, assuming \SI{5}{\percentpa} capital cost).
For complex energy carriers derived from hydrogen like methane, ammonia, methanol or \glsxtrlongpl{ftf}, imports from \glsxtrlong{ar} by ship to 
\glsxtrlong{de} are lower cost than closer exporters in the \glsxtrlong{eu} or \glsxtrlong{wana}.
For meeting hydrogen demand, direct hydrogen imports are more attractive than indirect routes using methane, methanol 
or ammonia imports and subsequent decomposition to hydrogen because of high capital investment costs and energetic losses 
of the indirect routes.
We make our model and data available under open licenses for adaptation and reuse.

%\linenumbers

\section{Introduction}

Climate change mitigation efforts are driving energy transitions across the world.
In these efforts alternatives for established fossil energy carriers are being sought.
These aspirations gained additional traction with the plans of major international players like 
China, the \gls{eu} and \gls{usa} to become climate neutral by the middle of this century.
With technologies for the electricity sector already existing, these plans require a shift of focus to
the industrial, heating and mobility sectors.
Today's and tomorrow's energy demand of these sectors will have to be met with climate neutral and sustainable alternatives.
The same requirements also hold for industrial feedstock which have to be de-fossilised.
For some countries producing chemical energy carriers and feedstock from domestic \gls{res} or other near-zero-carbon
energy sources may be an option.
For other countries this will prove challenging due to geographical, sociological or technological restrictions.
\glsxtrlong{de} can be considered such a country where limited potentials for domestic energy generation will presumably 
be insufficient to to meet energy demand for chemical energy carriers.
Nowadays \glsxtrlong{de} strongly relies on energy imports, which made up more than \SI{76}{\percent} 
(approximately \SI{13.5}{\exa\joule}) of \glsxtrlong{de}'s domestically handled energy in 2018~\cite{agenergiebilanzene.v.2020}.
Despite a high population density and mediocre \gls{res} potentials in a world-wide comparison,
\glsxtrlong{de} has commited itself to \gls{res} as a future source of energy.
With these limitations the continued import of energy should be investigated,
where fossil carriers are substituted by synthetic chemical energy carriers produced from \gls{res}.

Compared to electrical or heat energy, chemical energy carriers are easy to transport and store,
making them a preferred option for energy exports.
Aided by pre-existing infrastructure and experience from handling of fossil energy carriers,
extensive use and significant trade volumes of synthetic chemical energy carriers from \gls{res}
can be expected by 2050.
One option, hydrogen, is currently receiving renewed world-wide attention with an increasing number of nations adopting hydrogen strategies.
A convergence to a system with one single predominant chemical energy carrier might not be ideal:
Depending on the end use the adaptation of processes to a different chemical, e.g. hydrogen, will have to be weighed against 
substituting fossil chemicals with synthetic drop-in alternatives.
Adding to the complexity of this decision are the different chemical and energy carrier specific properties which influence
the conditions and behaviour of a chemical during transport and storage.
It therefore becomes important to gain insight into the costs, composition and interaction of steps inside
potential future chemical \glspl{esc}.

Previous works have already analysed possible schemes for sourcing chemical energy carriers.
Fasihi et al.~\cite{fasihi2020} conducted a world-wide analysis on how renewable energy sources 
may be combined with other technologies to locally provide electricity and hydrogen at baseload quality 
and determined possible cost developments for \numrange{2020}{2050}.
With a focus on synthetic fuels another study by Fasihi et al.~\cite{fasihi2017} analysed an \gls{esc} 
for \gls{ftd} and \gls{sng} (methane) from the Maghreb region to Europe.
Watanabe et al.~\cite{watanabe2010} and later Heuser et al.~\cite{heuser2019} modelled green 
liquefied hydrogen production and transport from Patagonia to Japan.
Ishimoto et al.~\cite{ishimoto2020} analysed the cost for transporting hydrogen and ammonia from Norway to Japan
and compared it with transport to the Port of Rotterdam in Europe.
Lanphen~\cite{lanphen2019} also looked into ship-based supply chain options for importing liquid hydrogen, ammonia
and methylcyclohexane from various exporting ports to the Port of Rotterdam.
Niermann et al.~\cite{niermann2019} explored the transport of hydrogen using a variety of \glspl{lohc} and compared 
their results against a hydrogen gas pipeline system to supply the energy carrier over a distance of \SI{5000}{\km} to \gls{de}.
Schorn et al.~\cite{schorn2021} compared shipping of methanol with shipping of \ce{H2 (l)} from \gls{sa} to \gls{de}
and the economic viability depending on \ce{H2} and \ce{CO2} feedstock costs.
More recently a number of global studies and analysis for importing hydrogen and other energy carriers from various
regions around the world to \gls{de} have been released~\cite{brandle2021,johnston2022,pfennig2022,staiss2022}.
With a stronger focus on downstream infrastructure and energy distribution to end users, Runge et al.~\cite{runge2020}
compared the costs for transport fuels in a well-to-wheel analysis for mobility services in \glsxtrlong{de}.
A global view on international hydrogen trade was taken by Heuser et al.~\cite{heuser2020} who modelled
transport by pipeline and ship to determine optimal global supply costs.
Later, \gls{irena}~\cite{irena2022a} gave an outlook on global energy trade scenarios for \gls{res}-based hydrogen 
and additionally ammonia.

While each case study provides important insights, it is difficult to compare these studies due to their different system 
boundaries, limited subsets of overlapping technologies, different energy carriers and regions investigated.

With this study we add several novel features to the existing literature.
First we provide a comprehensive comparison of multiple \glspl{esc} from several different countries
based on uniform assumptions and system boundaries.
Secondly we deduct local electricity demand from the renewable resource availability in exporting countries, 
so that the best resources may be used locally.
Thirdly we design our \glspl{esc} to work as islanded systems and be energy self-sufficient.
Fourthly we make our data and model available under open licenses to allow for reproduction, adaptation and reuse.

\section{Materials and methods}

In this section we first describe how our \glspl{esc} are structured.
The investment optimisation problem is outlined followed by a description of the assumed technologies and a motivation
for the countries selected.
We then illustrate how \gls{res} potentials and feed-in are derived domestic demand is considered.
We end this section by motivating our choice of \gls{wacc} and an overview of the technical model structure.
The most important equations on which this model builds are given in \ref{app:equations}.

\subsection{Design of \glsfmtlongpl{esc}}

\begin{figure}[ht]
    \centering
\includegraphics[width=.9\textwidth]{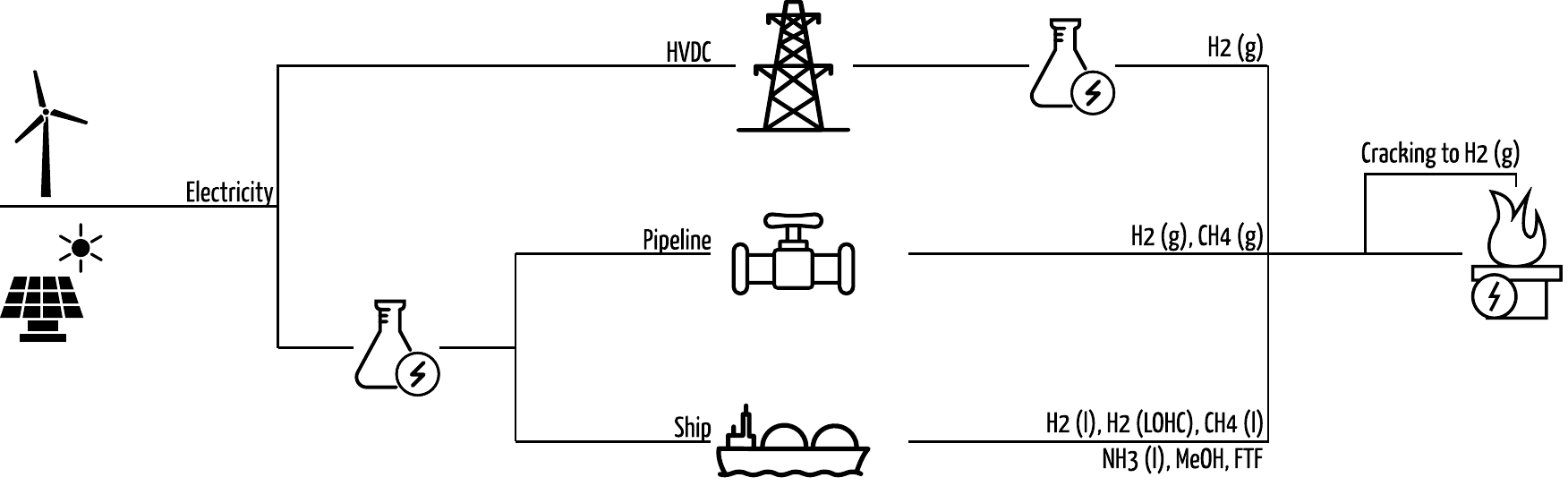}
    \captionwithlegend{Schematic representation of \glsfmtshortpl{esc} considered in this study.}{
        \glsfmtshortpl{esc} cover electricity generation from \glsfmtshort{res},
        intermediary buffer storage for electricity and chemicals,
        conversion to chemical energy carriers, conditioning and transport from exporter to importer.
        Each \glsfmtshortpl{esc} delivers one out of five chemical energy carriers.
        Detailed representations of each \glsfmtshort{esc} and all involved technologies are included in \ref{app:esc-visualisations}.
        License for icons:
        CC-BY-3.0 \cite{dmitrypodluzny2016, yushchenko2019, oliverguinfr2012, viralfaisalovers2018, iconika2018} and CC-0.
    }
    \label{fig:esc-scheme}
\end{figure}

We model and investigate \glspl{esc} for chemical energy carriers starting at the energy source in an exporting
country until the energy carriers are available in the importing country, in this analysis choosen to be \gls{de}.
The \glspl{esc} considered in this study are export of
  	\begin{enumerate*}[label=\alph*.)]
  		\item
            electricity by \gls{hvdc} with conversion to hydrogen in \gls{de},
        \item
            hydrogen gas by pipeline,
        \item
            methane gas by pipeline,
        \item
            liquid hydrogen by ship,
        \item
            liquid methane by ship,
        \item
            liquid ammonia by ship,
        \item
            liquid methanol by ship,
        \item
            hydrogen bound to \gls{lohc} \gls{dbt}~\cite{niermann2019} by ship,
        \item
            liquid \glspl{ftf} (kerosene-like) by ship.
  	\end{enumerate*}
\autoref{fig:esc-scheme} shows a schematic representation of all \glspl{esc}.
For \glspl{esc} transporting ammonia, methane and methanol, an optional cracking step to hydrogen is further included
for the case that the consumer needs pure hydrogen.
Key properties of the chemical energy carriers are listed in \autoref{tab:energy-carrier-properties}.

\begin{table*}
    \centering
    \small
    \begin{threeparttable}
        \captionwithlegend{Energy content of chemical energy carriers considered.}{}
        \label{tab:energy-carrier-properties}
        \begin{tabular}{llSS}
            \toprule
            Energy carrier                 & State of matter          & {Specific energy\tnote{a}} & {Hydrogen content}  \\
            & (normal conditions)      & {[\si{\MWh\per\tonne}]}    & {[\si{\wtpercent}]} \\
            \midrule
            Hydrogen \ce{H2} (g) / (l)     & Gas / Liquid (cryogenic) & 33.33                      & 100                 \\
            \gls{lohc} (\glsxtrshort{dbt}){\tnote{b}} & Liquid                   & 1.87{\tnote{c}}            & 5.6{\tnote{d}}      \\
            Methane \ce{CH4} (g) / (l)     & Gas / Liquid (cryogenic) & 13.89                      & 25                  \\
            Ammonia \ce{NH3}               & Gas / Liquid (cryogenic) & 5.17                       & 17                  \\
            Methanol {MeOH} (\ce{CH3OH})   & Liquid                   & 5.54                       & 12.5                \\
            \gls{ftf}                      & Liquid                   & 11.95                      & {- (var.)}          \\
            \bottomrule
        \end{tabular}
        \begin{tablenotes}
            \scriptsize
            \item Values based on \cite{wikipediacontributors2020}.
            \item[a] All values represent the \glsfmtshort{lhv}.
            \item[b] \glsfmtshort{lohc} chemical is not consumed and reused, hydrogen is the chemical energy carrier delivered to the importer.
            \item[c] \ce{H2} share.
            \item[d] \glsfmtshort{dbt} can hold up to \SI{6.2}{\percent} \ce{H2}, a depth-of-discharge for cycling at \SI{90}{\percent} is favourable for (de-)hydrogenation~\cite{runge2020}.
        \end{tablenotes}
    \end{threeparttable}
\end{table*}

The basic idea of the \glspl{esc} is shown in \autoref{fig:esc-scheme}.
Detailed representations for all components, energy and chemical flows considered in each \gls{esc} are included in \ref{app:esc-visualisations}.
In each \gls{esc} we consider
\begin{enumerate}[label=\alph*.)]
    \item
        sourcing of energy as electricity from \gls{res},
    \item
        sourcing of the major chemical feedstock for synthesis 
        i.e. water from seawater desalination, carbon-dioxide (\ce{CO2}) using \gls{dac} and nitrogen (\ce{N2}) using an \gls{asu} from ambient air,
    \item
        electrolysis of hydrogen and optional synthesis (hydrogenation) of the chemical energy carrier,
    \item
        necessary conversion of the energy carriers for transport, e.g. compression or liquefaction,
    \item
        back-conversion into the energy carriers' usable form at ambient conditions, e.g. relaxation or evaporation.
\end{enumerate}
If downstream processes require energy as input in addition to chemical feedstock, then this energy is provided from within the \gls{esc}.
Either electricity is used for processes on the exporter side or the currently available chemical energy carrier is used, 
e.g. the propulsion energy for shipping is provided using the energy carrier transported by the ship
and \gls{lohc} dehydrogenation uses part of the transported hydrogen.
We exclude the transport within the exporting and importing country and storage in the importing country, i.e. excluded is
\begin{enumerate}
    \item
        transport of electricity, feedstocks and chemical energy carriers between facilities and to the export terminal
    \item
        short-term or long-term storage at the importer in \gls{de}
    \item
        distribution and end use of chemical energy carriers within \gls{de}
\end{enumerate}
We exclude these steps because the range of possible options, such as the time patterns of the demand,
would create too many scenarios and detract from the generality of our analysis.
Such scenarios are better suited for investigation in specific case studies.

The \glspl{esc} are generally designed to not interact with any systems outside the \glspl{esc}.
This design decision excludes the possibility for use of secondary products such as process heat and cooling services 
from cryogenic carriers, the sale of chemical by-products such as \ce{O2}.
Also excluded are possible synergies by sector-coupling to heat and electricity systems on the exporter's and importer's sides,
e.g. of industrial waste heat into the \glspl{esc}.
Integration with other processes and commercial use of secondary products may provide grounds for business cases and lower
market prices for chemical energy carriers.

\subsection{Investment optimisation problem}

We model and optimise for least-cost investment of the essential components of \glspl{esc} to supply an annual energy demand.
The \glspl{esc} components include electricity generators and generation based on historic weather data, conversion processes, 
buffer storage and transport between countries.
The \glspl{esc} with their components, energy and mass flows are modelled using the open source modelling framework PyPSA~\cite{brown2018}.
We use a greenfield approach for our model and disregard existing infrastructure.
We justify this modelling decision as a comparable scale of infrastructure described by the \glspl{esc} does not yet exist anywhere.
Minimum annual investment costs are determined for each \gls{esc} based on the annualised cost $c_i$ of every
single component $i$.
The annualised cost represent cost for investment \gls{capex} $C_i$ and \gls{fom} and are annualised using the \gls{eac} method:
\begin{equation}
    c_i 		= C_i \cdot \left( \mathrm{A}_{i} + \mathrm{FOM}_i \right)
\end{equation}
where the component-specific annuity factor $A_i$ is
\begin{equation}   
    \mathrm{A}_{i} 	=  \frac{\left(1+r\right)^{\text{lifetime}(i)}\cdot r}{\left(1+r\right)^{\text{lifetime}(i)} - 1}
\end{equation}
The annual interest rate $r$ is assumed equal to the selected \gls{wacc} discussed below.
The objective function to be optimised for minimal investment is
\begin{align}
    \text{min} \left\{\qquad
    \smashoperator{\sum_{i \in \text{Generators}}} c_i\cdot G_i +
    \smashoperator{\sum_{i \in \text{Converters}}} c_i\cdot F_i +
    \smashoperator{\sum_{i \in \text{Storage}}} c_i\cdot H_i    
    \right\}
\end{align}
where generators (e.g. \gls{pv}) are considered with their nominal capacities $G_i$ (e.g. \si{\MW}),
converters (e.g. compressors, electrolysers) with their throughput capacity $F_i$ (e.g. tonne per hour \si{\tonne\per\hour}, 
\si{\MW}) and storage components (e.g. batteries, tanks) with their storage capacity $H_i$ (e.g. \si{\MWh}, \si{\meter\cubed}).
The optimisation is subject to constraints which ensure conservation of energy and mass flow and runs at hourly resolution.
Most components may be freely dispatched without restrictions on ramping rates and minimum must-run capacities,
as most technologies are usually flexible within a below hourly time-scale.
Must-run capacities are only enforced for the synthesis of methane, ammonia, methanol and \gls{ftf}, see the discussion
in the technologies section.
\gls{res} electricity feed-in is limited to the individual modelled time-series and excess electricity may be freely curtailed.

\gls{res} capacities may only be extended to the maximum potentials of their respective technology and resource
quality class determined through \gls{gis}-analysis, see the section on electricity supply below for details.
The nominal capacities of all other components may be freely extended without limit.
Capital costs for all components scale linearly with their capacity representing a situation where capacity expansion 
requires new facilities rather than extending existing ones.
The used capital costs and \gls{fom} already assume large scale facilities with respective economies of scale applied 
as well as exogenous learning rates for cost reductions between 2030 to 2050.
Process efficiencies are assumed constant for all years (see \ref{app:conversion-efficiencies}) due to the difficulty
of making well-founded estimates for future technological developments and improvements.
One exception is made for hydrogen electrolysis where efficiency is expected to increase across all available technology
options~\cite{irena2020, danishenergyagency2021}, an improvement which would affect all \glspl{esc} 
presented here.
Thus the electricity-to-hydrogen (\gls{lhv}) efficiency for electrolysis is assumed to improve from \SI{68}{\percent} in 2030 to 
\SI{71.5}{\percent} in 2040 and \SI{75}{\percent} in 2050.

\subsection{Choice of technologies and energy carriers}
The following section discusses the chemical energy carriers listed in \autoref{tab:energy-carrier-properties} and their simplified production pathways.
For most production pathways, alternative methods and integrated technologies with potential for efficiency improvement exist.
Integrated technologies and alternative technology options are beyond the scope of our investigation and not discussed further.

\subsubsection*{\glsfmtlongpl{res}}
We select utility \gls{pv}, on-shore and off-shore wind as the only sources of energy for our model.
We consider these technologies the only available ones with sufficient modularity and possibility for quick build up to high capacities
while delivering near-zero carbon electricity necessary for a sustainable deployment.
We consider the following other sources of energy unsuitable (with selected reasons):
\begin{enumerate}[label=\alph*.)]
    \item
        hydro power (prioritised for domestic demand,
        limited geographical locations, 
        significant environmental impact, 
        poor modularity and scalability),
    \item
        concentrated solar power (limited to specific geographical locations, low modularity and scalability),
    \item
        conventional nuclear fission 
        (intransparent cost, unmodular and difficult deployment, 
        sustainability issues with fuel and waste streams),
    \item
        nuclear fusion and unconventional nuclear fission 
        (insufficient technology readiness level with unclear techno-economic prospects).
\end{enumerate}

\subsubsection*{Hydrogen}
In all \glspl{esc} electricity is converted to the simplest chemical energy carrier, hydrogen, via \gls{ae}.
Compared to alternatives like \gls{pem} electrolysers, \glspl{ae} electrolysers were historically considered to be less suitable
to provide grid services due to their slow start-up times in the range of minutes~\cite{danishenergyagency2021}.
For large scale operations in chemical energy carrier production the slow start-up time may be neglected because
the electrolysis here does not need to provide grid services and the variability of \gls{res} feed-in can be managed using
for example battery storage.
The main advantages of \glspl{ae} are that it is a well established technology which has in comparison to \glspl{pem}
lower costs and does not require rare-earth-metals or platinum which may become bottlenecks for massive deployments in the future.
In the case of large-scale hydrogen production and hydrogen-derived chemical energy carriers investigated here,
we expect installations to run at high utilisation factors with a large number of modular units that can combine
into a large unit to create virtual flexibility, thus negating the main weakness of \gls{ae}.
The necessary water for electrolysis is produced through desalination of seawater via reverse osmosis to
achieve the required purity and conserve existing, potentially scarce fresh water resources.

Hydrogen as a chemical energy carrier is versatile and may also be used as a feedstock for chemical synthesis.
The physical properties of hydrogen make it difficult to transport as a gas due to its low energy density
and as a cryogenic liquid due to the very low temperature required and high energy demand for liquefaction (\SI{0.203}{\MWh\per\MWh\lhv}).

\subsubsection*{Methane}
Methane is an alternative gaseous energy carrier to hydrogen and is synthesised in established production processes 
through catalytic reaction of \ce{CO2} with \ce{H2} via the reverse water gas shift reaction:
\begin{equation}
    \ce{CO2 + 4 H2 <--> CH4 + 2 H2O}
\end{equation}
The synthesis is accompanied by adverse side reactions but can be run at high selectivity~\cite{gotz2016}.
Transport of methane is well established and infrastructure for pipeline transport or 
transport as \gls{lng} already exists on large scales were synthetic methane may be used as drop-in replacement.
Downsides are the emissions of \ce{CO2} from combustion and methane during handling (leakage, slippage)
and the high \gls{gwp} of methane.
Liquefaction of methane requires less energy (\SI{0.036}{\MWh\per\MWh\lhv}) than liquefaction of hydrogen and handling
of cryogenic \gls{lng} is less complex and better established compared to liquefied hydrogen.

\subsubsection*{Ammonia}
Ammonia is another option for an gaseous energy carrier and was already used as energy carrier in past applications.
Ammonia is synthesised by hydrogenation of nitrogen in the Haber-Bosch reaction:
\begin{equation}
    \ce{N2 + 3 H2 <--> 2NH3}
\end{equation}
Feedstock nitrogen gas \ce{N2} is available through extraction from the atmosphere via e.g. cryogenic \glspl{asu}.
The ammonia synthesis process is well-established and an increasing focus in research for direct energetic applications 
of ammonia can be seen.
In comparison to \ce{H2} or \ce{CH4}, \ce{NH3} has a lower specific energy but a high boiling point
at \SI{-33}{\degreeCelsius}, making it easier to handle as a liquid with a high volumetric energy density.
In addition industry has long-standing experience of transporting and handling ammonia in its various forms,
with around \SI{10}{\percent} of the global \SI{183}{\mega\tonne} annual ammonia production being traded~\cite{irena2022}.
Direct energetic use of ammonia is possible but poses a range of challenges including suppression of \ce{NO_x} emissions~\cite{macfarlane2020}.
Alternatively ammonia may be used as hydrogen carrier where the dehydrogenation of \ce{NH3} happens through thermal decomposition.
The dehydrogenation process has high energy requirements ($> \SI{25}{\percent}$ of \gls{lhv}~\cite{engie2020})
and the technology is not yet fully commercially mature.
Large specialised ammonia crackers are already in operation for the production of heavy water~\cite{irena2022}.

\subsubsection*{Methanol (MeOH)}
Methanol (MeOH) is a liquid organic compound with favourable properties for transport, storage and energetic use.
It is obtained by hydrogenation of \ce{CO2}:
\begin{equation}
    \ce{2 CO2 + 6H2 <--> 2CH3OH + 2 H2O}
\end{equation}
either in a direct catalytic methanolisation reaction or by an indirect route using a reverse water gas shift reactor to obtain 
syngas~\cite{bazzanella2017}.
With an annual production volume of ca. \SI{100}{\mega\tonne}~\cite{methanolinstitute2022} methanol synthesis
is a well-established process with industry experience in handling, storage and transport.
Methanol is used either as a industrial feedstock, for power and heat generation, as a mobility fuel~\cite{dolan2019,irena2021a}
and also be considered a member the \gls{lohc} family with the dehydrogenated form being \ce{CO2}~\cite{niermann2019}.
Methanol-to-hydrogen cracking (steam methanol reforming) is mature on small industrial scales~\cite{caloricanlagenbaugmbh2020}
and practical where methanol logistics are easier than logistics for alternative hydrogen feedstocks like e.g. natural gas.

\subsubsection*{\glsfmtfullpl{ftf}}
Liquid fuels like \glsxtrlong{ftd} or kerosene can be produced from \ce{CO2} and \ce{H2} via Fischer-Tropsch synthesis.
These fuels can be used as drop-in replacements for today's fossil based fuels and are simple to transport and store due to their
liquid nature.
The catalytic Fischer-Tropsch synthesis is not very selective and yields a mixture of hydrocarbon products,
requiring post-processing to yield \glspl{ftf}~\cite{konig2015}.
We neglect gaseous outputs like fuel gases which constitute approximately \SI{20}{\percent} of the output~\cite{danishenergyagency2021} 
in our analysis and assume the products to have an average \gls{lhv} of \SI{11.95}{\MWh\th\per\tonne} which is
similar to regular diesel and aviation kerosene.
Fischer-Tropsch synthesis is well established process using fossil syngas and infrastructure as well as experience
from fossil hydrocarbon handling can be directly applied to \glspl{ftf}.

\subsubsection*{\glsfmtfull{lohc}}
\gls{lohc} is the last chemical energy carrier we consider which allows for piggyback transport of hydrogen.
We choose \gls{dbt} as a representative of the variety of \glspl{lohc} available~\cite{niermann2019}.
Between its dehydrogenated (\enquote{unloaded}, \textsc{H0DBT}) and hydrogenated (\enquote{loaded}, \textsc{H18DBT}) form it may 
be loaded with up to \ce{9 H2}:
\begin{equation}
    \ce{9 H2 + C21H20 <--> C21H38}
\end{equation}
For repetitive cycling and favourable (de-) hydrogenation a depth of discharge of \SI{90}{\percent} is more 
favourable~\cite{runge2020} corresponding to \SI{5.6}{\wtpercent} \ce{H2}.
One significant advantage of \gls{dbt} is that its properties do not change significantly between its hydrogenated and 
dehydrogenated form, allowing for the same infrastructure to be reused to achieve a closed \gls{lohc} cycle.
The cost for the \gls{lohc} chemical \gls{dbt} is assumed to be \SI{2264}{\EUR\of{2015}\per\tonne}.
The \gls{lohc} can be easily handled and stored with infrastructure similar to that of commonly traded liquid carbohydrates.
To access the hydrogen stored it has to be dehydrogenated which requires about \SI{28}{\percent} of the hydrogen content 
as energy~\cite{niermann2019} as we assume the necessary heat has to be provided by the \gls{esc} itself and 
is not provided from an external source.

\subsubsection*{\texorpdfstring{\ce{CO2}}{CO2} feedstock}
Methane, methanol and \glspl{ftf} require carbon dioxide (\ce{CO2}) as feedstock for synthesis.
In our model all \ce{CO2} feedstock is sourced from atmospheric \ce{CO2} using \gls{dac}, thus creating a closed carbon 
cycle via the atmosphere between energy carrier synthesis and use.
We do not consider the alternative approach of \gls{cc} at the location of use and back-transport of pure \ce{CO2} to 
the location of energy carrier synthesis via a dedicated \ce{CO2} infrastructure.
This approach would require guaranteed capture of \ce{CO2} from all use cases, which will prove complicated for applications
where concentrated point sources are not available, like in aviation and individual mobility.
Another obstacle to a perfect carbon cycle via \gls{cc} is carbon leakage from imperfect \gls{cc} which would need to
be compensated through \gls{dac} infrastructure to ensure atmospheric carbon neutrality.
Finally the infrastructure for recirculation of \ce{CO2} would for most \glspl{esc} require additional dedicated
infrastructure for \ce{CO2} transport, incurring additional costs and complexity.
While the costs of carbon capture at the location of use and back-transport may not be prohibitive, but such closed 
carbon cycles would require detailed analysis that is out of the scope of this paper.
For the \gls{lohc} \gls{esc} recirculation is considered here as the infrastructure for recirculation of \gls{dbt} 
in the \gls{lohc} \gls{esc} is the same as required for delivery of the hydrogen-loaded energy-carrying \gls{lohc}.

\subsubsection*{Battery and chemical storage}
Storage technologies are essential for balancing the variable nature of \gls{res} electricity, buffering chemical feedstock 
and  storing chemical energy carriers before export.
Storage technologies smoothen the utilisation of downstream processes by buffering
variable upstream processes like \gls{res} or hydrogen production in \gls{res}-follow mode.
Storage capacity expansion is an alternative way to increase process capacities by additionally increasing downstream utilisation 
rates and thus leading to lower \gls{lcoe} if the investment into storage capacity is lower than into process capacity expansion.
In our model electricity from \gls{res} may be stored in a battery buffer storage.
\ce{CO2} as a feedstock gas may be stored in liquefied form.
Hydrogen and methane may be stored for short term buffer storage in a compressed form as their liquefaction process is energy and capital intensive.
Larger amounts of any chemical are only stored in liquid form to reduce the necessary storage volume.
For hydrogen and methane this requires energy intensive and well insulated tanks to store both liquids at cryogenic temperatures.
Underground storage like salt caverns for gases are not considered to keep our \glspl{esc} independent of location and geological conditions.
In comparison ammonia liquefaction and storage is significantly easier as its boiling point is only \SI{-33}{\degreeCelsius} and thus
ammonia may be stored in liquefied form.
Storage tanks for methanol, \gls{lohc} and \gls{ftf} are straightforward as they correspond to today's technologies used for light and heavy hydrocarbons.
The storage technology options available for each \gls{esc} are as shown in \ref{app:esc-visualisations}.
Storage capacities are endogenously determined by the model and represent optimal capacities under the given constraints minimising the objective function.

\subsubsection*{Flexibility of synthesis processes}
In addition to the economic perspective of operating synthesis processes at a high utilisation rate,
some synthesis processes may be designed for continuous operation from a chemical process point of view 
\cite{gotz2016,danishenergyagency2021} and not be suited for flexible operation or standby.
We consider this by assuming a must-run capacity for the methanation and ammonia synthesis processes of \SI{30}{\percent} each,
based on what could potentially be feasible for the methanation~\cite{gotz2016} and Haber-Bosch~\cite{danishenergyagency2021} processes.
Methanol and \gls{ftf} synthesis are assumed to run at a minimum of \SI{94.25}{\percent} capacity corresponding
to a maximum of 3 weeks downtime for e.g. maintenance per year.
The must-run capacity is assumed for the aggregated availability factor of the whole respective process plant, e.g. a must-run capacity of \SI{94.25}{\percent}
translates to a maximum of \SI{5.75}{\percent} of the facilities capacities being unavailable for maintenance or other reasons at the same time (see \ref{app:equations}).

\subsubsection*{Transport: Transmission lines, pipelines and ships}
Transfer of the energy between exporting and importing country plays a significant role due to the different characteristics between transport modes.
\gls{hvdc} transmission lines for transfer of electricity and pipelines for hydrogen and methane gas are already deployed technologies.
These technologies allow for a continuous export-import supply stream.
We consider here average costs, energy demand and distance-related losses for \gls{hvdc} transformers, transmission lines, pipelines and pipeline compressors
for above-ground if geographically possible.
For imports from \gls{ar} and \gls{au} subsea connectors are unavoidable and we consider average subsea \gls{hvdc} transmission line 
and subsea pipeline costs.
The pipeline compressor costs in these few cases are the same as for above-ground pipelines due to a lack of reference numbers
for long-distance deep-sea pipeline connectors.
The numbers may thus prospectively underestimate real compressor costs.
Shipping is the third class of transport modes considered and characterised by a non-continuous and delayed transfer of the energy carrier.
We take this shipping characteristics into account by not allowing for concurrent use of loading and unloading infrastructure
by different (groups of) ships at the same time.
Rather than having ships travel at their (maximum) average cruise speeds and wait for the (un-) loading terminals to become available
we create ex ante shipping schedules with lowered cruise speeds such that the arrival and departure of (groups of) ships does not overlap.
The shipping distance along sea routes affect the shipping duration in the shipping schedules and the energy demands for propulsion, 
optional onboard refrigeration of boil-off gases and finally the necessary number of ships.

\subsection{Countries investigated}

\begin{figure}[!htbp]
    \centering
\includegraphics[width=1.0\linewidth]{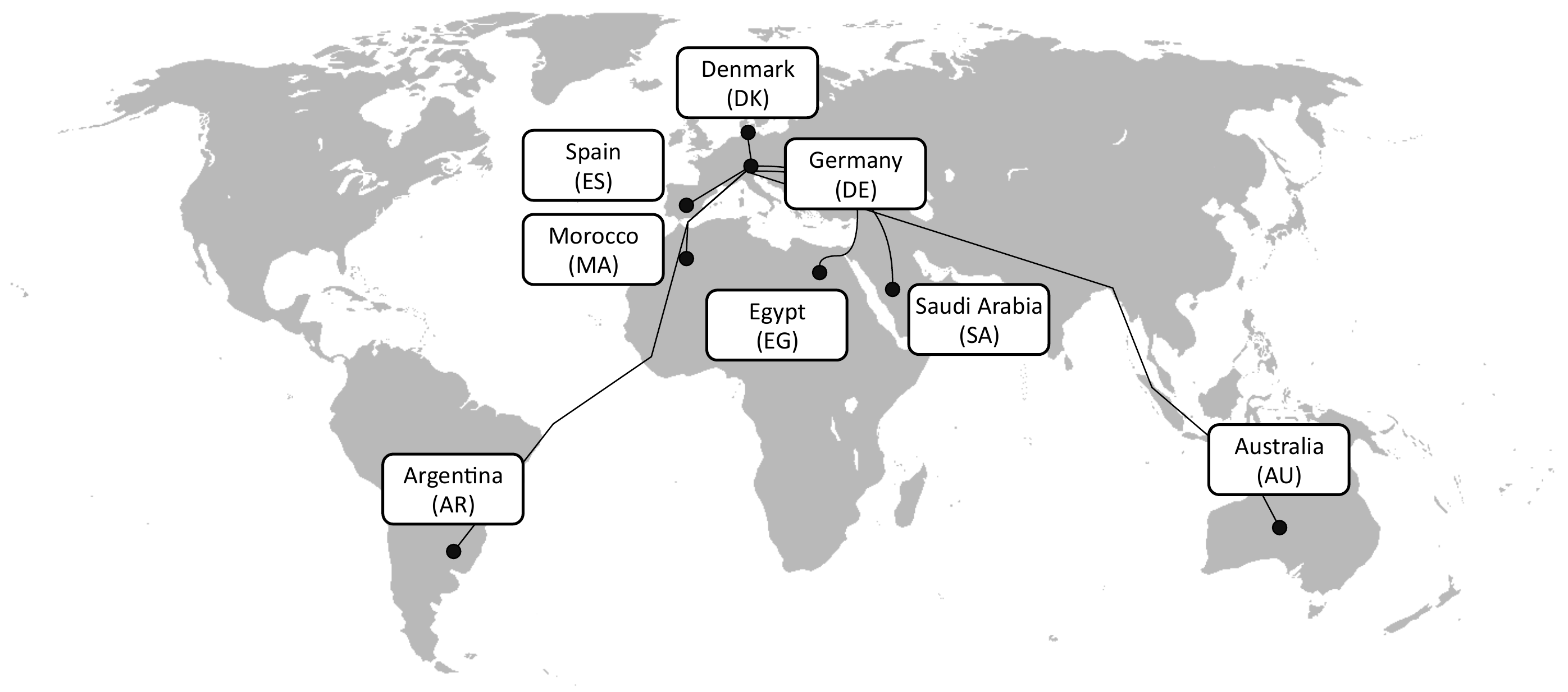}
    \captionwithlegend{Countries considered for export.}{Exports of chemical energy carriers from countries shown to \glsfmtfull{de} are modelled and investigated for nine different \glsfmtfullpl{esc}.}
    \label{fig:map}
\end{figure}

In our study we investigate large scale energy imports to \glsxtrfull{de} from the various countries shown in \autoref{fig:map}.
Germany as a country is interesting, as it heavily relies on energy imports today,
is phasing out nuclear power and has less abundant renewable energy resources compared to other countries.
We assume an energy import volume of \SI{120}{\TWh} based on the estimated hydrogen energy demand (\gls{lhv}) for 2030 in the 
German Hydrogen Strategy~\cite{bundesministeriumfuerwirtschaftundenergie2020}.
On the one hand this number may seem ambitious given the high German fossil energy imports today and 
that its hydrogen production currently is mostly captive or merchant hydrogen from fossil sources without carbon capture.
On the other hand we also presume an increase of this volume by 2040 and 2050 as not only fossil energy carriers but also
industrial feedstock will have to be substituted by hydrogen or other chemicals.
The import volume in our model is considered as annual demand since hydrogen uses and their demand patterns are yet to be known.
This way it is decoupled from a specific demand pattern such with constant baseload or seasonally changing hydrogen demand
and their case-specific buffer storage needs.
As a reference we model and include chemical energy carriers produced from domestic resources in \gls{de}.
The other countries included in our study are:
\begin{enumerate}[label=\alph*.)]
    \item \glsxtrfull{es}, an \gls{eu} country with high solar potentials,
    \item \glsxtrfull{dk}, an \gls{eu} country in close proximity to \gls{de} with high wind potentials,
    \item \glsxtrfull{ma}, \glsxtrfull{eg} and \glsxtrfull{sa}, representative countries in relative proximity to the \gls{eu} with low population densities and high renewable potentials,
    \item \glsxtrfull{ar}, a country repeatedly investigated by similar studies for exports of hydrogen to Japan,
    \item \glsxtrfull{au}, a country discussed for energy exports to \gls{de} and as a possible future powerhouse for Asian countries.
\end{enumerate}

The distance of an export-import route {\gls{esc}} influences the associated costs.
With increasing distance investment costs, transportation energy demand, losses and duration (for shipping) increase.
The lengths of pipelines and transmission lines are based on the as-the-crow-flies distances between the country centres
scaled by different detour factors for transmission lines (\num{1.2}) and pipelines (\num{1.4}).
For \gls{ar} and \gls{au} a land-only connection is not possible.
We therefore estimate the share of distance to be traversed with submarine technologies based on the shortest 
cross-continental distances and scale them with the same detour factors as for land connections.
A comparable approach with detour factors does not work for shipping routes as it does not account for land and water bodies.
Instead we opt to use a freely available data source~\cite{searates2020} to measure the shortest shipping routes.
All resulting distances used are shown in \autoref{tab:region-distances}.

\begin{table*}[htbp]
    \small
    \centering
    \begin{threeparttable}
        \sisetup{uncertainty-separator={\enspace}}
        \captionwithlegend{
            Distances between exporting countries and \glsfmtshort{de} assumed for each \glsfmtshort{esc}.
        }{}
        \label{tab:region-distances}
        \begin{tabular}{llSSSS[table-format=5]}
            \toprule
            {From} & {Code\tnote{a}} & {Distance\tnote{b}} & {\gls{hvdc}~line\tnote{c}} & {Pipeline\tnote{d}} & {Ship\tnote{e}}\\
             & & {[\si{\km}]} & {[\si{\km}]} & {[\si{\km}]} & {[\si{\km}]}\\
            \midrule
            Argentina    & AR & 12280 (3000) & 14736 (3600) & 17192 (4200) & 13056 \\
            Australia    & AU & 14450 (3000) & 17340 (3600) & 20230 (4200) & 20284 \\
            Denmark      & DK & 580 (0)      & 696 (0)      & 812 (0)      & 812   \\
            Egypt        & EG & 3220 (0)     & 3864 (0)     & 4508 (0)     & 6605  \\
            Germany      & DE & 0 (0)        & 0 (0)        & 0 (0)        & 0     \\
            Morocco      & MA & 3330 (0)     & 3996 (0)     & 4662 (0)     & 2938  \\
            Saudi Arabia & SA & 4240 (0)     & 5088 (0)     & 5936 (0)     & 12174 \\
            Spain        & ES & 1600 (0)     & 1920 (0)     & 2240 (0)     & 3587 \\
            \bottomrule
        \end{tabular}
        \begin{tablenotes}
            \scriptsize
            \item Bracketed values indicate the share of submarine distances considered.
            \item[a] Based on ISO 3166-1 alpha-2.
            \item[b] As-the-crow-flies distance between region centres~\cite{google2012}, measured in Google Maps.
            \item[c] Distance times a detour factor \num{1.2}, own estimate.
            \item[d] Distance times a detour factor \num{1.4}, based on~\cite{timmerberg2019}.
            \item[e] Shortest sea route, determined with~\cite{searates2020}.
        \end{tablenotes}
        
    \end{threeparttable}
\end{table*}

\subsection{Electricity supply, demand and supply curves}
\label{sec:res-supply}
For each country we model \gls{res} potentials and time-series based on results from a \gls{gis}-analysis and historical 
weather data using the \gls{gegis} model~\cite{mattsson2021}.
Eligible areas for \gls{pv} and wind installations are determined on a \SI{1}{\km\squared} grid resolution
by exclusion of protected areas, unsuitable land types and areas of high population density.
Areas not within a \SI{400}{\kilo\meter} radius of a \gls{gdp} density of \SI{100000}{\USD\per\km\squared}
are further excluded where the threshold serves as a proxy to grid access and location accessibility.
For a detailed description we refer to~\cite{mattsson2021}, for maps of the resulting eligible area see \ref{app:country-masks}.
Annual capacity factors are determined for all eligible grid cells and types of \gls{res} using
ERA5 reanalysis weather data and data from the \gls{gwa} for 2013 as representative weather year.
Grid cells are then categorised into one of 100 quality classes (\SIrange{0}{100}{\percent} \gls{cf}) for each \gls{res} 
technology based on their annual capacity factor.
From the categorisation the potential of each quality class for each of the three \gls{res} technologies is calculated 
assuming a deployable potential of \SI{1.45}{\MW\per\km\squared} (\gls{pv}) and \SI{3}{\MW\per\km\squared} 
(on-shore and off-shore wind).
This potential is a compromise between technical potential, accessible potentials and social acceptance
used in another study for the European electricity grid~\cite{horsch2018}.
For \gls{pv} these potentials may be conservative for less population dense regions and closer to
the equator where others consider \SI{75}{\MW\per\km\squared} (\gls{pv}) and \SI{8.4}{\MW\per\km\squared} (wind) feasible 
~\cite{fasihi2020}.
For large continuous wind farms power densities may need to artificially be reduced to lessen wake effect penalties~\cite{deutschewindguardgmbh2018}.
These influences become more relevant for exporters with already high potentials and flat supply curves
and therefore should therefore not contribute a major influence on this studies' supply side.

In addition to the potentials we derive hourly generation time-series for each technology and quality class, 
resulting in a total of up to 300 independent \gls{res} time-series for each exporter.

\begin{table*}[!htbp]
    \centering
    \small
    \begin{threeparttable}
        \captionwithlegend{
            Main technology assumptions for \glsfmtshort{res} and electrolysis.
        }{}
        \label{tab:renewables-technology-assumptions}
    \begin{tabular}{llllll}
        \toprule
        Technology     & Lifetime  & CAPEX                            & FOM                 & Density                     & Efficiency \\
        2030/2040/2050 & {[years]} & {[\si{\EUR\of{2015}\per\kW}]}    & {[\si{\percentpa}]} & {[\si{\MW\per\km\squared}]} & {[\%]}     \\
        \midrule
        PV (utility)   & 40/40/40  & 376/329/302                      & 1.9/2/2.1           & 1.45                        & ---        \\
        wind onshore   & 30/30/30  & \num{1035}/978/963               & 1.2/1.2/1.2         & 3                           & ---        \\
        wind offshore  & 30/30/30  & \num{1573}/\num{1447}/\num{1416} & 2.3/2.3/2.3         & 3                           & ---        \\
        electrolysis   & 30/32/35  & 450/330/250                      & 2/2/2               & ---                         & 68/71.5/75 \\
        \bottomrule
    \end{tabular}
    
    \begin{tablenotes}
        \scriptsize
        \item Assumptions based on~\cite{danishenergyagency2020} and ~\cite{danishenergyagency2021} used for \numlist{2030;2040;2050}.
              \gls{capex} reflects the \glsfmtlong{epc} price.
              A full list of all technology assumptions is included in \ref{app:technology-assumptions}.            
    \end{tablenotes}
    \end{threeparttable}
\end{table*}

\begin{figure}
    \centering
\includegraphics[width=1.0\linewidth]{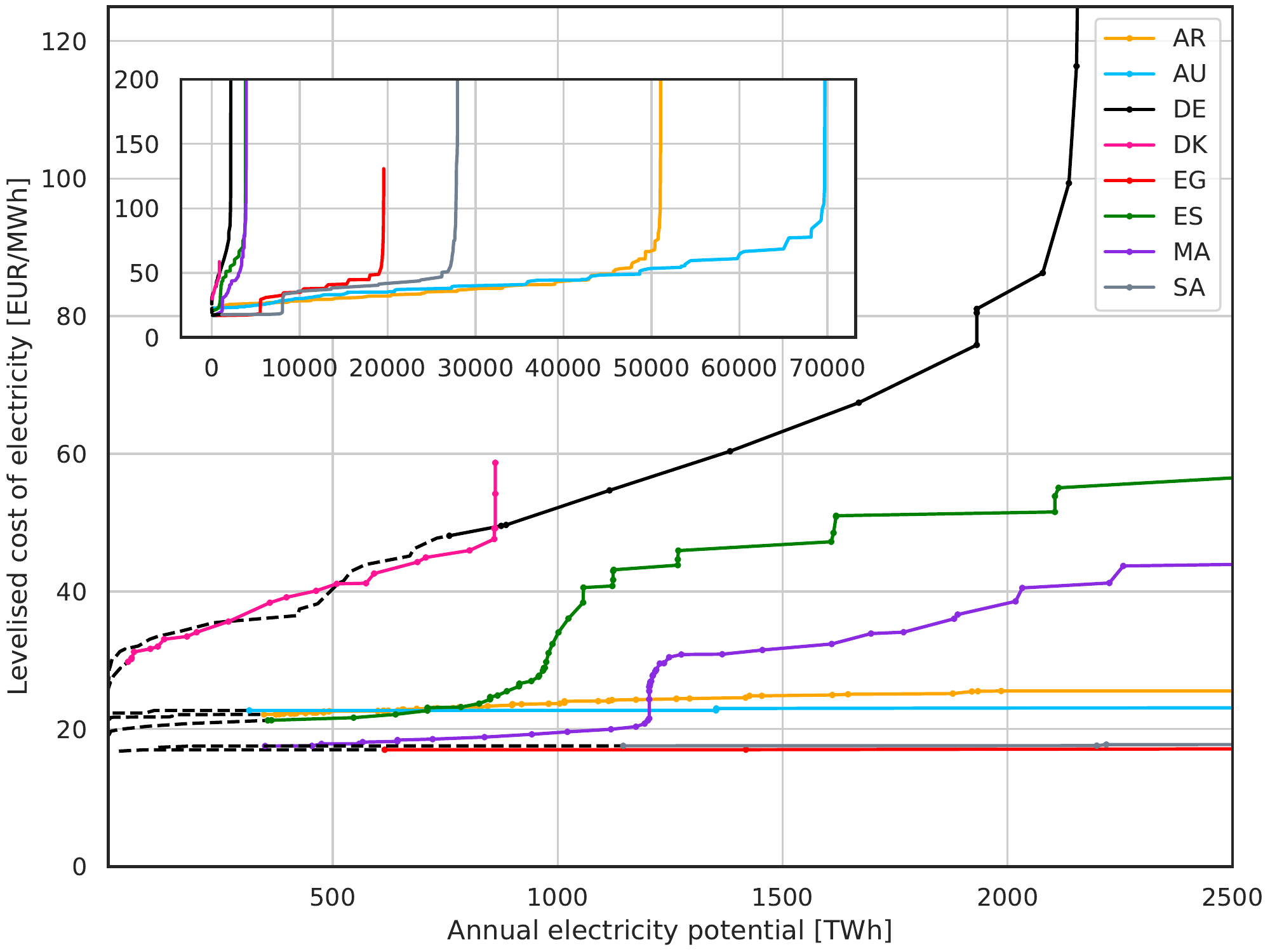}
    \captionwithlegend{
        Modelled electricity supply curves for 2030 at \SI{10}{\percentpa} \glsfmtshort{wacc}.
       }{
        Dashed black parts are reserved for meeting domestic electricity demand and unavailable for export.
        The inlet contains the same plot on a larger scale.
        The visible step-wise increase in \glsfmtshort{lcoe} for \glsfmtshort{es} and \glsfmtshort{ma} is where
        the cheapest electricity potentials from low cost \glsfmtshort{pv} are exhausted and the onshore 
        and offshore wind enter the supply curve.
    }
    \label{fig:supply-curves}
\end{figure}

From the annual generation for each quality class we can calculate each classes respective \gls{lcoe} following
  	\begin{align}
        \text{\gls{lcoe} (\gls{res})}
            &= \frac{\text{Annualised cost}}{\text{Annual generation}}
            &= \frac{\mathrm{CAPEX}}{G(2013)} \cdot  \left[\mathrm{FOM} + \frac{r}{1 - \left(1+r\right)^{-t}}\right]
\end{align}
assuming technology specific parameters (\autoref{tab:renewables-technology-assumptions}), as well as $r=\mathrm{\gls{wacc}}$ 
and the modelled annual electricity generation $G$ in 2013.
By ordering the class potentials based on their \glspl{lcoe} we obtain country specific electricity supply curves.
\autoref{fig:supply-curves} shows examples using \SI{10}{\percentpa} \gls{wacc} and technology (cost) assumptions for 2030.

Using the supply curve we account for projected domestic electricity demand:
We generate electricity demand projections with a machine learning approach implemented by \gls{gegis}~\cite{mattsson2021}.
The demand projections are based on global datasets for GDP, calendar days, temperature from ERA5 and the 
{SSP2-34} \enquote{Middle of the Road} scenario~\cite{riahi2017} for 2050.
This approach extrapolates past demand into the future and cannot account for structural changes like increasing demand through electrification.
The projections are thus to be considered conservative estimates for electricity demand.
The projected demands are shown in \autoref{tab:demand_projections}.

\begin{table}[!htbp]
    \centering
    \small
    \captionwithlegend{
        Used future (2050, projected) electricity demand and actual (2018) for reference.
    }{}
    \label{tab:demand_projections}
    \begin{threeparttable}
    \begin{tabular}{lS[table-format=6.0]S}
        \toprule
        Country      & {demand 2018 (actual)\tnote{a}} & {demand used 2050 (projected)} \\
                     & {[\si{\GWh}]}                   & {[\si{\GWh}]}                  \\
        \midrule
        \glsxtrfull{ar} & 125030                          & 346904                         \\
        \glsxtrfull{au} & 234278                          & 314389                         \\
        \glsxtrfull{dk} & 32865                           & 44854                          \\
        \glsxtrfull{eg} & 150579                          & 615351                         \\
        \glsxtrfull{de} & 533177                          & 759065                         \\
        \glsxtrfull{ma} & 29678                           & 122419                         \\
        \glsxtrfull{sa} & 322373                          & 1145638                        \\
        \glsxtrfull{es} & 245426                          & 355416                         \\
        \bottomrule
    \end{tabular}
    \begin{tablenotes}
        \item[a] Source: \cite{u.s.energyinformationadministrationeia2018} .
    \end{tablenotes}
    \end{threeparttable}
\end{table}

We consider the domestic electricity demand by removing the equivalent volume and \gls{res} with the 
lowest cost \gls{res} supply from our model.
This corresponds to reserving the capacities with lowest expected \glspl{lcoe} for domestic use.
The respective volumes are marked in the supply curves by the black dashed lines for the shown example 2030 and \SI{10}{\percentpa} \gls{wacc}.
Changes to the \gls{res} technology costs (year assumption) and \gls{wacc} assumption affect the order of \gls{res}
and therefore change \gls{res} with associated time-series available for export.
Considering the shape of the supply curves in \autoref{fig:supply-curves}, 
this approach noticeably affects the \glspl{mcoe} for \gls{de} and \gls{dk}.

\subsection{Choice of \glsfmtshort{wacc}}

The choice of the costs of capital influences all cost calculations and is therefore crucial for meaningful results.
\cite{schyska2020} showed how configurations of a cost-optimised European electricity system
change significantly from changes to \gls{wacc} assumptions, especially non-homogeneous, country-specific assumptions.
At the same time we are aware of the possible bias this might introduce, cf.~\cite{egli2019}.

To retain comparability we assume time and technology independent \gls{wacc}.
We further choose to assume \SI{10}{\percentpa} \gls{wacc} for all investments within all \glspl{esc} independent of the exporting country.
This choice is founded on the assumptions used by \gls{irena}, where the authors used inhomogeneous \gls{wacc} assumptions of
\SI{10}{\percentpa} for non-\glsxtrshort{oecd} countries and \SI{7.5}{\percentpa} for \glsxtrshort{oecd} countries and China in~\cite{irena2020a}.
\gls{wacc} for local \gls{res} projects usually depend on the technologies used and on individual project as well as country-specific risks~\cite{steffen2020}.
It will therefore be interesting to see how \gls{wacc} will develop for highly vertically integrated, multi-national and mixed-technology \glspl{esc} as 
presented here.

\subsection{Technical model structure}

We use a multi-step workflow hard-linked using \texttt{snakemake}~\cite{molder2021}.
In the first part of the workflow we utilise \gls{gegis}~\cite{mattsson2021} to determine potentials for \glspl{res}, 
\glspl{res} generation time-series and predict electricity demand.
In the second part of the workflow we implement the \glspl{esc} in PyPSA~\cite{brown2018} and combine them with
the \glspl{res} potentials, \glspl{res} time-series and demand predictions into one dedicated PyPSA model for each combination of \gls{esc} and exporting country.
The structure is also visualised in \autoref{app:graphical-overview}.

\section{Results}

Results are compared with a focus on their \glsxtrlong{lcoe} and \glsxtrlong{lcoh}.
The \glspl{lcoe} represent the costs for delivering \SI{1}{\MWh} of energy in the form of the energy carrier 
of the respective \gls{esc}, i.e. \ce{H2 (g)}, \ce{CH4 (g)}, \ce{NH3 (g)}, methanol or \gls{ftf}.
For the \glspl{lcoh}, costs are compared for delivering \SI{1}{\MWh} of \ce{H2 (g)} to the importer.

We first present \glspl{lcoe} for all \glspl{esc} and exporting countries for 2030.
Then we show the development of \glspl{lcoe} based on technology cost projections up to 2050.
Cost compositions for the \glspl{esc} are examined and main cost drivers discussed.
We then continue by looking at the \glspl{lcoh} and by discussing sensitivities of the \glspl{esc}
based on an sensitivity analysis for two selected scenarios.
The sensitivity analysis shows an expected strong dependence to the choice of \gls{wacc} on 
two selected \glspl{esc} from \gls{es} to \gls{de}.
We therefore also present \glspl{lcoe} and \glspl{lcoh} for \numrange{2030}{2050} under
a more optimistic choice for \gls{wacc} of \SI{5}{\percentpa}.
Additional results focusing on supply chain efficiency, curtailment rates and installed \gls{res} capacities
are discussed in the \ref{app:esf-curtailment-lcoe} and \ref{app:res-capacity-and-generation}.
Presented results for \glspl{lcoe} and \glspl{lcoh} are included as tabular form in \ref{app:results-lcoes} 
and \ref{app:results-lcohs}.

\subsection{Energy import costs for 2030 to 2050}

\begin{figure*}[!htbp]
    \centering
\includegraphics[width=1\linewidth]{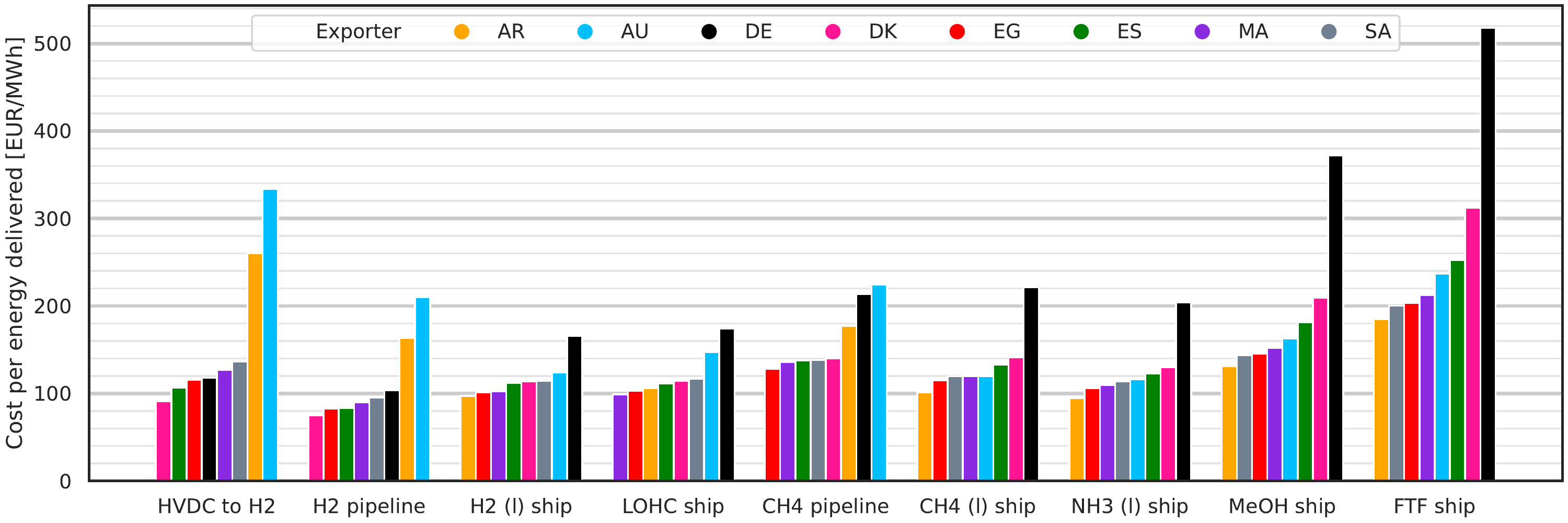}
    \captionwithlegend{
        \glsfmtshort{lcoe} in 2030 assuming \SI{10}{\percentpa} \glsfmtshort{wacc} by \glsfmtshort{esc} and exporter.
    }{
        \glsfmtshortpl{lcoe} are per \si{\MWh\th} delivered to \glsfmtshort{de}.
        Lowest cost options are imports via \glsfmtshort{hvdc} with subsequent electrolysis in \glsfmtshort{de} and \ce{H2} pipelines.
        For more complex and energy intensive \glsfmtshortpl{esc} on the right the order of preference is different compared to the static
        import options with imports from \glsfmtshort{ar} in all but one case being the cheapest.
    }
    \label{fig:lcoenergyperesc-exp}
\end{figure*}

\glspl{lcoe}, i.e. total system cost per \si{\MWh\th} delivered to \gls{de}, are shown in \autoref{fig:lcoenergyperesc-exp}
for 2030, \SI{10}{\percentpa} \gls{wacc} and all exporting countries.
The lowest cost options for import are by \ce{H2} pipeline from \gls{dk} at \SI{75}{\EURpMWh} 
and at \SI{83}{\EURpMWh} from \gls{eg} and \gls{es}.
All three \glspl{esc} take advantage of the low losses and investments associated with \ce{H2} pipelines
as static transport option and the short to medium transport distances to \gls{de}.
With costs of \SI{104}{\EURpMWh} domestic \ce{H2} production in \gls{de} is less attractive than these imports.
\ce{CH4} imports by pipeline are the least cost attractive option of the three (\gls{hvdc} to \ce{H2}, \ce{H2} \& \ce{CH4} pipeline)
static transport connection \glspl{esc}.
Methane in particular may be imported at lower costs as \ce{CH4 (l)} by ship rather than by \ce{CH4 (g)} pipeline.
Cost performances of the shipping \glspl{esc} for \ce{H2 (l)}, \gls{lohc} and \ce{NH3 (l)} are 
similar to those for \ce{CH4 (l)}.
They are followed by ship-based imports of methanol and finally imports of \gls{ftf} by ship.
There are some outliers for \gls{ar} and \gls{au} for the static \gls{hvdc} and pipeline connection \glspl{esc}.
Reasons are the long transport distances for both exporters and the related high investment required
for the \gls{hvdc} and pipeline infrastructure.

An order of preference for the exporting countries can be identified:
For statically connected \glspl{esc} short and mid-distanced exporters like \gls{dk}, \gls{es} and \gls{eg} are preferable.
For ship-based \glspl{esc}, \gls{ar} offers the lowest cost imports of energy to \gls{de},
making use of its good-quality \gls{res}, followed by imports from \gls{eg} again.
The least favourable position is taken by domestic production in \gls{de}:
Generally for each \gls{esc} exists an alternative where the same energy carrier can be sourced for \SI{75}{\percent} of
the cost of domestic production in \gls{de}.
This development is driven by the low quality \gls{res} with high electricity costs and the approach
we used to reserve \gls{res} capacities for domestic electricity demand.
The approach reserved all \gls{pv} potentials and only left wind resources for chemical production 
(cf.~\ref{app:res-capacity-and-generation}).
Wind energy in Germany suffers from low output between June and September which the model compensates by 
over expanding wind capacities to keep supplying the synthesis processes.
This is cheaper than using storage by batteries, which are not economical for storing more than a 
few hours worth of electricity demand, and there is no longer term electricity storage in the model.
The remaining time of the year excess electricity is curtailed, causing high curtailment rates (cf.~\ref{app:esf-curtailment-lcoe}) for \gls{de}.
This leads to domestic production in \gls{de} to be competitive with imports only if \ce{H2} is produced without further processing.
Under this condition the absence of need to transport chemicals internationally can compensate for the higher production cost in \gls{de}.

\begin{figure*}[!htbp]
    \centering
\includegraphics[width=1.\linewidth]{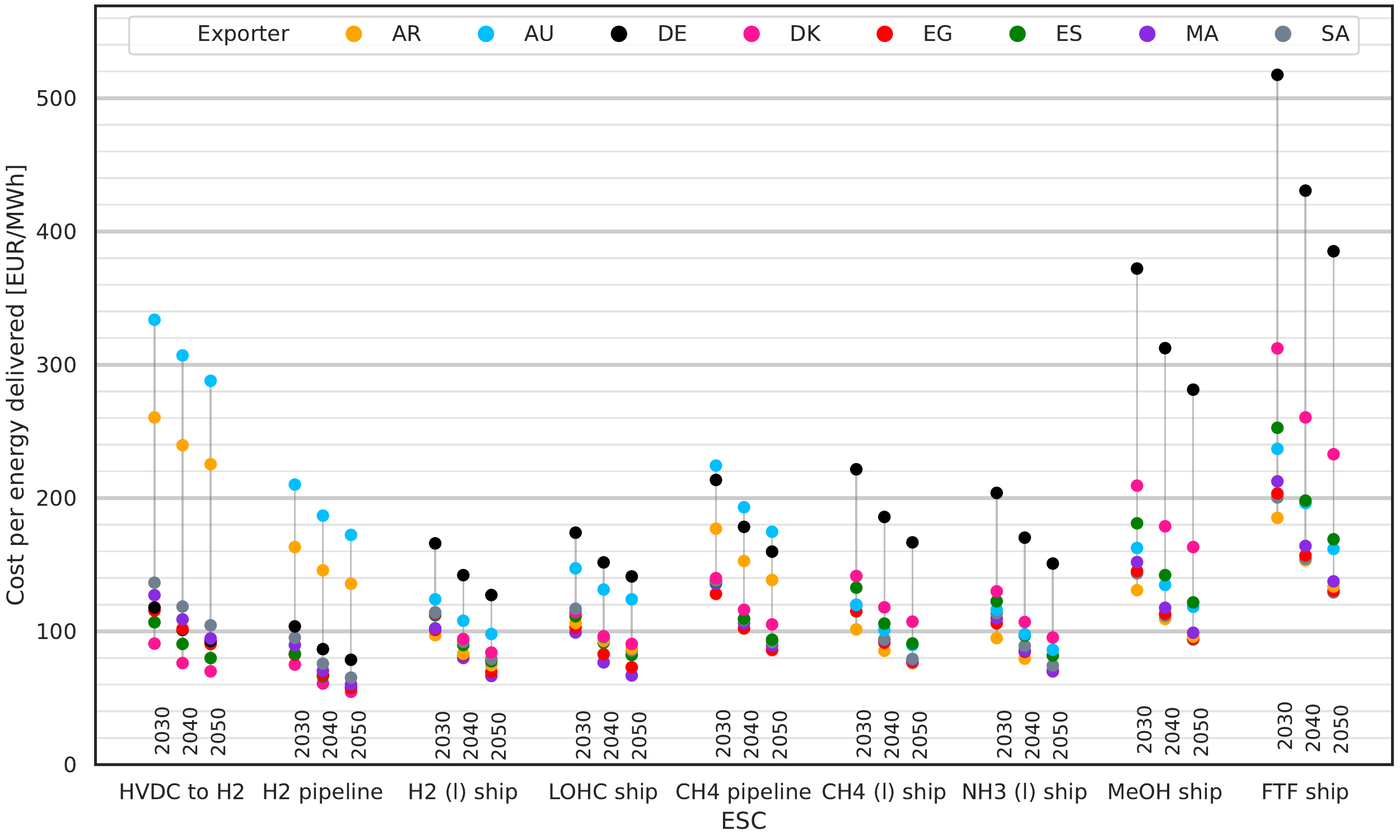}
    \captionwithlegend{
        \glsfmtshort{lcoe} in \numrange{2030}{2050} assuming \SI{10}{\percentpa} \glsfmtshort{wacc} by \glsfmtshort{esc} and exporter.
    }{
        \glsfmtshortpl{lcoe} are per \si{\MWh\th} delivered to \glsfmtshort{de}.
        The lowest cost import options are \ce{H2 (g)} imported by pipeline and electricity imports by \glsfmtshort{hvdc} with subsequent electrolysis.
        Methanol and \glsfmtshort{ftf} imports experience the strongest cost decrease linked to the anticipated cost reduction for synthesis and \glsfmtshort{dac}.
    }
    \label{fig:lcoenergyperesc2030-2040-2050}
\end{figure*}

In \autoref{fig:lcoenergyperesc2030-2040-2050} the \glspl{lcoe} are shown declining in accordance with decreasing technology cost by 2050.
Some of the projected \glspl{lcoe} decrease more strongly than others, most notable for methanol and \gls{ftf}.
While \ce{CH4 (l)} transport is cost competitive with \ce{CH4} pipeline transport due to technological developments 
in the \gls{lng} industry over the past decades, for hydrogen the more complex of the transport chain 
and higher energy demands for liquefaction continue to make \ce{H2} pipelines the preferred option for \ce{H2} imports in the future.
Noteworthy are the spreads between different \glspl{esc} and years.
Neglecting domestic production in \gls{de}, the spread of \glspl{lcoe} for \ce{H2 (l)} or \gls{lohc} ship imports
is low compared to the spread of methanol and \gls{ftf} imports.
It is also worth noting that the exporting country preference order does not change much between the years.
Highest and lowest cost exporters stay the same and only some reordering in the cost mid-field can be seen where some countries benefit from anticipated cost 
developments more than others, see the mid-field options for methanol or \ce{NH3} shipping.
This observation translates to no significant changes in the cost compositions of the \glspl{esc},
but rather just a general more or less homogeneous decrease of the total costs due to the cost technology reductions and electrolysis efficiency gains by 2050.

\subsection{Cost composition and drivers}
\begin{figure*}[!htbp]
    \centering
\includegraphics[width=1.\linewidth]{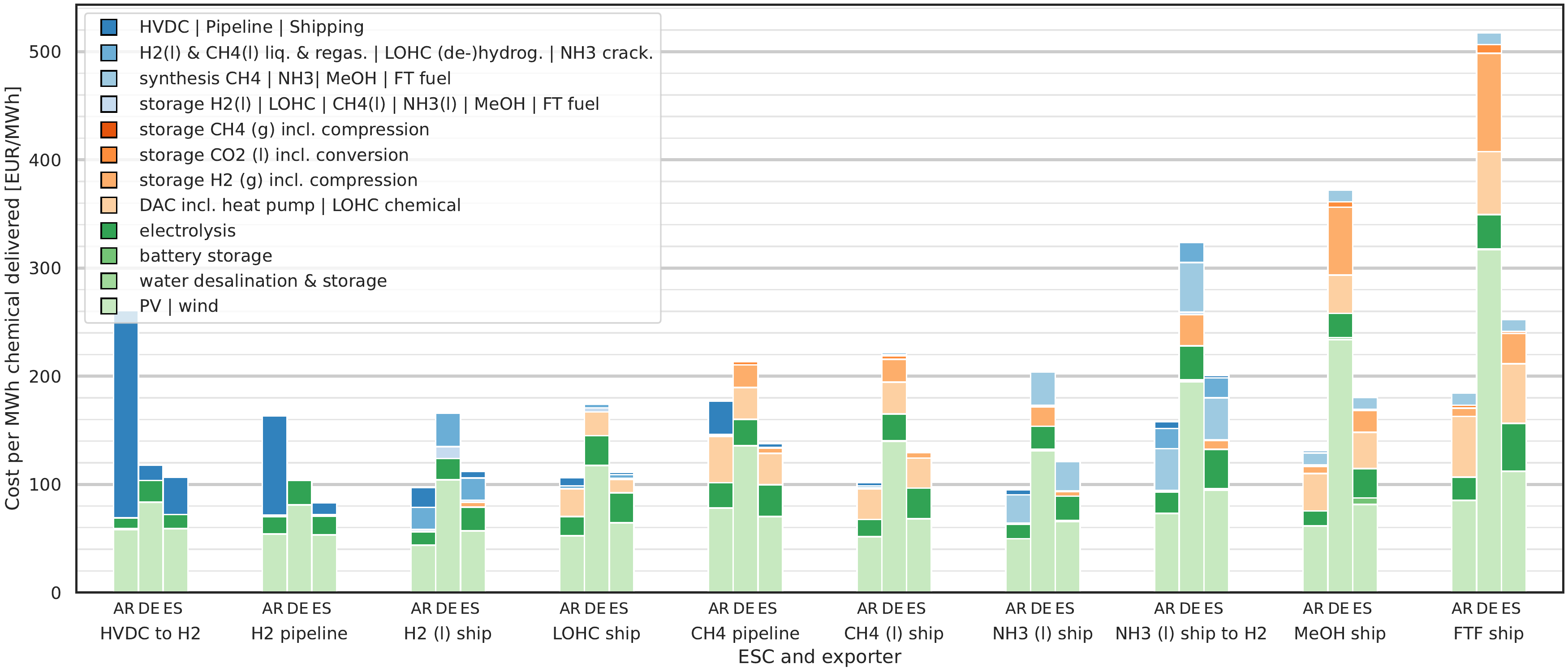}
    \captionwithlegend{
        Cost composition of selected \glspl{esc} for 2030 at \SI{10}{\percentpa} \glsfmtshort{wacc}.
    }{
        Simpler, less energy intensive \glsfmtshortpl{esc} are favourable for short distances, 
        e.g. domestic sourcing from \glsfmtshort{de} and imports from \glsfmtshort{es}, where low costs associated with 
        transport can compensate high \glsfmtshort{res} costs.
        More complex, energy intensive \glsfmtshortpl{esc} allow long distances exports, e.g. from \glsfmtshort{ar},
        to become cost competitive due to lower \glsfmtshort{res} costs and trading more complex molecules with 
        higher energy intensity for lower transport cost.
    }
    \label{fig:cost-compositionsselected-esc-exp}
\end{figure*}

Cost drivers can be individually identified for all \glspl{esc} by investigating their cost compositions.
\autoref{fig:cost-compositionsselected-esc-exp} shows the cost compositions for three selected countries,
comparing exports from \gls{ar} and \gls{es} with domestic production in \gls{de}.
Additionally a tenth \gls{esc} is included where ammonia is decomposed back into hydrogen for investigating hydrogen import costs discussed in the next section.
The costs shown are the annualised cost per component.
Energy costs are not attributed to the components and instead cause higher upstream investments
for conversion steps or \gls{res} capacities to keep the \glspl{esc} self-sufficient.

\gls{res} generally make the single largest cost contribution between \SIrange{39}{55}{\percent} (interquartile distance).
The specific cost contribution depends on the \gls{esc}, exporter, the local energy demand and conversion processes involved.
Electrolysis plays only a minor role with on average \SI{5}{\percent} of the costs in the shown cases.
The costs for synthesis or liquefaction processes and \gls{dac} (for \ce{CO2}-based \glspl{esc}) have a higher contribution than electrolysis.
Related to the synthesis processes is the necessity for chemical feedstock storage required
to operate the synthesis processes at and above their must-run capacities.
The contribution from \ce{H2} and \ce{CO2} feedstock above-ground steel tank storage is higher for most \glspl{esc} than the contribution from electrolysis.
Some feedstock storage is also deployed for domestic provisioning of \ce{H2 (l)} in \gls{de} but it cannot be seen for all \ce{H2 (l)} \glspl{esc}.
The costs associated with shipping are negligible in comparison to \gls{hvdc} and pipeline costs.

From \autoref{fig:cost-compositionsselected-esc-exp} we can also identify cost drivers for the bad performance of \gls{ar} as 
an exporter with the static pipeline and \gls{hvdc} to \ce{H2} \glspl{esc}:
The \gls{hvdc} and \ce{H2} pipeline connections have high investment costs due to the long distance between \gls{ar} and \gls{de}.
For the \ce{CH4} pipeline connection the investment costs are only a third of the \ce{H2} pipeline, but the \gls{lcoe}
is driven by energy demand of the \ce{CH4} pipeline which is assumed to use the transported \ce{CH4 (g)} as energy source.
This energy demand for transport requires additional capacities for \ce{CH4} synthesis, \ce{H2} electrolysis and \ce{CO2} capture
as well as \gls{res} capacities, thus making pipeline transport less attractive across medium and long distances compared
to \ce{H2} and \ce{CH4} shipping.
For the \gls{lohc} \gls{esc} the long shipping distance increases shipping time and thus makes larger volumes of 
the \gls{lohc} chemical necessary which is reflected in the share of cost of the \gls{lohc} chemical.
In comparison to \gls{es} as an exporter the necessary \gls{lohc} chemical investment volume is \num{2} times higher for \gls{ar} as exporter.
Despite this the \gls{lcoe} for imports from \gls{ar} by \gls{lohc} are lower than from \gls{es} due to the country-specific 
differences in available \gls{res}:
While the \gls{lohc} investment volume for \gls{es} is lower compared to \gls{ar},
the \gls{lcoe} in the electricity supply curve are higher for \gls{es} than for \gls{ar}.
For domestic production in \gls{de} the \gls{lcoe} for electricity from \gls{res} are driving the \glspl{esc} \gls{lcoe} even higher.
Furthermore the lower \gls{cf} of \gls{res} in \gls{de} lead to higher capacities for electrolysis needed during peak production 
but with lower overall utilisation rate.
The case of \gls{lohc} is the only ship-based \gls{esc} where imports from \gls{ar} do not offer the lowest \gls{lcoe}.
Comparing the \gls{lohc} \gls{esc} and the other carbon-based shipping \glspl{esc} with the \ce{NH3 (l)} \gls{esc}
shows the advantages of an energy carrier with low investment costs on the synthesis molecule (\ce{N2}) compared
to the high cost for the \gls{lohc} chemical and \ce{CO2} sourcing and handling.
Inflexibilities in the synthesis processes do not directly become apparent from the cost composition figure but
show up indirectly by the need for increased buffer storage capacities and overcapacities of components upstream
in the \glspl{esc}, e.g. \gls{res}.
A good example for this are the \gls{ftf} and \ce{CH4 (l)} \glspl{esc} where the more inflexible \gls{ftf} synthesis
leads to higher \gls{res} investments per \si{\MWh} with higher curtailment and a different mix of \gls{res} capacities 
(see \ref{fig:esfvslcoenergy-and-curtailment-10pwacc-2030}).

Battery storage is usually only deployed with limited capacities, able to sustain the \gls{esc} only for a few hours.
In a few cases larger capacities of battery storage are deployed with a notable influence on the \gls{lcoe}.
Battery deployment can mainly be seen for \gls{hvdc}, methanol and \gls{ftf}-based \glspl{esc}, e.g. the \gls{hvdc} to \ce{H2} \gls{esc} 
for \gls{es} in 2040 (see \ref{app:cost-composition-other-scenarios-years}).
Investment into battery storage in the model coincides with higher shares of \gls{pv} in the generation mix to increase the 
utilisation factor of downstream infrastructure, e.g. \gls{hvdc} links, and to provide continuous electricity supply for must-run methanol and
\gls{ftf} synthesis processes.
While sea water desalination is an important aspect for ensuring a sustainable production environment
its costs and electricity needs do not contribute in a significant way to the final energy carrier cost.

Leverages for decreasing overall costs lie in reduction of storage costs by either direct reduction of the investment costs
or usage of different technologies, e.g. cavern storage instead of steel-tank storage for \ce{H2}.
An indirect leverage is the flexibility of synthesis processes as the must-run processes drive the storage volumes in our results,
something which was shown for methanol by \cite{chen2021}.
Increasing flexibility of processes and enabling lower must-run capacities as well as hot-standby would decrease the required storage volumes.
Simultaneously it would increase the cost share of the synthesis process, such that cost improvements of the said processes 
could have a higher impact in lowering total product costs.

\subsection{Hydrogen import costs for 2030 to 2050}
In addition to comparing the costs of energy delivered, we can also compare the \glspl{esc} based on 
their import costs for delivering hydrogen.
For this we calculate the levelised cost of delivering \SI{1}{\MWh} \ce{H2 (g)} using adapted versions of 
the previously discussed \glspl{esc}.
Comparing the cost of hydrogen rather than the cost of energy is useful for applications which either 
require pure hydrogen like hydrogen fuel cells or processes requiring hydrogen as an industrial feedstock.

The \glspl{esc} used are identical where the \glspl{esc} already delivered \ce{H2 (g)}.
The \ce{CH4}, \ce{NH3} and \ce{MeOH} \glspl{esc} are extended by cracking processes for converting the energy carrier
to \ce{H2 (g)} with their respective energy demand and investment costs.
The additional cracking processes are \gls{smr}, \gls{msr} and ammonia cracking.
Additional water demand of the cracking processes is neglected.
The extra process steps and their location in the \glspl{esc} are shown in \ref{app:esc-visualisations}.
We exclude cracking of \gls{ftf} as methanol can be considered an equivalent choice for the purpose 
of being a liquid, carbon-based hydrogen carrier under ambient conditions but with easier synthesis.
\Glspl{ftf} are better suited as drop-in fuel replacements.

\begin{figure*}[!htbp]
    \centering
\includegraphics[width=1.\linewidth]{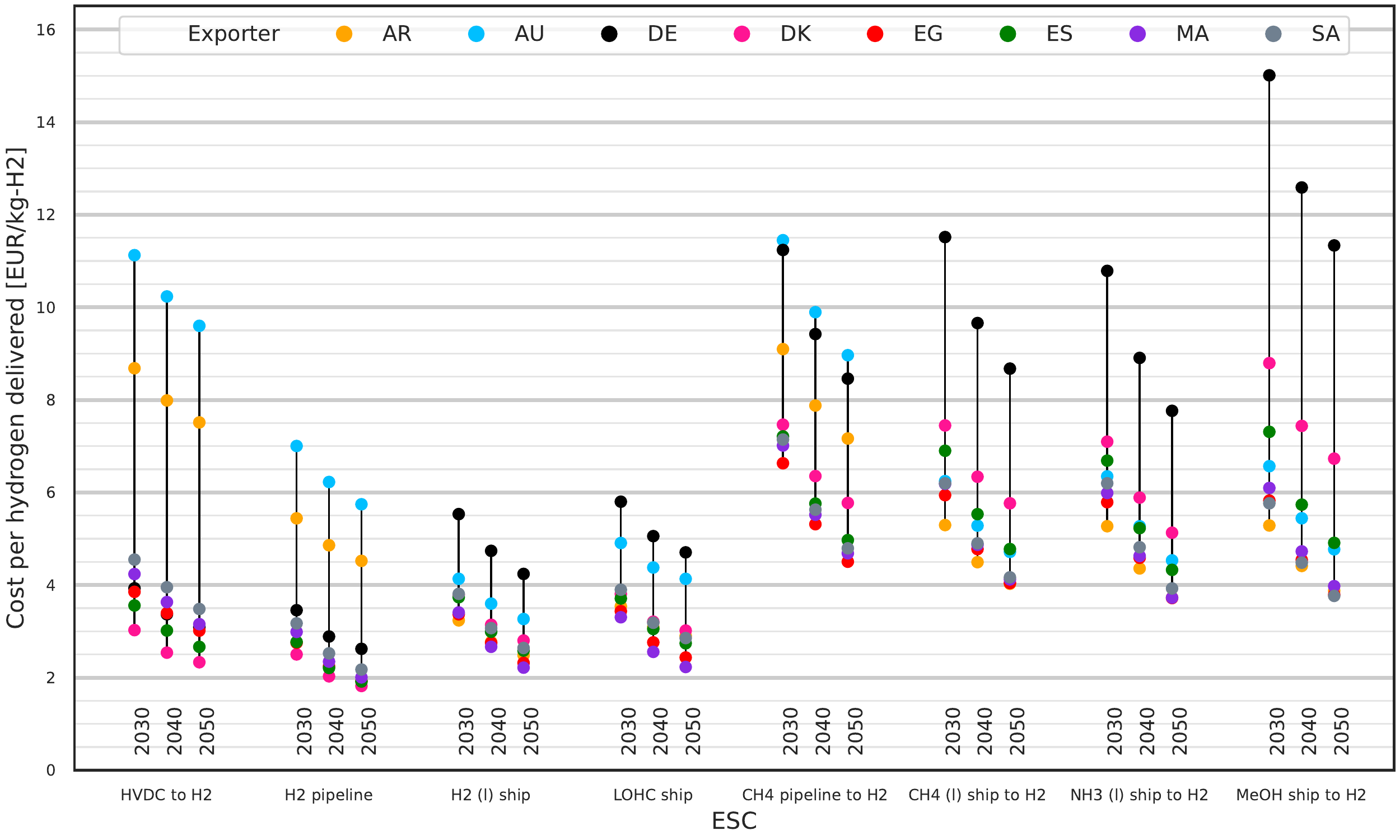}
    \captionwithlegend{
        \glsfmtfull{lcoh} for \numrange{2030}{2050} using extended \glsfmtshortpl{esc}.
    }{
        Direct hydrogen imports have lower \glspl{lcoh} than the alternative \glsfmtshortpl{esc} using 
        methane, methanol or ammonia as energy carrier.
        The extended \glspl{esc} include extra process steps for cracking of the energy carrier to deliver hydrogen.
        All costs assuming \SI{10}{\percentpa} \glsfmtshort{wacc}.
    }
    \label{fig:lcohperesc2030-2040-2050}
\end{figure*}

Resulting \glspl{lcoh} for 2030 to 2050 are shown in \autoref{fig:lcohperesc2030-2040-2050}.
There is a clear cost advantage for the four \glspl{esc} which import hydrogen directly (left side) compared to 
the four \glspl{esc} requiring an additional cracking step (right side).
Lowest cost imports are from \gls{dk} by \ce{H2 (g)} pipeline at \SI{2.5}{\EURpkgHydrogen} in 2030
and \SI{1.8}{\EURpkgHydrogen} in 2050.
Other exporters located in close proximity, i.e. \gls{es} and the \gls{wana} countries, offer hydrogen at only
slightly higher \gls{lcoh} and can thus be considered comparable alternatives.
For non-static imports via ship \ce{H2 (l)} and \gls{lohc} \glspl{esc} exports from \gls{ar} and to some extent 
from \gls{au} are also attractive.
By 2050 as most exporting countries show similar \glspl{lcoh} the question of exporter makes no big difference
from a techno-economic point of view as there does not seem to be an inherent advantage to a specific exporter.
For the remaining non-\ce{H2}-based \glspl{esc} the additional conversion steps required for \ce{CO2}-based and 
the \ce{NH3} \glspl{esc} drive their energy demand and investment costs, leading to less attractive \gls{lcoh} 
starting at \SI{5.3}{\EURpkgHydrogen} in 2030 decreasing to \SI{3.7}{\EURpkgHydrogen} in 2050 from various 
exporters via \ce{NH3} and methanol.

If long-term storage were additionally taken into account, the ammonia and methanol \glspl{esc} 
would have an advantage over the \ce{CH4}-based \glspl{esc} due to easier long-term bulk storage.
This advantage would translate into lower \glspl{lcoh} in comparison to the other \glspl{esc} with 
the advantage increasing with the storage duration.
Examining the results for alternative exporter options, it shows that in general for each lowest cost option
of an \gls{esc} an alternative exporter with similar or slightly higher \gls{lcoh} exists.
Neglecting the extreme outliers for \gls{ar}, \gls{au} and \gls{de}, the spreads for pipeline based and 
direct hydrogen imports are lower than for ship based \ce{NH3}, \ce{CH4 (l)} and methanol imports.
The higher spreads translate to a higher uncertainty of hydrogen import costs when trade relations change and 
a switch in exporting country become necessary.
Choosing a chemical energy carrier and \gls{esc} for scale up based on lower spreads in \gls{lcoh} and 
existing low cost exporter alternatives ensures the opportunity for increased competition and potential 
exporter substitution in the future.

\subsection{Sensitivity analysis}

\begin{figure*}[!htbp]
    \begin{subfigure}{.5\textwidth}
        \centering
\includegraphics[width=1.\linewidth]{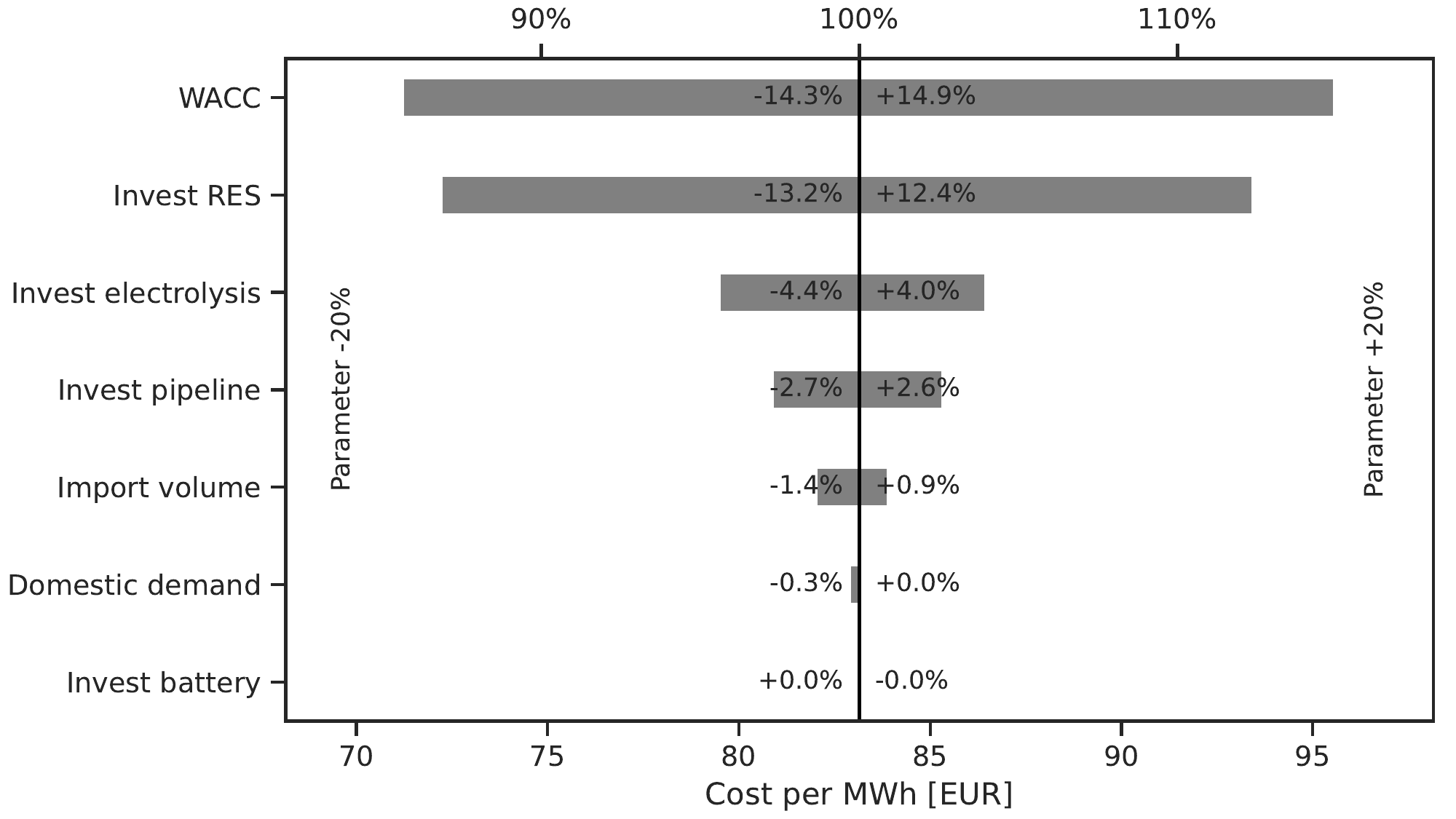}
\end{subfigure}
\begin{subfigure}{.5\textwidth}
        \centering
\includegraphics[width=1.\linewidth]{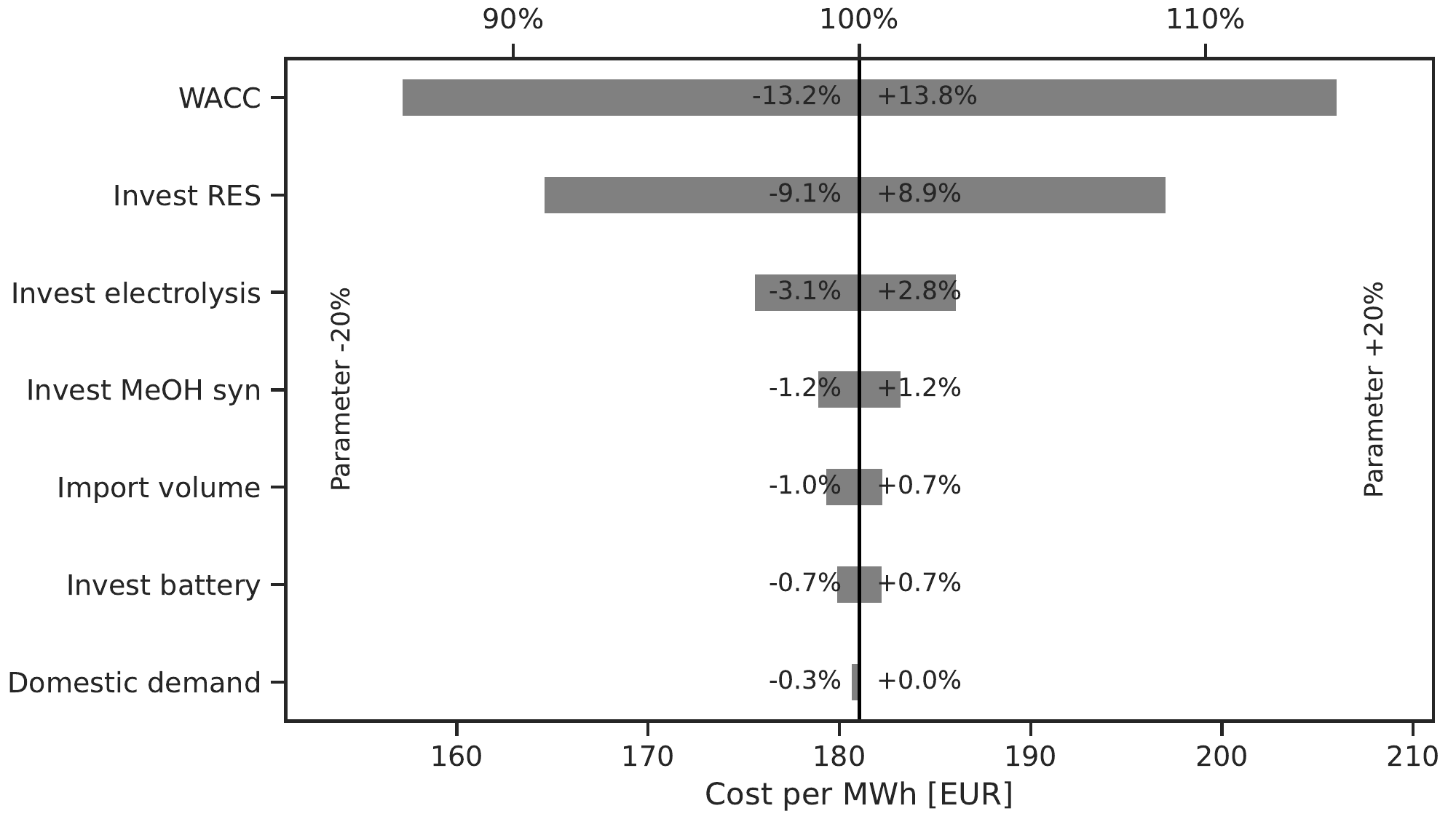}
\end{subfigure}
    \captionwithlegend{
        Sensitivities for two \glspl{esc} (left) \ce{H2 (g)} pipeline and (right) methanol by ship from \gls{es} to \gls{de} for 2030.
    }{
        Parameters listed on the left y-axis were varied by \SI{\pm 20}{\percent}.
        The x-axis shows the resulting \gls{lcoe} after variation and the relative change.
        The highest impact on the \gls{lcoe} can be attributed to the choice of \gls{wacc}
        followed by the cost of renewables and hydrogen electrolysis.
    }
    \label{fig:sensitivities2030}
\end{figure*}

The sensitivities to exogenous parameter changes are visualised in \autoref{fig:sensitivities2030}.
Sensitivities are dependent on the scenario (year, \gls{esc}, exporter) by model design.
We present quantified results for two selected scenarios of imports from \gls{es} to \gls{de} in 2030,
by \ce{H2 (g)} pipeline and shipping of methanol.
The exogenous parameters to be considered in the sensitivity analysis were pre-selected based on 
which parameters were expected to have the highest influence on the \gls{lcoe}.
For both \glspl{esc} one additional major cost contributor based on the cost composition analysis 
was added, i.e. \gls{capex} for \ce{H2} pipeline including compressors and methanol synthesis.

The results show the highest influence on \gls{lcoe} to be from the choice of \gls{wacc},
followed by variations to \gls{capex} of \gls{res} and then the \gls{capex} of electrolysis 
or the \gls{esc} specific contributor (investment in pipeline or \ce{MeOH} synthesis).

Reduction of \gls{capex} for batteries has no effect for the pipeline-based \gls{esc} and only a small 
symmetric effect on the methanol \gls{esc}.
Variations of domestic demand and import volume are negligible.
This is to be expected because the relevant supply curve range does not show considerable changes 
to the \gls{lcoe} in the relevant area around \SI{500}{\TWh} of electricity demand.

\subsection{Lower \glsfmtshort{wacc} scenario}

\begin{figure*}[!htbp]
    \centering
\includegraphics[width=1.\linewidth]{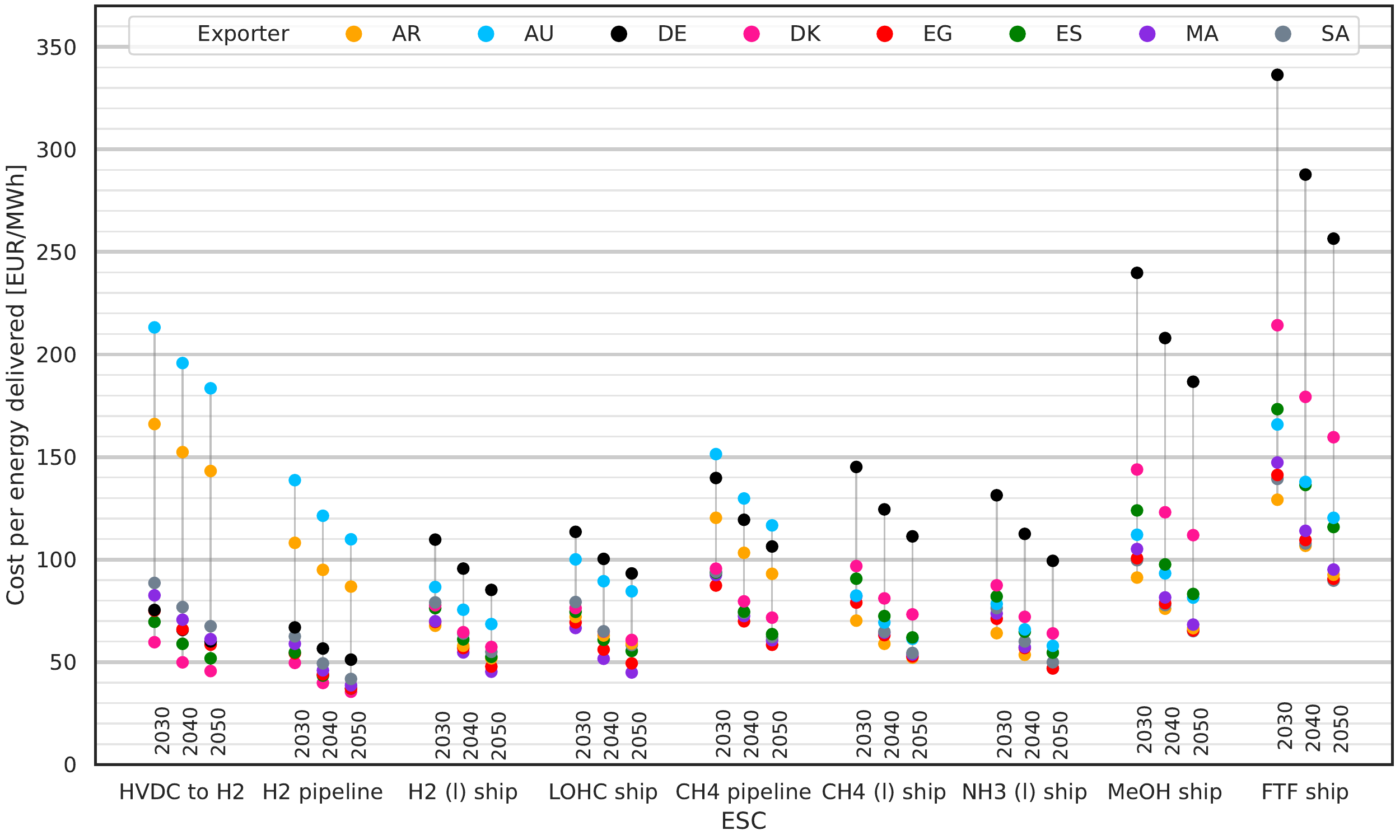}
    \captionwithlegend{
        \glsfmtshortpl{lcoe} for \numrange{2030}{2050} at a reduced \glsfmtshort{wacc} of \SI{5}{\percentpa}.
    }{
        Lowering the \glsfmtshort{wacc} reduces \glsfmtshortpl{lcoe} by around \SI{35}{\percent}.
    }
    \label{fig:lcoenergyperesc2030-2040-2050lowhomogeneous}
\end{figure*}

\begin{figure*}[!htbp]
    \centering
\includegraphics[width=1.\linewidth]{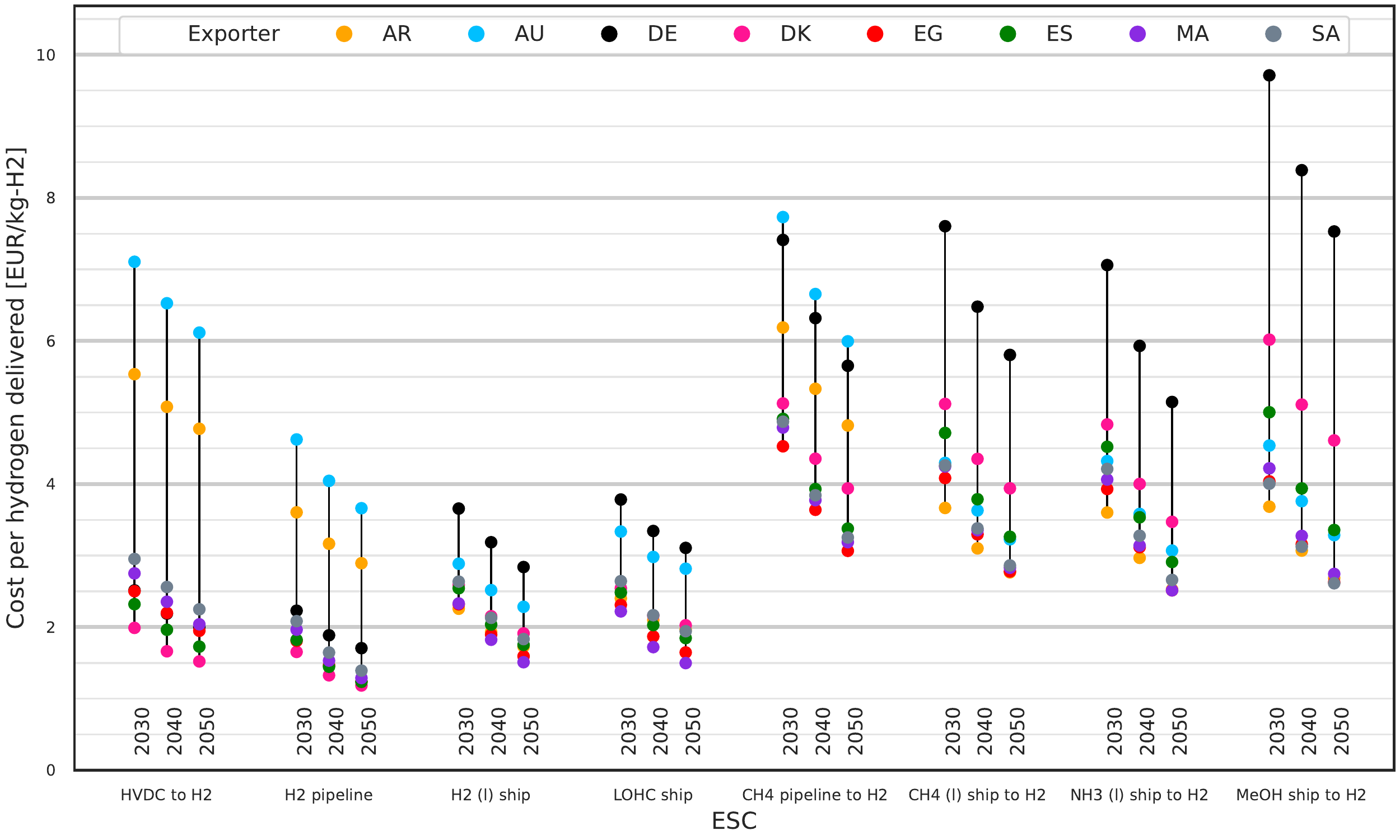}
    \captionwithlegend{
        \glsfmtshortpl{lcoh} for \numrange{2030}{2050} at a reduced \glsfmtshort{wacc} of \SI{5}{\percentpa}.
    }{
        Lowering the \glsfmtshort{wacc} reduces \glsfmtshortpl{lcoh} by around \SI{35}{\percent}.
    }
    \label{fig:lcohperesc2030-2040-2050lowhomogeneous}
\end{figure*}

As indicated by the sensitivity analysis, the choice of \gls{wacc} has the highest influence on 
the two \glspl{esc} analysed.
Following this result we revisit the earlier analysis of \glspl{lcoe} and \glspl{lcoh} and rerun 
our model assuming \SI{5}{\percentpa} \gls{wacc} ($=\SI{-50}{\percent}$).
Resulting \glspl{lcoe} are shown in \autoref{fig:lcoenergyperesc2030-2040-2050lowhomogeneous} 
and \glspl{lcoh} in \autoref{fig:lcohperesc2030-2040-2050lowhomogeneous}.
The lower assumption is optimistic in comparison to other investigations like~\cite{irena2020a},
but plausible and in line with recent reports like~\cite{irena2021,iea2021} and in the face of similar or lower 
\gls{wacc} reported for e.g. \gls{pv} projects~\cite{steffen2020}.
For large scale projects the assumption is also more reasonable as the projects include arguably high national interests and therefore support.
The lower \gls{wacc} leads to reductions of \glspl{lcoe} and \glspl{lcoh} each by around \SI{35}{\percent}.
This exceeds most of the original scenario reductions seen between \numrange{2030}{2050} due to technological learning.
Under these more optimistic \gls{wacc} assumptions one can find within each \gls{esc} one exporter and for each
exporter one \gls{esc} with \gls{lcoe} below \SI{100}{\EURpMWh}.
Similarly options for \glspl{lcoh} below \SI{4}{\EURpkgHydrogen} are available for all \glspl{esc} and exporters.
What is left unchanged is the order of preference for exporting countries within each \gls{esc};
given that the change to \gls{wacc} affects all exporters the same way this result is not surprising.

Lowest \glspl{lcoe} are \SI{50}{\EURpMWh} in 2030 to \SI{36}{\EURpMWh} in 2050 by \ce{H2} pipeline from \gls{dk}
which are also the lowest \glspl{lcoh} at \SI{1.7}{\EURpkgHydrogen} in 2030 to \SI{1.2}{\EURpkgHydrogen} in 2050.
By 2050 more than 20 \glspl{esc} offer options for imports of hydrogen at \SI{2}{\EURpkgHydrogen} (\SI{60}{\EURpMWh}) or lower,
most of them being by a static transport connection via \ce{H2} pipeline or \gls{hvdc} but also including some shipping options.

\subsection{Comparison with today's commodity prices}

\begin{figure*}[!htbp]
    \centering
\includegraphics[width=1.\linewidth]{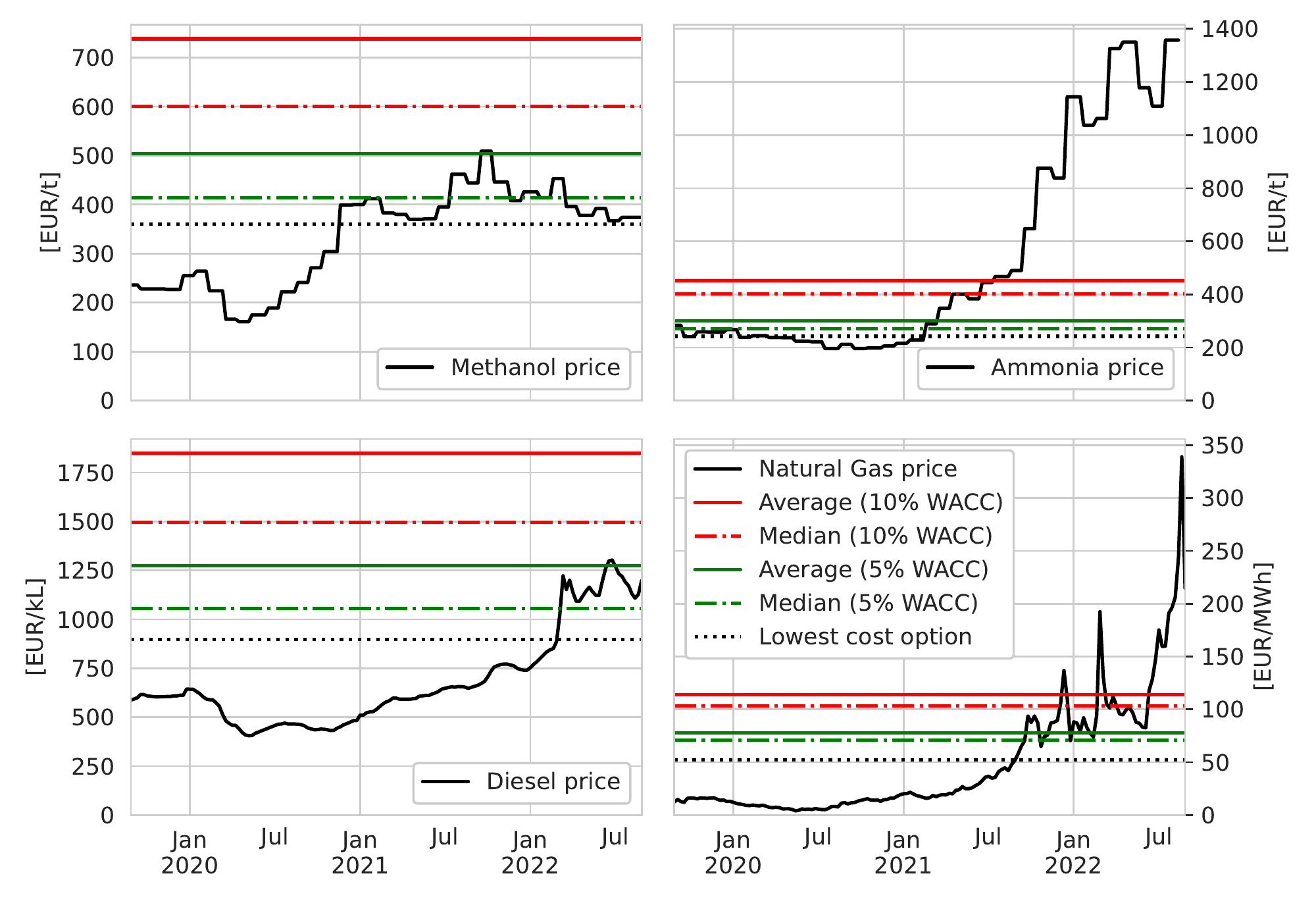}
    \captionwithlegend{
        Comparison of market prices of current fossil-based commodities with modelled costs for synthetic \gls{esc}-based alternatives.
    }{
        Market prices are for 
        Methanol based on MMSA Europe Spot FOB~\cite{methanolinstitute2022},
        Ammonia based on German export prices for ammonia~\cite{statistischesbundesamt2022},
        Natural gas based on Dutch TTF C1 future~\cite{investing.com2022},
        \glsfmtshort{ftf} based on EU Diesel prices without taxes~\cite{europeancommission2022}
        .
        Statistics for natural gas are based on the costs for the \ce{CH4} pipeline and shipping \ce{CH4 (l)} \glspl{esc}.
        The median value thus represents four exporter options for \gls{ftf}, ammonia and methanol.
        For methane the median value represents eight exporter options.
    }
    \label{fig:commodity-cost-comparison}
\end{figure*}

We evaluate the competitiveness of our \glspl{esc} by comparing the future cost scenarios against 
today's market prices for the fossil-based counterpart commodities.
The comparison between market prices and costs is not strictly valid because it ignores price formation in 
markets, but it can give a useful indication for the economic attractiveness of the \glspl{esc}.
\autoref{fig:commodity-cost-comparison} shows commodity prices relative to the average and median \glspl{lcoe}
under 2050 technology assumptions with \SI{5}{\percentpa} and \SI{10}{\percentpa} \gls{wacc}.

In our results methanol becomes available starting at \SI{360}{\EUR\per\tonne} from \gls{sa} which is around \SI{50}{\percent}
above 2020 market prices but within the range of market prices from 2021 and later.
The current market prices are also within the range of the median methanol cost, indicating at least four exporter options
for methanol within that cost range.
Under less favourable financing conditions, cost differences exceed \SI{100}{\percent} and even the highest historical market prices
are well below the \SI{600}{\EUR\per\tonne} median for methanol at \SI{10}{\percentpa} \gls{wacc}.
The constellation is similar for imports of \gls{ftf} which hit break-even with the European average Diesel 
2022 market price under favourable \gls{wacc} assumptions but are not attractive compared to 2021 and earlier market prices
or under higher \gls{wacc}.

Market prices of ammonia and natural gas are strongly correlated with natural gas, it being the major cost-driving feedstock
for today's ammonia production.
Until the middle of 2021 prices for natural gas and ammonia were moderate and about \SI{50}{\percent} below the costs of the
lowest cost option available from our \glspl{esc}.
In the second half of 2021 market prices started to increase and reached up to ten-fold of previous levels during the energy
crisis in Europe.
In the long-term market prices in Europe for natural gas and therefore also for ammonia can be expected to stay well above
pre-2021 levels as pipeline-based gas imports are substituted with \gls{lng}.
At higher market prices our modelled import costs are within the cost range of being attractive.
Even median import costs for the less favourable \SI{10}{\percentpa} \gls{wacc} scenario are below currently seen market prices.

The differences between costs and market prices can also be bridged through \ce{CO2} prices on the direct emissions of fossil-based energy carriers,
which may also make higher \glspl{lcoe} with unfavourable financing conditions (\SI{10}{\percentpa} \gls{wacc}) more attractive.
Bridging e.g. a cost difference of \SI{500}{\EUR\per\kilo\litre} for diesel/\gls{ftf} would be equivalent to a \ce{CO2} price of \SI{187}{\EURptCD}.
A difference of \SI{100}{\EUR\per\tonne} on methanol could be compensated by a \ce{CO2} price of \SI{73}{\EURptCD}.
For natural gas a \ce{CO2} price of \SI{126}{\EURptCD} could cover a difference of \SI{25}{\EURpMWh} which is the difference
seen between mid July 2021 spot market prices and median \glspl{lcoe} from our scenarios.
Such \ce{CO2} prices are realistic when compared to the estimated \ce{CO2} emission prices for 2030 of \SI{129}{\EURptCD} 
under stricter \gls{eu} regulation \cite{pietzcker2021} and are well within the range of the estimated social 
cost of carbon from \ce{CO2} emissions in \gls{de} of \SIrange{195}{250}{\EURptCD} (\numrange{2020}{2050}) \cite{astridmatthey2020}.

This shows that under fast-track technology deployment that makes the 2050 cost projections appear earlier and 
under favourable financing conditions leading to low \gls{wacc}, the differences in \gls{lcoe} to the fossil chemical energy 
carrier alternatives could in some cases be overcome in the near future.
It also shows the important role of enabling an environment of lower financing cost (\gls{wacc}) to reduce the costs.
Until the cost gap between synthetic \gls{res}-based and fossil energy carriers is closed,
a purely cost-driven switch from fossil to synthetic fuels is improbable.
Supporting policies, e.g. \ce{CO2} prices and fuel quotas, or softer factors which incentivise change, e.g. adjusted customer preferences,
exporter independence and diversification or business self-regulation following e.g. \glsfmtfull{csr} or \glsfmtfull{esg} criteria,
will be required to drive the change.

\section{Discussion}

\subsection{Limitations}
The results we presented in this study are subject to a range of limitations.
Some general limitations are due to the nature of the methods used while others arise from our model assumptions.

\gls{gis}-based analysis at low resolutions for determining \gls{res} potentials can over- or underestimate real potentials 
as it ignores local siting limitations below the resolution limit or not contained in the data input, e.g. terrain.
The \gls{res} potentials and generation are likewise affected by using 2013 as representative weather year which 
disregards multi-year weather and climate change induced variations on \gls{res} generation and costs.
By including exogenous technology development and learning for costs which use fixed projections for technology capacities
being deployed in the future, we ignore the influence our \glspl{esc} capacities could have on world-wide learning and development.
This limitation similarly extends to process efficiencies which are exogenously fixed and learning independent.
For cost assumptions the validity and uncertainties cannot be assessed ex ante, a pitfall famously
encountered by cost projections for \gls{res} technologies like \gls{pv} which in hindsight were often too high \cite{victoria2021}.
Also some technologies might realise technological maturity faster or slower than others leading to earlier or later 
cost decreases in some \gls{esc} thus influencing the time-horizon of the results presented.
There could also be path dependencies where existing infrastructure influences the choice of \glspl{esc}, such as
existing \gls{lng} terminals leading to an advantage for \gls{sng} imports.
Lastly assuming identical \gls{wacc} for all countries and \glspl{esc} increases comparability but is not realistic;
\glspl{wacc} are generally country- and project-specific.
More specific to our model we conducted investment optimisation for greenfield conditions in an islanded system.
This assumption may not be appropriate where significant infrastructure for reuse or co-use already exists.
Such cases are to be expected e.g. for electrical transmission lines and pipeline systems in the \gls{eu}.
Integrating energy exports with existing systems rather than deploying islanded systems can open opportunities
e.g. for reuse of existing heat sources at lower costs than sourcing of the energy from inside an \gls{esc}.
Reuse of existing infrastructure and integration both would lower the costs along an \gls{esc}.
A similar effect should be expected from integration of technologies along the \glspl{esc}, 
e.g. heat generated by electrolysis used for \gls{dac}, which was not considered either.
Further we use simplified transport assumptions:
Energy transport properties only scale by distance using representative average values which ignore e.g. topography,
and we neglect shipping terminal costs.
For forwarding electricity and chemicals between production locations and centralised facilities we assume
copperplate-like transport without connection costs or losses on the exporter side and likewise for distribution systems 
on the importer side.
On the demand side we presume an annual energy demand for domestic electricity as well as chemical energy carrier demand 
without specific demand pattern.
This may cause unrealistically high amounts of low-cost, highly correlated electricity from \gls{pv} to be used for 
domestic demand rather than some of its peak supply for the production of chemical energy carriers.
The lack of demand pattern for the chemical energy carriers waives the need for short-term or long-term storage prior to 
end use where non-\ce{H2} energy carriers may be better suited than any pure \ce{H2} option.
For the demand we further exclude differences in conversion efficiencies from energy carriers to energy services for 
better comparison.
Potential competing demand by imports of other countries which may compete for the same \glspl{esc} is omitted as well
as a to be expected increasing demand in \gls{de} exceeding \SI{120}{\TWh}.
Aggregated demand may exceed available supply potential from some countries, e.g. \gls{dk}, or regions, e.g. \gls{eu}, cf.~\autoref{fig:supply-curves}.
The decisions for transport, distribution and energy demand are expected to lead to an underestimation of \glspl{lcoe}.
Another boost to \glspl{lcoe} can be expected as other nations start to import equally significant volumes 
of energy carriers, thus driving \glspl{mcoe} in the supply curve and competition for best \gls{res} sites,
unequally affecting \glspl{lcoe} from some exporters more, e.g. \gls{dk}, than others, e.g. \gls{au}, \gls{ar}.
Attempting a generalisation for diverse distribution and end-use cases appears inadequate to us as it
may lead to potentially misleading or easily misunderstood results.
Instead we offer basic results on which further studies using sector or project specific parameters may be undertaken
for which our results and model can act as a starting point.

In the broader picture we have considered imports to a single country, Germany, in our study.
In reality there is to be expected a complex web of trade and competition between importing and exporting countries and individual actors.
Our study is a first step into this direction.

\subsection{Comparison with other studies}
Opportunities to compare our modelling results with other studies are limited to the various different 
system boundaries common in literature.
\cite{fasihi2017} investigated import of \gls{ftd} and \gls{sng} from the Maghreb region.
For 2030 they arrived at \SI{85.3}{\EURpMWh\lhv} for \gls{ftd} and for regasified \gls{sng} \SI{98.8}{\EURpMWh\lhv} assuming \SI{5}{\percentpa} \gls{wacc} without \ce{O2} sale benefits.
Compared to our results with \SI{5}{\percentpa} (and \SI{10}{\percentpa}) \gls{wacc} for imports from \gls{ma} in 2030,
their costs are lower than ours reaching \SI{147}{\EURpMWh} (\SI{213}{\EURpMWh}) for \gls{ftf}
and within a similar range for \ce{CH4 (l)} by ship at \SI{82}{\EURpMWh} (\SI{120}{\EURpMWh}).
Differences can be found in \cite{fasihi2017} assuming \gls{dac} \gls{capex} 
to be a third of what we assumed and use a lifetime of 30 years compared to 20 years.
They also assumed \SI{37}{\percent} lower \gls{capex} for \gls{ae}, low cost cavern hydrogen storage rather than steel tanks,
higher \gls{ae} efficiencies and made use of heat integration into their process.
Fasihi et al.~\cite{fasihi2021} estimated 2050 costs for local ammonia production to be 
\SIrange{260}{300}{\EUR\per\tonne\ammonia} for the best sites including \gls{ar}, \gls{au} and the \gls{wana} region.
For \enquote{most habitable regions of the world} they estimate costs of around \SI{450}{\EUR\per\tonne\ammonia}.
The costs we arrived at in our study covers a similar range with costs (exclusing \gls{de}) between \SIrange{242}{330}{\EUR\per\tonne\ammonia} 
for the optimistic \gls{wacc} case of \SI{5}{\percentpa} and \SIrange{360}{492}{\EUR\per\tonne\ammonia} for the conservative \gls{wacc} of \SI{10}{\percentpa}.
The main differences between the studies are are the different assumptions on \gls{wacc} with \cite{fasihi2021} assuming \SI{7}{\percentpa}
and their significantly lower \gls{capex} assumptions for hydrogen and battery storage.
In terms of \gls{res} system compositions the authors saw a \gls{pv} dominated system with complementary onshore wind deployed, while we see
in our study quite different combinations of \gls{pv}, offshore and onshore wind being deployed depending on the exporter and \gls{esc}.

Results from Niermann et al.~\cite{niermann2021} for hydrogen imports from Algeria by \gls{hvdc}, ship with \ce{H2 (l)} and 
the \gls{lohc} \gls{dbt} show significantly higher import costs at \SI{6}{\percentpa} \gls{wacc} when compared with our 
results for \gls{ma} in 2030 with \SI{5}{\percentpa} \gls{wacc}:
They report \SI{11.5}{\EURpkgHydrogen} for \gls{lohc} shipping (we: \SI{2.2}{\EURpkgHydrogen}),
\SI{13.2}{\EURpkgHydrogen} for \ce{H2 (l)} shipping (we: \SI{2.3}{\EURpkgHydrogen})
and \SI{15.6}{\EURpkgHydrogen} for \gls{hvdc} imports (we: \SI{2.8}{\EURpkgHydrogen})
The stark cost difference is driven by their assumptions of a different electrolysis technology (\gls{pem}) with different technological parameters,
fixed electricity costs of \SI{50}{\EURpMWh\el} and inclusion of seasonal storage as well as distribution costs,
i.e. assumptions which are expected to drive costs for pure hydrogen storage (gaseous or liquid) and 
investment costs for the \gls{lohc} chemical substantially.

Lastly compared to \cite{heuser2020} we see comparably flat supply curves for regions with large \gls{res} potentials.
The lowest costs for \ce{H2 (l)} exports from Patagonia (\gls{ar}) seen by Heuser et al.~\cite{heuser2020} are at 
\SI{3.06}{\EURpkgHydrogen} excluding shipping cost.
The authors further estimate the shipping costs to add another \SI{0.53}{\EURpkgHydrogen}.
These results are significantly higher than our results for 2050 with \SI{2.5}{\EURpkgHydrogen} and \SI{1.7}{\EURpkgHydrogen} 
at \SI{10}{\percentpa} and \SI{5}{\percentpa} \gls{wacc} respectively.
The difference in cost can be partially attributed to the inclusion of the collection infrastructure on the exporter side 
necessary for transporting the energy carrier from the distributed points of production to the coast for exports.

Concluding the comparison with existing literature yields consistent results where comparison is possible within a broader range.
Differences to literature can be tracked and explained based on significantly different technology, cost and model assumptions.

\section{Conclusion}

In this paper we modelled large scale, islanded production and export of chemical energy carriers
to \glsxtrfull{de} from \glsxtrfull{au}, \glsxtrfull{ar}, \glsxtrfull{dk}, \glsxtrfull{eg}, \glsxtrfull{es}, 
\glsxtrfull{ma} and \glsxtrfull{sa}.
We compared these \glsxtrfullpl{esc} with their equivalents for domestic production of chemical energy carriers in \glsxtrlong{de}.
For the nine different ESCs we minimised greenfield investment costs to determine the \glsxtrfullpl{lcoe} per \si{\MWh\th}
of the specific chemical energy carriers as well as \glsxtrfullpl{lcoh} per \si{\kg\hydrogen} after dehydrogenation of the energy carriers.
We determined and used local \glsxtrfull{res} potentials via \glsxtrshort{gis}\nobreakdash-analysis, considered the influence of local electricity demand 
on exporters' supply curves and included the intermittency of \gls{res} sources via modelled hourly time-series.

In all investigated scenarios domestic sourcing of the individual chemical energy carriers in \gls{de} is among the most 
expensive options.
Sourcing through \glspl{esc} from other countries is in almost all cases cheaper.
Under conservative assumptions of \SI{10}{\percentpa} \gls{wacc} we find the lowest \glspl{lcoe} (\glspl{lcoh}) for 2030 to be
\SI{75}{\EURpMWh} (\SI{2.5}{\EURpkgHydrogen}) from \gls{dk} by \ce{H2 (g)} pipeline.
Under optimistic assumptions of \SI{5}{\percentpa} \gls{wacc} the costs reduce to \SI{50}{\EURpMWh} (\SI{1.7}{\EURpkgHydrogen}).
With technological development, costs further decrease by 2050 to \SI{55}{\EURpMWh} (\SI{10}{\percentpa} \gls{wacc})
and \SI{36}{\EURpMWh} (\SI{5}{\percentpa} \gls{wacc}).
Other than \glsxtrlong{dk}, imports from \glsxtrlong{es} and \glsxtrlong{wana} by hydrogen pipeline are also attractive,
e.g. at between \SIrange{37}{42}{\EURpMWh} (2050, \SI{5}{\percentpa} \glsxtrshort{wacc}).
With optimistic \SI{5}{\percentpa} \gls{wacc} assumptions, import options for any to all of the investigated chemical energy 
carriers and from all exporters become available at cost of \SI{100}{\EURpMWh} or lower.
Importing \glspl{ftf} by ship is the most expensive \gls{esc} with the largest cost range we investigated,
with the costs ranging \SIrange{90}{256}{\EURpMWh} under best assumptions (2050, \SI{5}{\percentpa} \gls{wacc}).
The lowest cost options and exporters for importing energy are also the lowest cost options for imports of hydrogen.
While better financing conditions and fast-track technology development lead to lower costs for imported energy and hydrogen, 
they do not lead to changes in in the order of preference regarding technology and exporter if they affect all exporters and 
\gls{esc} similarly.

As a rule of thumb \glspl{lcoe} increase with the complexity and energy intensity of an \gls{esc}.
Costs for \gls{res} are the major cost driver usually contributing \SIrange{39}{55}{\percent} to the total cost of an \gls{esc}.
Electrolysis costs play only a minor role with on average \SI{5}{\percent} of the costs of an \gls{esc}.
Costs for \glspl{esc} using ammonia, methane, methanol and \glspl{ftf} are additionally driven by the costs for synthesis 
and costs for hydrogen storage using above-ground steel tank, which is used to ensure feedstock availability for the inflexible synthesis processes.

Our findings support the notion that no best one-size-fits-all solution exists for chemical energy carrier imports.
In fact no single exporter or \gls{esc} shows a unique techno-economic advantage over the others,
leading to the conclusion that preference for one energy carrier, \gls{esc} and exporter should not be given 
solely based on small differences in \gls{lcoe} and \gls{lcoh}.
This allows supply of energy carriers and feedstocks to be sourced from a diverse selection of countries, without affecting costs too strongly.
For large volume imports of energy and pure hydrogen, imports by \ce{H2 (g)} pipeline or electricity by \gls{hvdc} with domestic electrolysis
are favourable.
Alternatives incur only slightly higher costs.
Our results support the notion that the choice of chemical energy carrier should be based on the requirements for end-use including short-term 
and long-term storage necessity, energy service conversion efficiency and distribution logistics rather than the cost alone.
Accordingly future analysis of full \glspl{esc} could provide more insights by using specific demand assumptions 
and distinguishing between specific energy service or chemical needs.
Such analysis could further include local production (captive) scenarios compared to our nation-level analysis.
Finally, qualities of \glspl{esc} like reliability of supply, long-term cost predictability, local value chains and
employment generation, which are generally not represented by the investment costs, need to be investigated and considered.

To overcome the limitations of this study and increase insight into energy imports, future research should focus on
differentiating \gls{wacc} based on exporter and technologies, include spatially resolved \gls{res} and collection 
infrastructure, include additional relevant energy carriers and \gls{esc} designs like \ce{CO2}-cycling and import 
of secondary, energy-intense products like refined-iron or other valuable hydrocarbons and investigate sensitivities
of results also to technical aspects like process flexibilities and synthesis must-run conditions.

\section{Acknowledgements}
We thank Niclas Mattsson of Chalmers University of Technology, Sweden, for the valuable exchange and help with \glsxtrfull{gegis}.
We also like to express our gratitude to Herib Blanco and the anonymous reviewer, for their time and effort in reviewing our 
manuscript and providing us with very valuable and helpful comments and suggestions.

\printglossary[title={Acronyms}, type=\acronymtype, nonumberlist]

\section{Supporting information}
\newcommand\siref[1]{%
    \textbf{\autoref{#1}.~\nameref{#1}\newline}
}
\label{LastPageBeforeSI}

\siref{app:esf-curtailment-lcoe}
\siref{app:graphical-overview}
\siref{app:equations}
\siref{app:esc-visualisations}
\siref{app:res-capacity-and-generation}
\siref{app:cost-composition-other-scenarios-years}
\siref{app:results-lcoes}
\siref{app:results-lcohs}
\siref{app:technology-assumptions}
\siref{app:conversion-efficiencies}
\siref{app:shipping-assumptions}
\siref{app:country-masks}
\siref{app:data-and-code-availability}

\clearpage

%\nolinenumbers % Disable linenumbers for supporting information

\newcommand\modifysubsectionname[1]{\renewcommand{\thesubsection}{S \arabic{subsection} #1}}

\modifysubsectionname{Appendix}
\subsection{\glsfmtshort{esf}, \glsfmtshort{lcoe} and curtailment}
\label{app:esf-curtailment-lcoe}
\setcounter{page}{1}
\rfoot{\thepage/\pageref{app:esf-curtailment-lcoe_end}}
\lfoot{}

Additional characteristics of the \glspl{esc} can become apparent by looking into the energy required
for each \gls{esc} for synthesis, conversion and transport.
We use the \gls{esf} (Input electricity required per unit of energy delivered) instead of the energy efficiency
(Share of energy delivered per unit of electricity generated) for this purpose.
In \autoref{fig:esfvslcoenergy-and-curtailment-10pwacc-2030} the \glspl{esf} are shown on the left versus the \gls{lcoe} 
for each \gls{esc} and exporter.

The \glspl{esf} for imports range from \num{1.4} (\ce{H2 (g)} pipeline and \gls{hvdc} with subsequent electrolysis)
to \num{5.4} (\ce{CH4 (g)} pipeline to \ce{H2} from \gls{au}).
The simpler and lower cost \glspl{esc} using \gls{hvdc} and hydrogen pipelines have as a rule of thumb lower \glspl{esf}.
The majority of shipping options are between \SIrange{100}{200}{\EUR\per\MWh\th} at \glspl{esf} ranging from \numrange{1.8}{3.8}.
\gls{ftf} ship-based imports are clustered around an \gls{esf} of \num{3.2}.
Notable outliers are the domestic production of methanol and \gls{ftf} in \gls{de} which we will discuss further below.
Comparing the \glspl{esf} can provide insight into the necessary of \gls{res} capacities and thus 
e.g. land requirements involved between different \glspl{esc}.

The right side of \autoref{fig:esfvslcoenergy-and-curtailment-10pwacc-2030} shows the \glspl{esf} versus the share of electricity curtailed.
Curtailment in islanded system as in our scenarios may only be avoided with investments into storage capacities
or increasing the flexibility of involved synthesis processes (\ce{CH4}, \ce{NH3}, methanol and \gls{ftf}).
We find a wide range of curtailment levels with the majority being below \SI{20}{\percent} and no apparent correlation with
\glspl{esc} and exporter.

Looking at the curtailment we again find the \glspl{esc} for methanol, methanol to \ce{H2} and \gls{ftf}
with domestic production in \gls{de} as outliers.
Closer inspection of these \glspl{esc} links the high curtailment rate to the available \gls{res} potentials for \gls{de}
where, after domestic electricity demand was given priority in the supply curve, only onshore and offshore wind remain
as \gls{res} and no \gls{pv} is available to the \glspl{esc}.
The high temporal correlation of the wind quality classes cause in combination with the inflexible synthesis processes 
for these chemical energy carriers the need to deploy high volumes of feedstock storage (\ce{H2} and \ce{CO2})
as well as electrolysis capacities with low \gls{cf} to be build.
This in turn causes the investment costs and therefore \glspl{lcoe} to increase dramatically and
significant curtailment on days of high wind speeds.
This issue does not appear with any other exporter, as all other exporters have some \gls{pv} potentials available,
see \autoref{app:res-capacity-and-generation}.

\begin{minipage}{\textwidth}
    \captionsetup{type=figure}
    \centering
\includegraphics[width=1\textwidth]{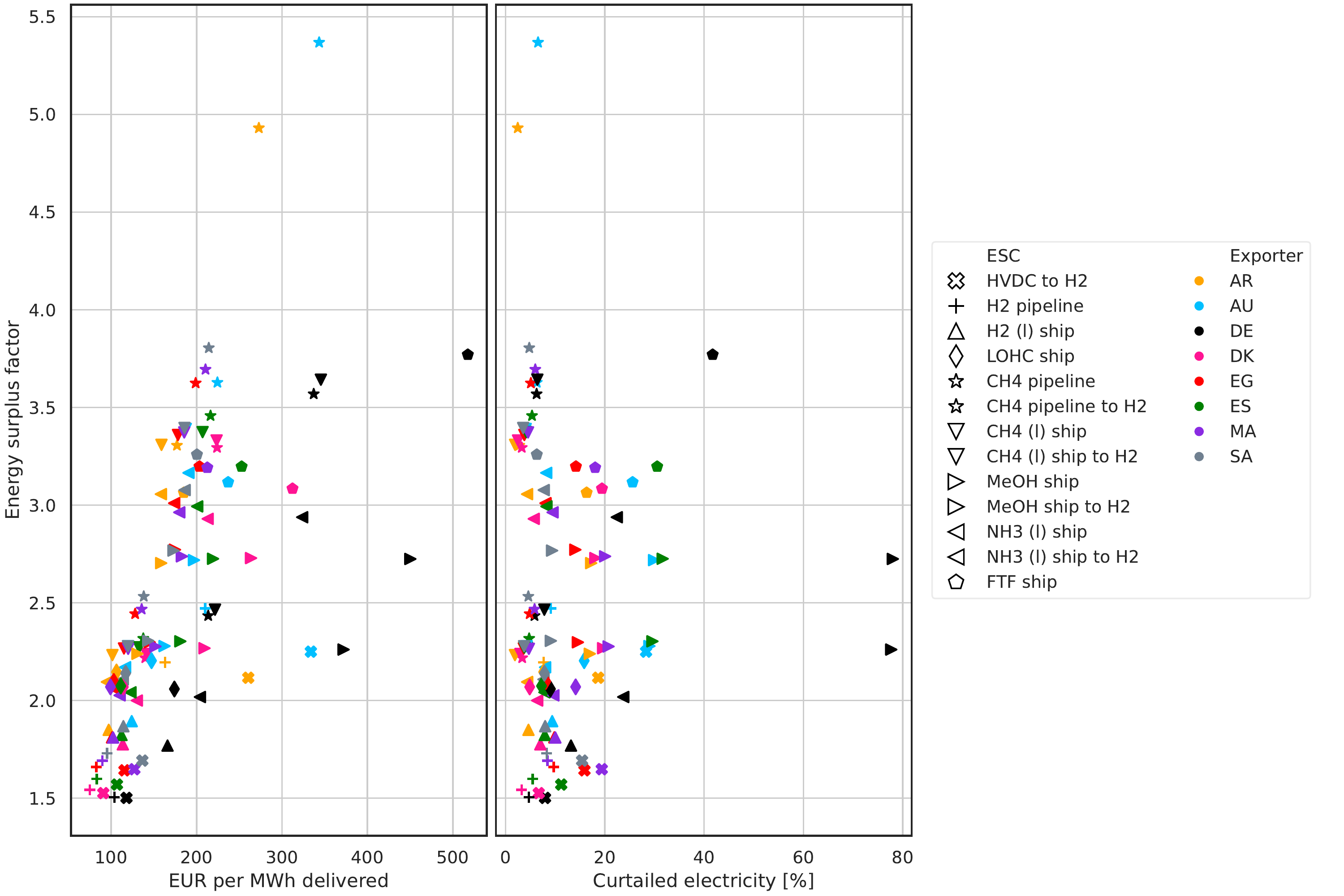}
    \caption{
        \textbf{\glspl{esf} for \glspl{esc} and exporters displayed against \gls{lcoe} (left) and electricity curtailment (right).
        Values shown are for 2030 and \SI[detect-weight=true]{10}{\percentpa} \gls{wacc} scenarios.}
        Simplicity (lower \gls{esf}) and costs of \glspl{esc} increase to some extent together
        while curtailment varies greatly between exporter and \gls{esc} driven 
        by the available \gls{res}, inflexibility of synthesis processes and storage deployment.
    }
    \label{fig:esfvslcoenergy-and-curtailment-10pwacc-2030}
\end{minipage}

\begin{minipage}{\textwidth}
    \captionsetup{type=figure}
    \centering
\includegraphics[width=1\textwidth]{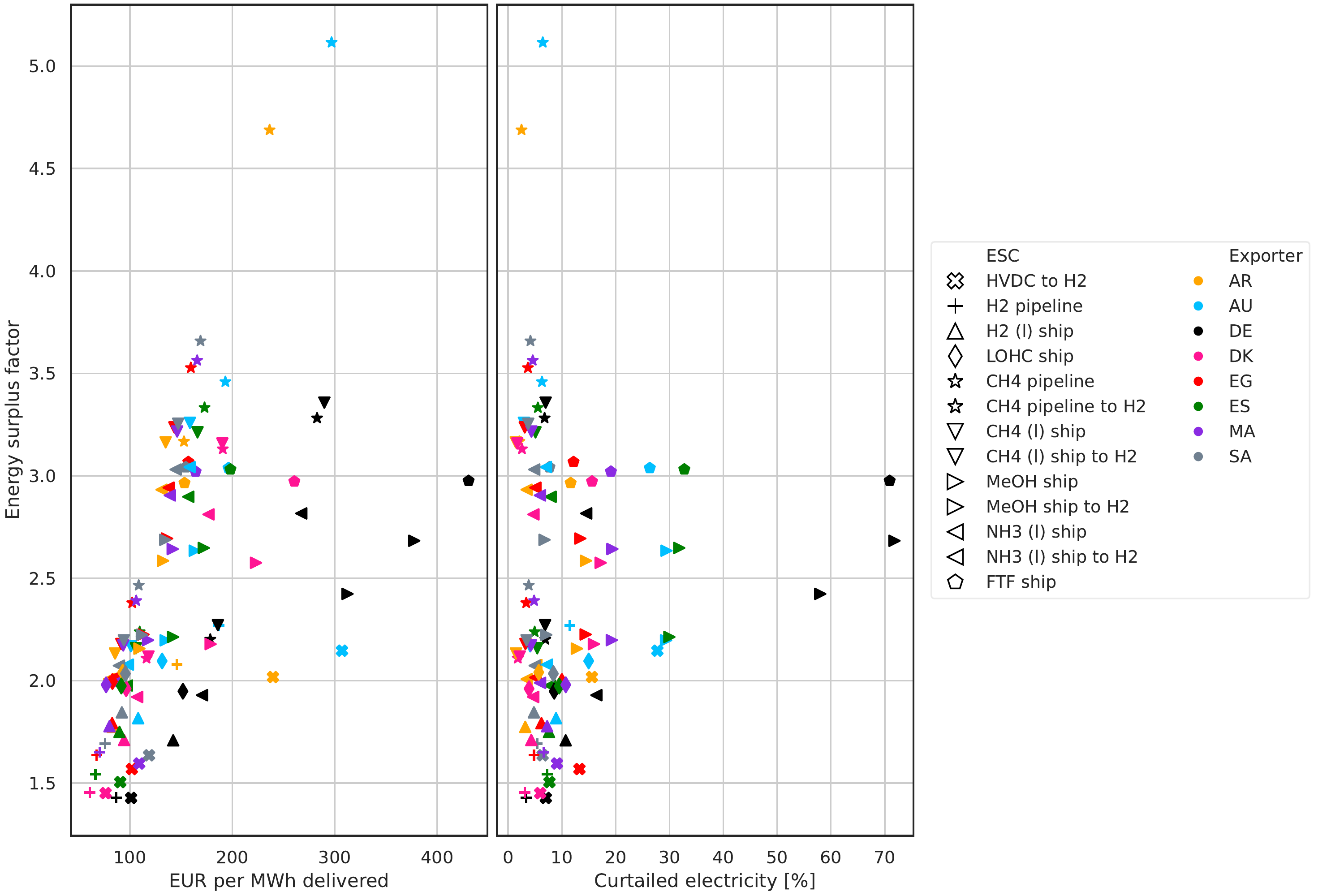}
    \caption{
        \textbf{\glspl{esf} vs. \gls{lcoe} and electricity curtailment for 2040 and \SI[detect-weight=true]{10}{\percentpa} \gls{wacc} scenarios.}
    }
    \label{fig:esfvslcoenergy-and-curtailment-10pwacc-2040}
    \end{minipage}

\begin{minipage}{\textwidth}
    \captionsetup{type=figure}
    \centering
\includegraphics[width=1\textwidth]{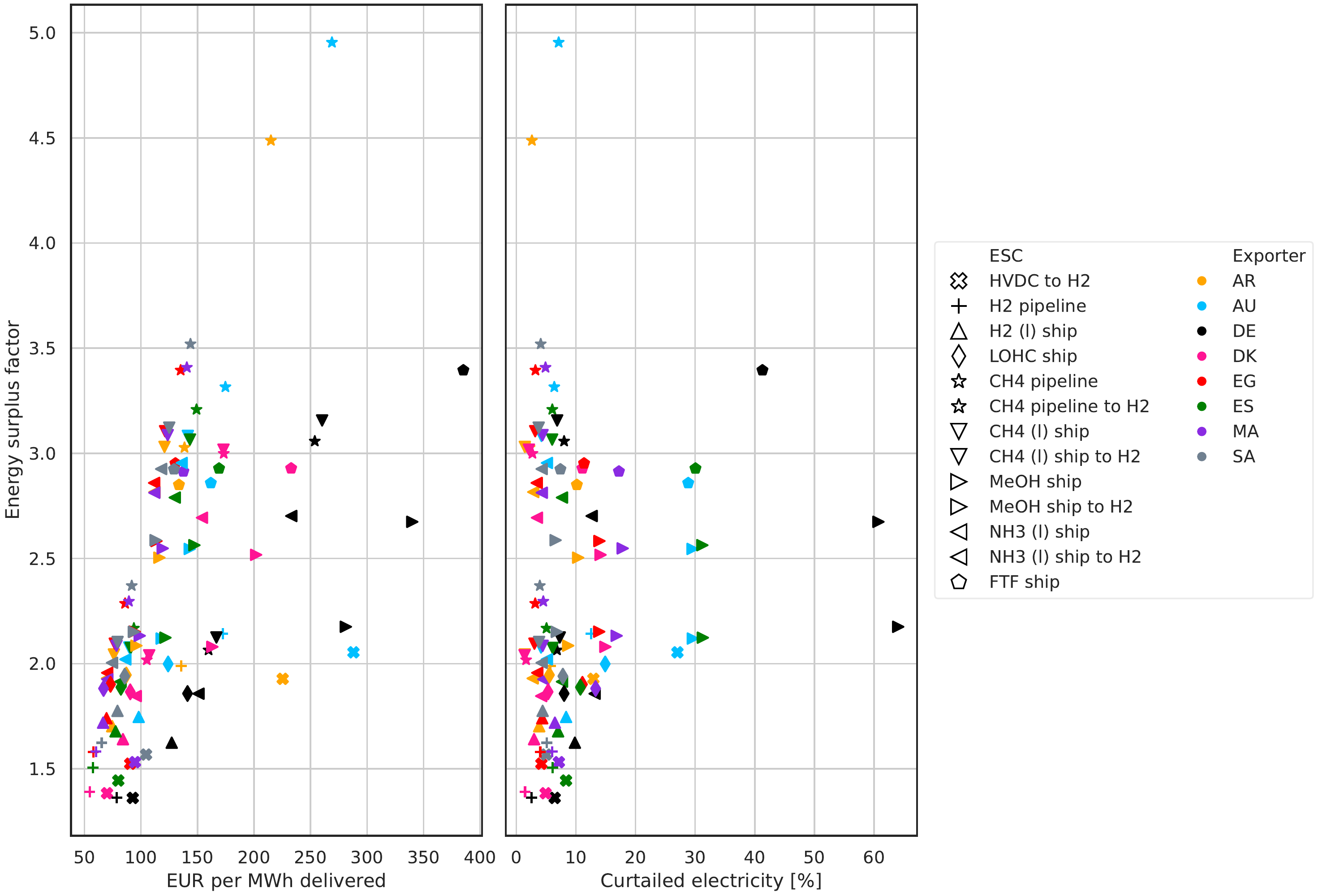}
    \caption{
        \textbf{\glspl{esf} vs. \gls{lcoe} and electricity curtailment for 2050 and \SI[detect-weight=true]{10}{\percentpa} \gls{wacc} scenarios.}
    }
    \label{fig:esfvslcoenergy-and-curtailment-10pwacc-2050}
\end{minipage}

\begin{minipage}{\textwidth}
    \captionsetup{type=figure}
    \centering
\includegraphics[width=1\textwidth]{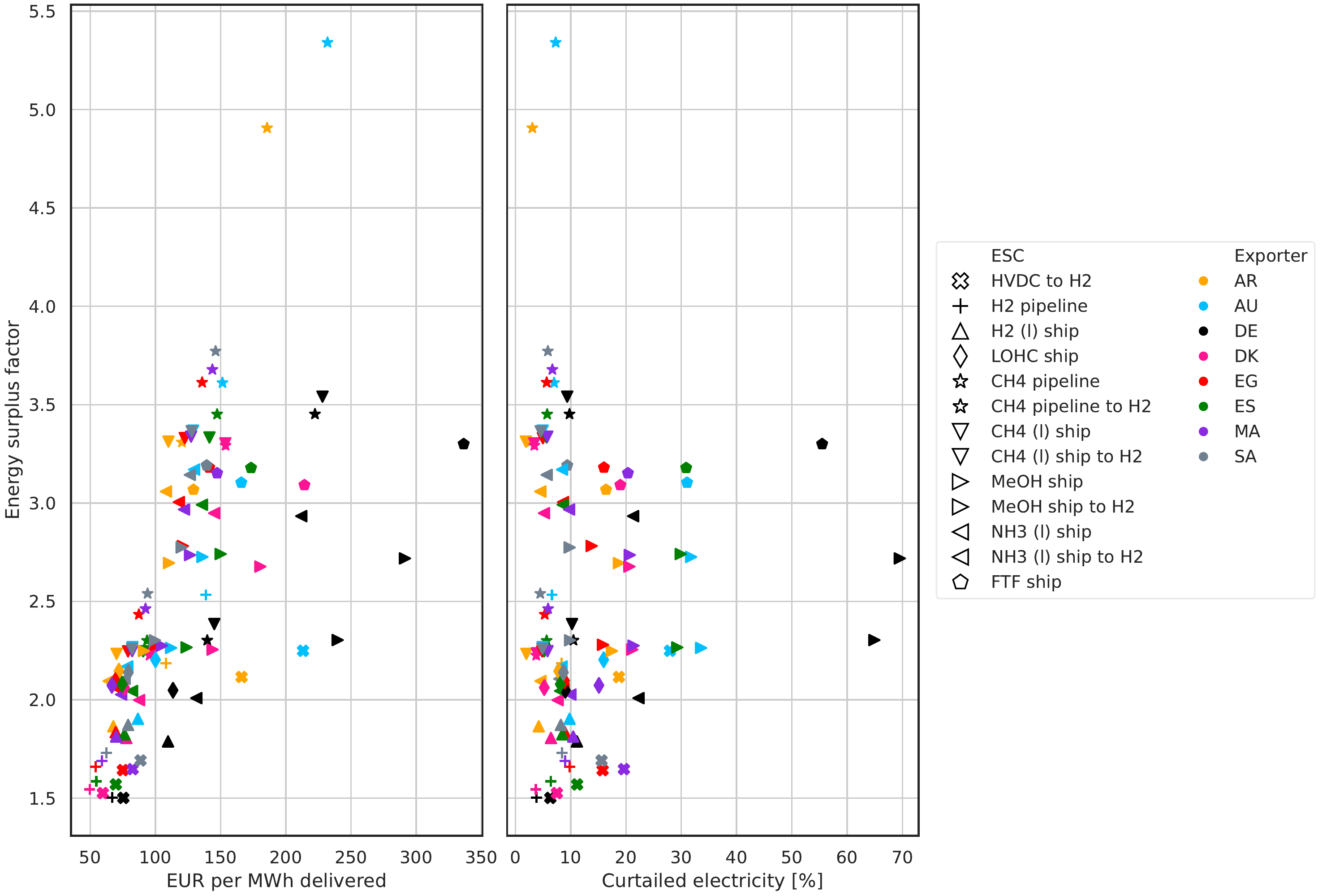}
    \caption{
        \textbf{\glspl{esf} vs. \gls{lcoe} and electricity curtailment for 2030 and \SI[detect-weight=true]{5}{\percentpa} \gls{wacc} scenarios.}
    }
    \label{fig:esfvslcoenergy-and-curtailment-5pwacc-2030}
\end{minipage}

\begin{minipage}{\textwidth}
    \captionsetup{type=figure}
    \centering
\includegraphics[width=1\textwidth]{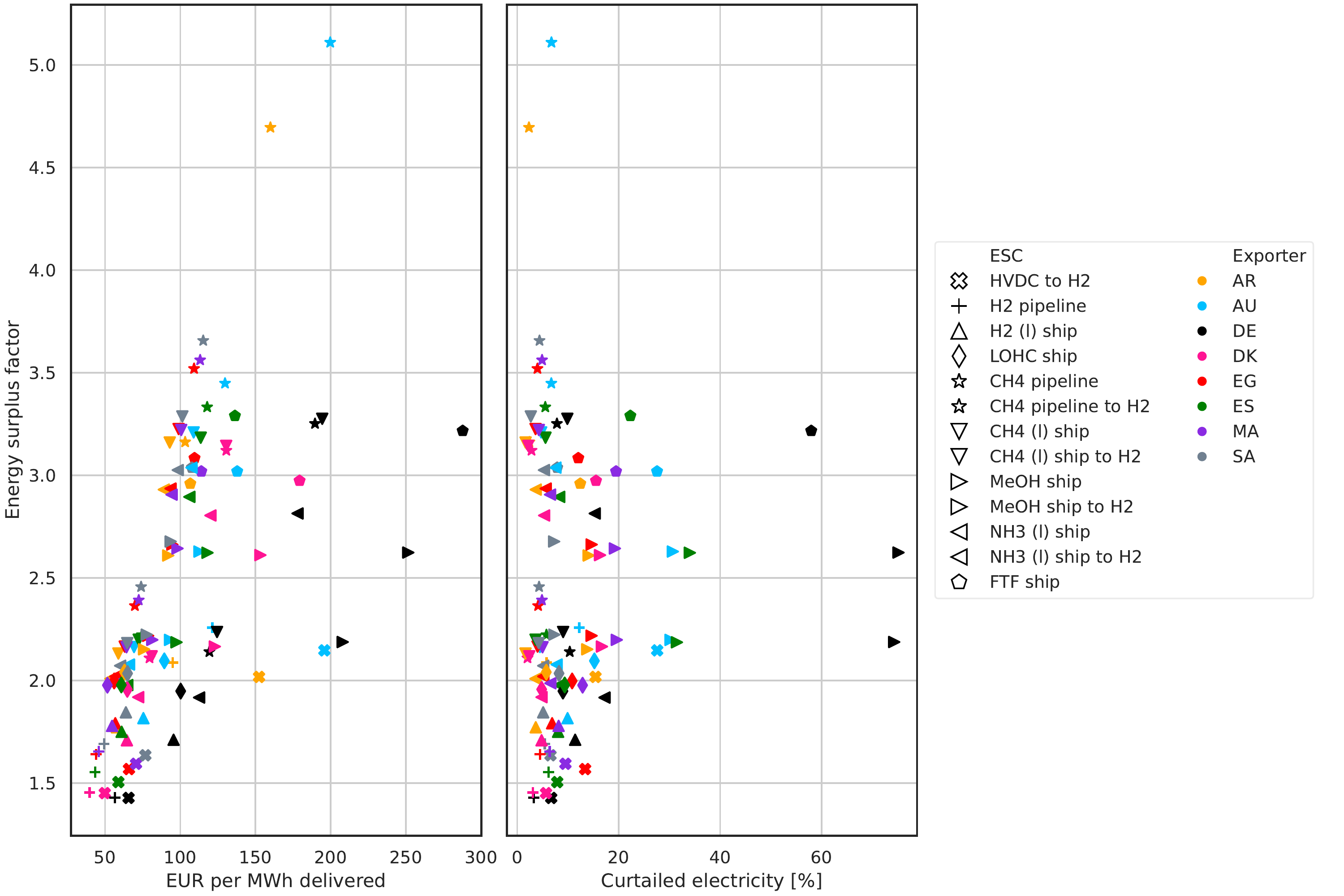}
    \caption{
        \textbf{\glspl{esf} vs. \gls{lcoe} and electricity curtailment for 2040 and \SI[detect-weight=true]{5}{\percentpa} \gls{wacc} scenarios.}
    }
    \label{fig:esfvslcoenergy-and-curtailment-5pwacc-2040}
\end{minipage}

\begin{minipage}{\textwidth}
    \captionsetup{type=figure}
    \centering
\includegraphics[width=1\textwidth]{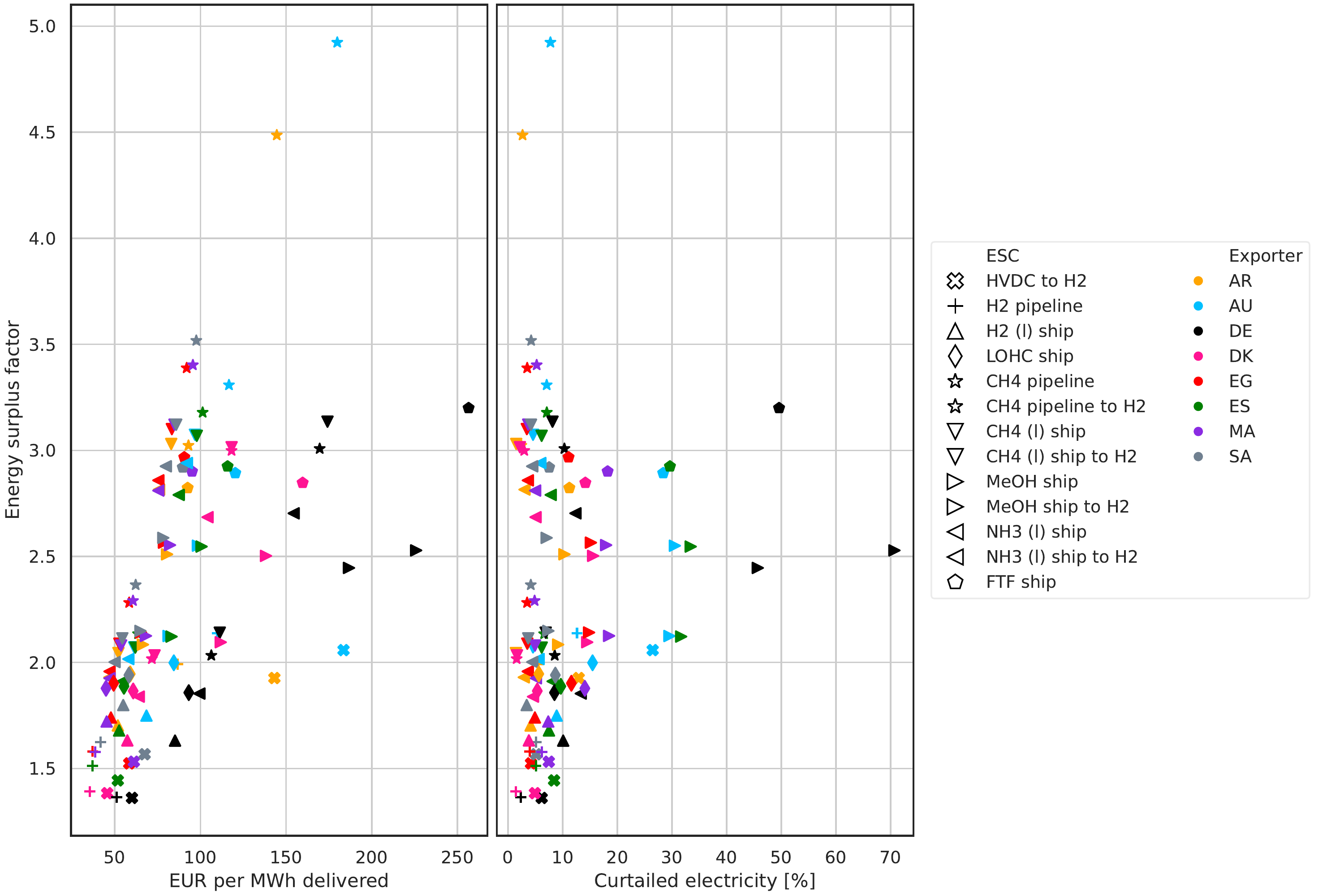}
    \caption{
        \textbf{\glspl{esf} vs. \gls{lcoe} and electricity curtailment for 2050 and \SI[detect-weight=true]{5}{\percentpa} \gls{wacc} scenarios.}
    }
    \label{fig:esfvslcoenergy-and-curtailment-5pwacc-2050}
\end{minipage}

\label{app:esf-curtailment-lcoe_end}

\clearpage
\modifysubsectionname{Appendix}
\subsection{Technical model structure}
\label{app:graphical-overview}
\setcounter{page}{1}
\rfoot{\thepage/\pageref{app:graphical-overview_end}}

\begin{minipage}{\textwidth}
    \captionsetup{type=figure}
    \centering
\includegraphics[width=1\textwidth]{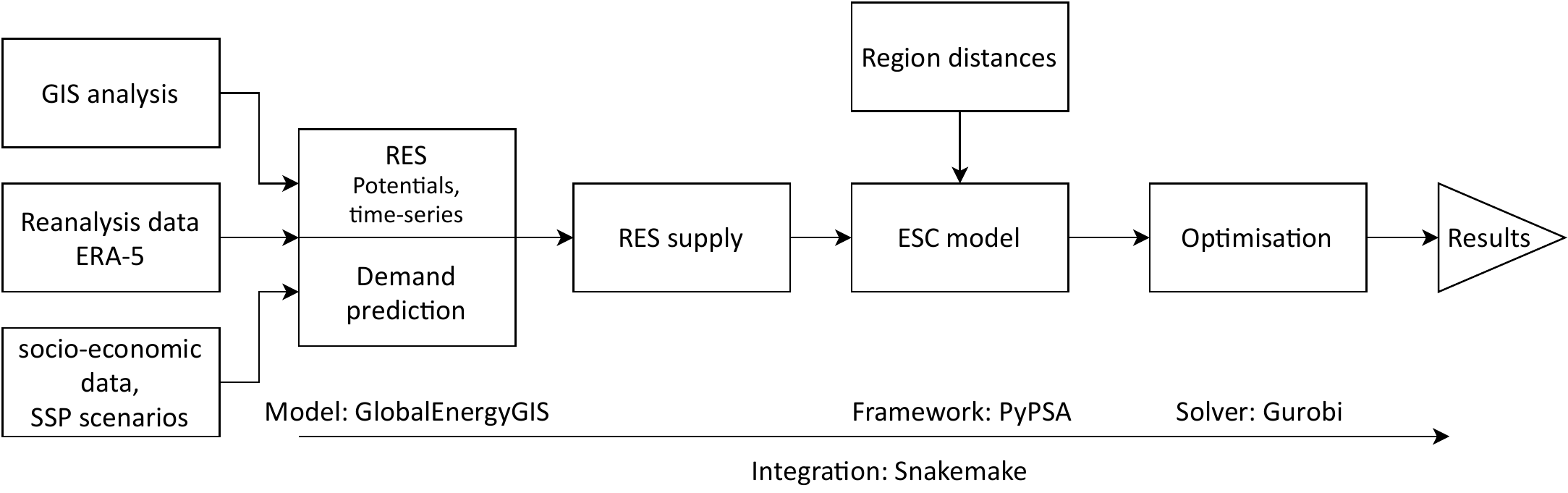}
    \caption{
        Model structure, underlying workflow and software used for this study.
    }
\end{minipage}

\label{app:graphical-overview_end}

\clearpage
\modifysubsectionname{Appendix}
\subsection{Model equations}
\label{app:equations}
\setcounter{page}{1}
\rfoot{\thepage/\pageref{app:equations_end}}

The underlying model equations for each scenario are constructed through the open source framework PyPSA.
PyPSA is a framework for building power and energy system models for running capacity expansion optimisation among other features.
A detailed description of the mathematical background can be found in \cite{brown2018si} and in the online documentation at \href{https://pypsa.readthedocs.io/en/latest/index.html#}{pypsa.readthedocs.io}.

PyPSA models are built using elemental components, such as generators, links, loads, stores and buses.
Buses are the fundamental nodes to which all other components attach.
In this study buses represent intermediary steps in the conversion between materials and energy carriers 
and different locations during the transport of materials or energy.
Each bus is associated with one specific type of energy or material.
Energy and materials enter a model through generators or links with efficiencies $\eta > 1$.
Links connect different buses and represent in this study conversion steps or the transport of materials or energy.
Each link's inflows and outflows are linked to individual effficiencies.
Loads represents sinks for energy and materials where they are consumed, i.e. they leave a model.

From the interconnected components PyPSA generates a network (graph) model for which then energy and material flows 
are conserved~\cite{brown2018si}, i.e. for at each bus $n$ for each timestep $t$
\begin{equation}
    \sum\limits_{r} g_{n,r,t} + \sum\limits_s h_{n,s,t} + \sum\limits_l \alpha_{l,n,t} f_{l,t} = d_{n,t} \qquad\forall i,t
\end{equation}
where $g_{n,r,t}$ is the inflow from generator $r$ at bus $n$ at timestep $t$,
$h_{n,s,t}$ is the flow from or into store $s$ attached to bus $n$ at timestep $t$,
$f_{l,t}$ incoming or outgoing flows through link $l$,
$\alpha_{l,n,t}$ the incidence matrix and $d_{i,t}$ the outflow through loads attached to the bus.
The incidence matrix takes on the values
\begin{equation}
    \alpha_{l,n,t} = \begin{cases}
        -1           & \text{if $l$ starts at $n$}\\
        \eta_{l,n,t} & \text{if $l$ ends at $n$}
    \end{cases}
\end{equation}
with $\eta_{l,n,t}$ being the specific efficiency for the link $t$ ending in $n$.

The pipeline efficiency $\eta^{\text{pipe}}$ is determined from the pipeline efficiency per \SI{1000}{\kilo\meter} $\eta^{\text{pipe}}_0$
(losses and energy consumption for pressure boosting compressors) and the pipeline length $d$ in $[\si{\kilo\meter}]$
\begin{equation}
    \eta^{\text{pipe}} = 1 - {\eta^{\text{pipe}}_0}^{\tfrac{d^{\text{pipe}}}{1000}}
\end{equation}
The efficiencies for \gls{hvdc} connections are calculated analogously.

The shipping efficiency $\eta^{\text{ship}}$ is determined by the boil-off and the propulsion 
energy demand as ships are assumed to use their cargoed energy carrier also as propulsion fuel.
The lower efficiency, i.e. higher losses due to propulsion or boil off, are used as total efficiency.
The total propulsion energy demand for outbound and return journey is determined by the
shipping distance $d^\text{ship}$ in $[\si{\kilo\meter}]$,
the ship's specific energy demand $e$ in $[\si{\mega\watt\hour\per\kilo\meter}]$ and
the ship's total cargo capacity $c$ in $[\si{\mega\watt\hour}]$.

Boil-off is only considered for the outbound journey as we assume the absolute boil-off during
the return journey with near empty cargo hold to be minimal.
The boil-off for the outbound journey is determined by the specific boil-off rate
$b^{\text{ship}}$ in $[\si{\percent\per\hour}]$ and the adjusted outbound journey travel time $t^\text{ship}_\text{outbound, adjusted}$ in $[\si{\hour}]$.

The total shipping efficiency is thus calculated as
\begin{equation}
    \eta^{\text{ship}} = 
    \mathrm{min}\left\{
    1 - 2d^\text{ship} \cdot \frac{e}{c}, \left(1 - b^\text{ship} \right)^{t^{\text{ship}}_{\text{outbound, adjusted}}} \right\}
\end{equation}

The outbound journey travel time $t^\text{ship}_\text{outbound}$ is determined by the ship's average speed
$v^\text{ship}$ and shipping distance $d^\text{ship}$
\begin{equation}
    t^\text{ship}_\text{outbound} = \frac{d^\text{ship}}{v^\text{ship}}
\end{equation}

The round-trip time for shipping is further affected by the time required for loading and unloading
the ship $t^\text{ship}_\text{(un-)loading}$
\begin{equation}
    t^\text{ship}_\text{round-trip} = 2t^{text{ship}}_{\text{outbound}} + 2t^{\text{ship}}_{\text{(un-)loading}}
\end{equation}

The adjusted outbound journey time $t^{\text{ship}}_{\text{outbound, adjusted}}$ is then determined by stretching 
the round-trip journey time.
Stretching the journey time keeps the number of trips per year the same while reducing the number of 
hours a ship is not engaged in transporting cargo to a minimum.

\begin{equation}
    t^\text{ship}_\text{outbound, adjusted} 
    = t^\text{ship}_\text{outbound} 
    + \mathrm{round}\left(\frac{1}{2} \left\lceil \frac{t^\text{ship}_\text{gap}}{n^\text{ship}_\text{journeys}}\right\rceil\right)
\end{equation}

Here $\mathrm{round}(\ldots)$ is rounding to the next integer,
\begin{equation}
    t^\text{ship}_\text{gap} = \SI{8760}{\hour} \cdot \mathrm{mod}\,t^\text{ship}_\text{round-trip}
\end{equation}
the time a ship would be idle per year (\SI{8760}{\hour}) if the travel time was not adjusted
and
\begin{equation}
    n^\text{ship}_\text{journeys} = \left\lceil\frac{\SI{8760}{\hour}}{t^\text{ship}_\text{round-trip}} \right\rceil
\end{equation}
the number of journeys a ship can undertake per year.
The shipping process also suffers from cargo losses during the loading and unloading process of the 
ship $l$ in \si{\percent}.
These losses are accounted for separately and are not part of the shipping efficiency.
Technical shipping parameters are listed in \ref{app:shipping-assumptions}.

For specific synthesis processes (methanation, Haber-Bosch synthesis, methanol synthesis, Fischer-Tropsch synthesis)
a must-run constraint is implemented in the model which forces a minimum capacity of synthesis capacitiy of each \gls{esc}
to be online at all times.
The constraint is
\begin{equation}
    p_\text{nom} (t, \text{plant}) \geq p_\text{nom, min} (t, \text{plant}) \qquad \forall t
\end{equation}
with $p_\text{nom}$ being the normalised output of a synthesis in timestep $t$ of a specific synthesis plant
and $p_\text{nom, min}$ the normalised lower limit must-run availability, e.g. \num{0.9425}.

\label{app:equations_end}
\clearpage
\modifysubsectionname{Figs}
\subsection{\glsfmtlongpl{esc} visualisations}
\label{app:esc-visualisations}
\setcounter{page}{1}
\rfoot{\thepage/\pageref{app:esc-visualisations_end}}

\begin{minipage}{\textwidth}
    \captionsetup{type=figure}
    \centering
\includegraphics[width=.7\textwidth]{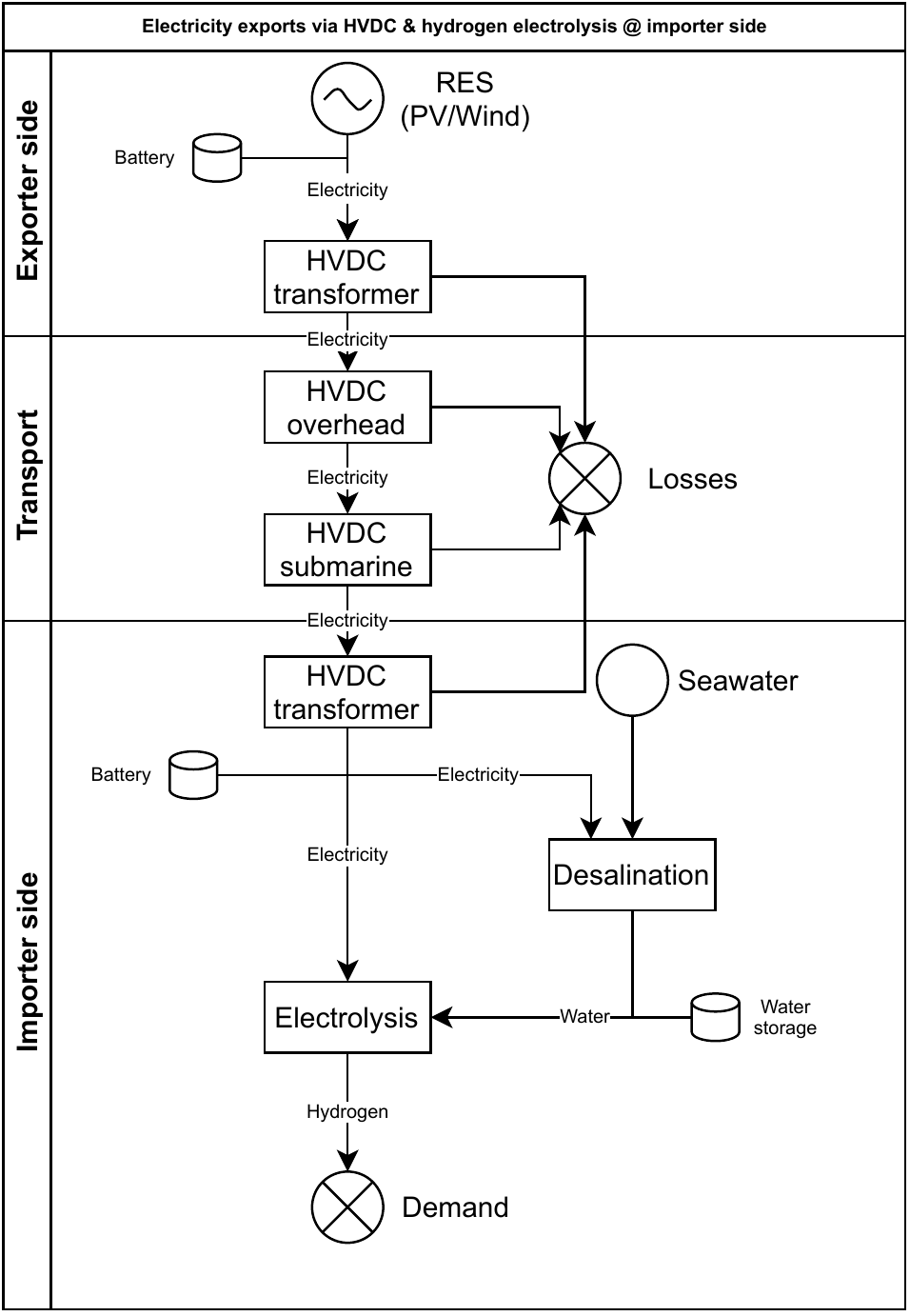}
    \caption{\glsfmtshort{esc} schematic for \glsfmtshort{hvdc} electricity imports and domestic hydrogen electrolysis.}
    \label{app:fig:esc_hvdc}
\end{minipage}

\begin{minipage}{\textwidth}
    \captionsetup{type=figure}
    \centering
\includegraphics[width=.7\textwidth]{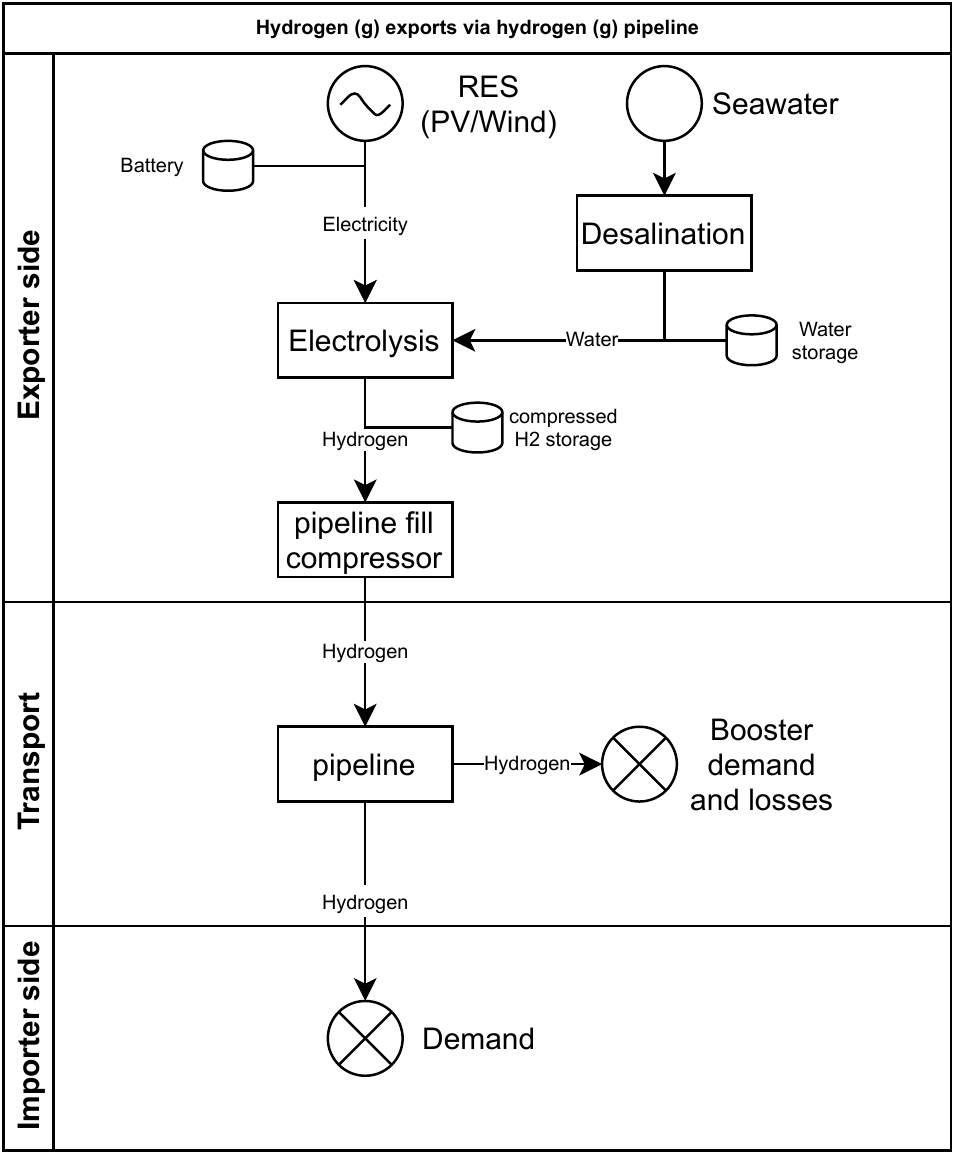}
    \caption{\glsfmtshort{esc} schematic for imports of hydrogen gas by pipeline.}
    \label{app:fig:esc_hydrogengas}
\end{minipage}

\begin{minipage}{\textwidth}
    \captionsetup{type=figure}
    \centering
\includegraphics[width=.7\textwidth]{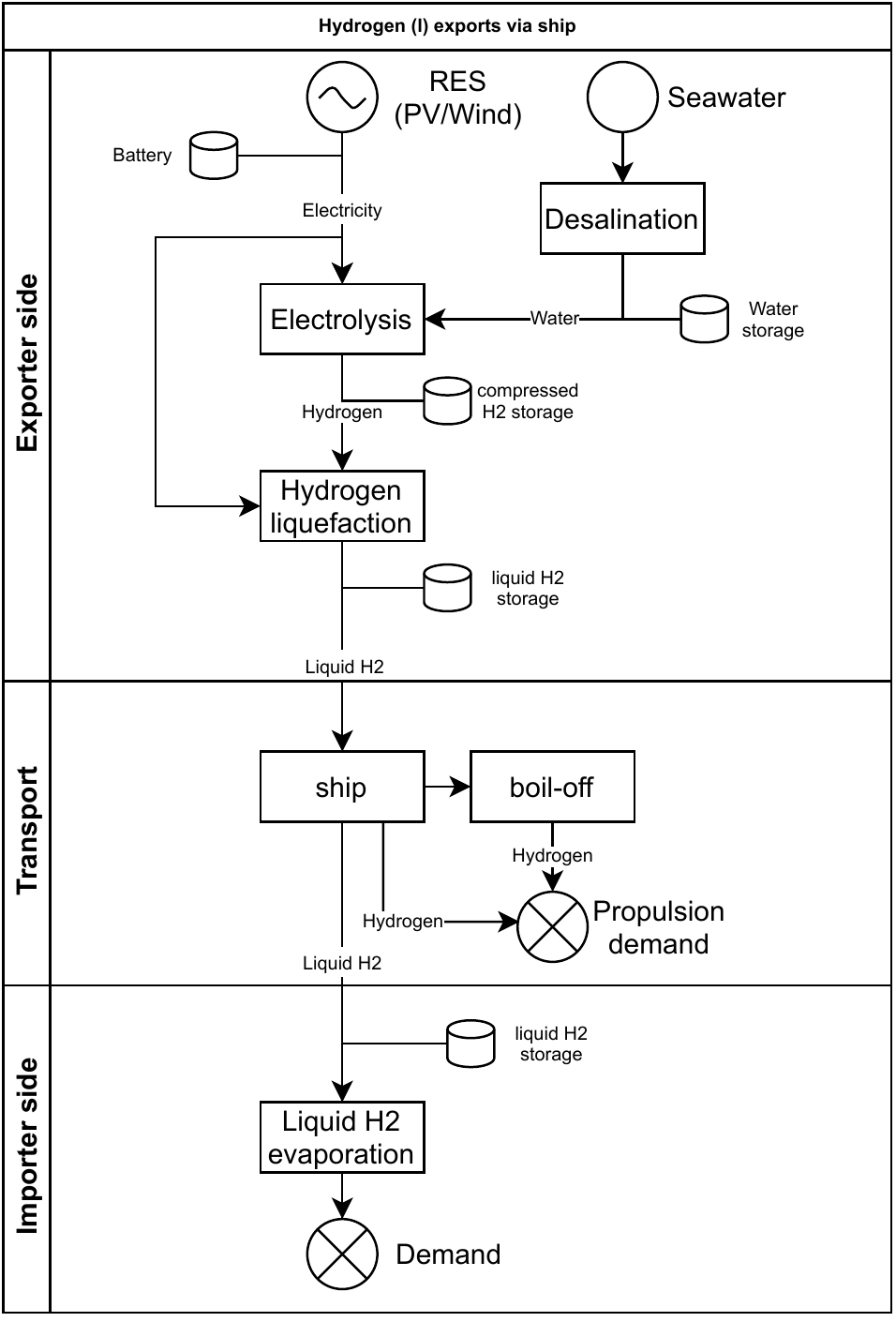}
    \caption{\glsfmtshort{esc} schematic for liquid hydrogen imports by ship.}
    \label{app:fig:esc_hydrogenliquid}
\end{minipage}

\begin{minipage}{\textwidth}
    \captionsetup{type=figure}
    \centering
\includegraphics[width=.7\textwidth]{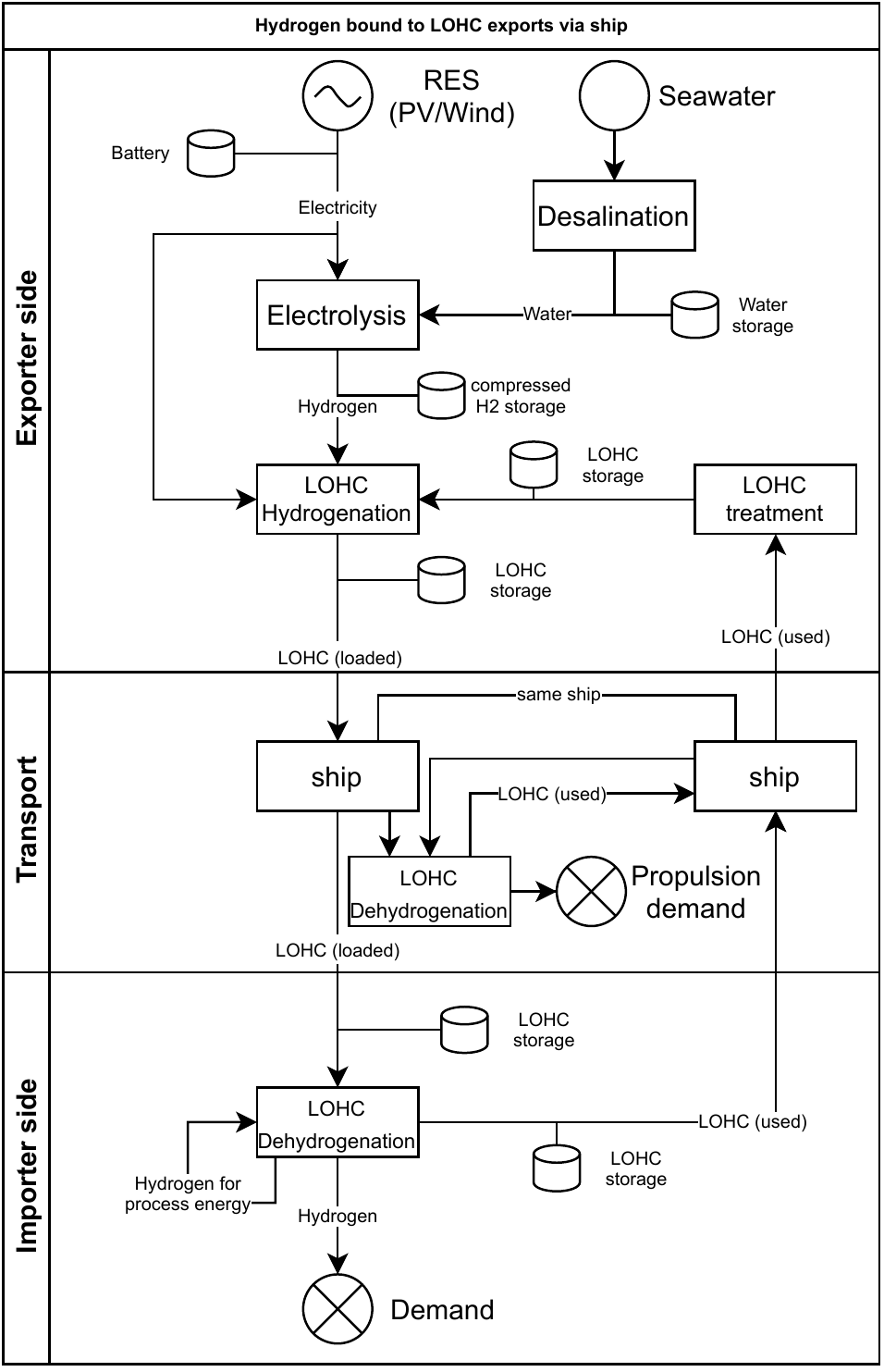}
    \caption{\glsfmtshort{esc} schematic for hydrogen imports using \glsfmtshort{lohc} by ship.}
    \label{app:fig:esc_lohc}
\end{minipage}

\begin{minipage}{\textwidth}
    \captionsetup{type=figure}
    \centering
\includegraphics[width=.7\textwidth]{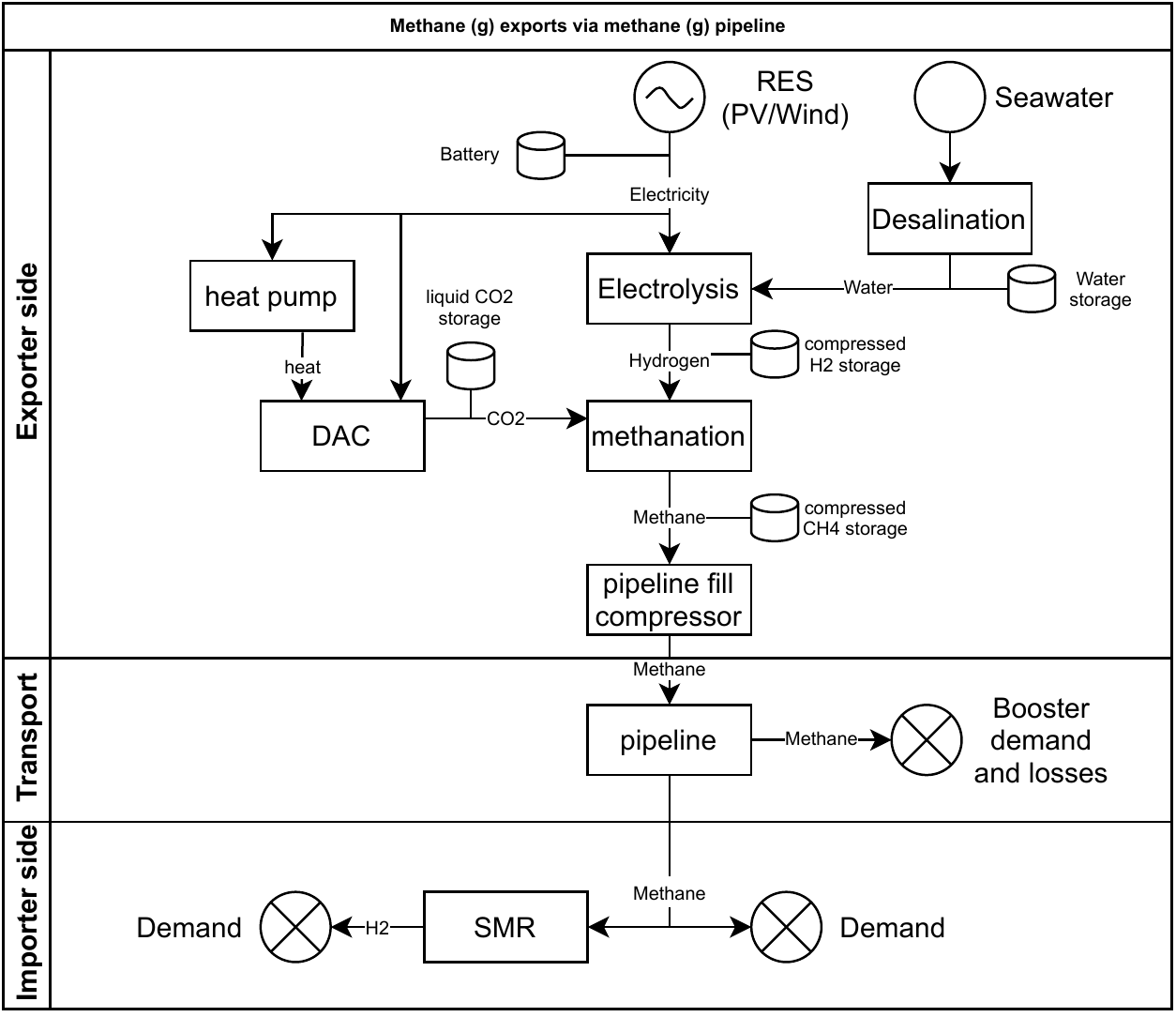}
    \caption{
        \glsfmtshort{esc} schematic for imports of methane gas by pipeline.
        To serve an optional demand of hydrogen, methane may be split via \glsfmtlong{smr}.
    }
    \label{app:fig:esc_methanegas}
\end{minipage}

\begin{minipage}{\textwidth}
    \captionsetup{type=figure}
    \centering
\includegraphics[width=.7\textwidth]{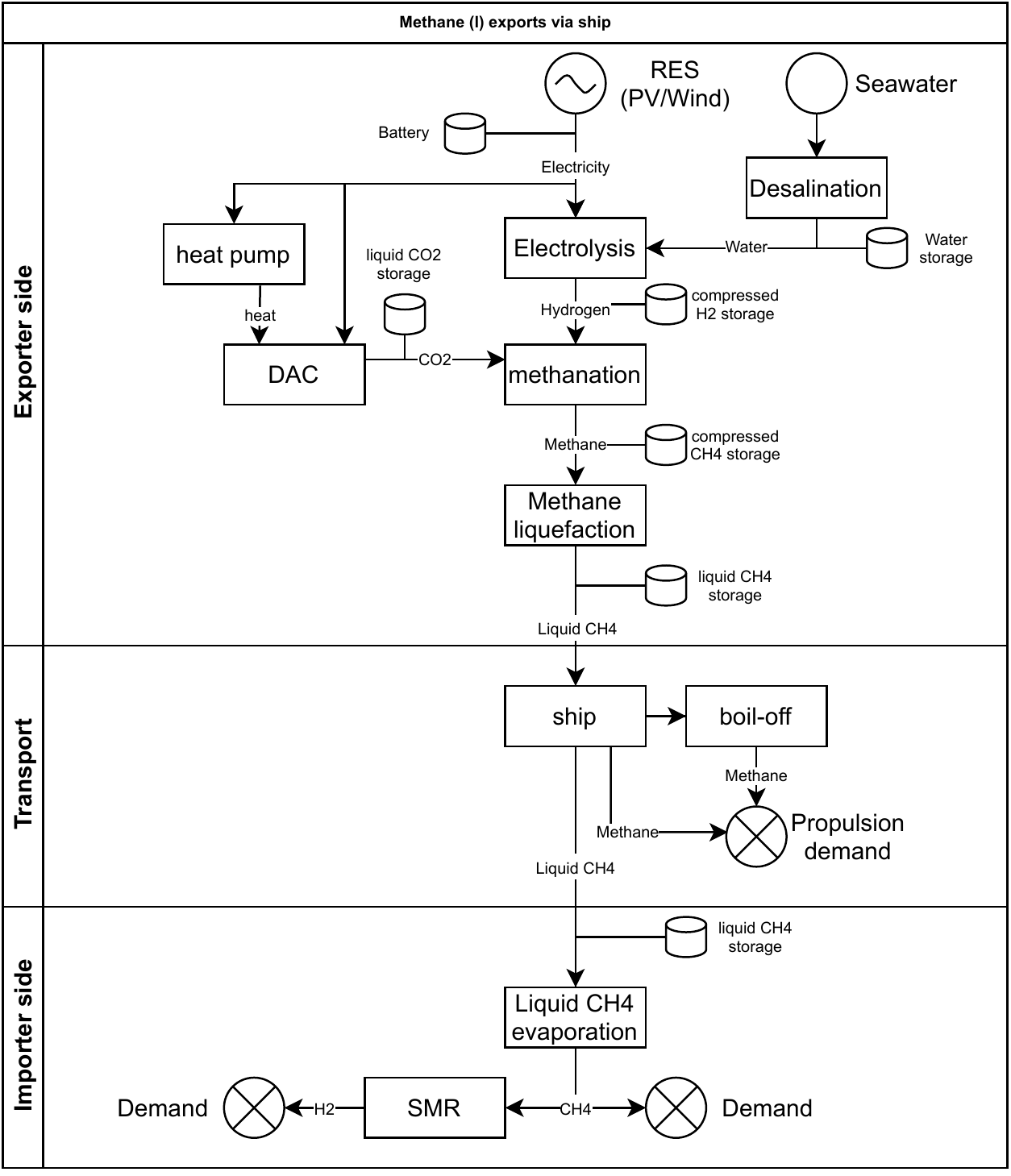}
    \caption{
        \glsfmtshort{esc} schematic for liquid methane imports by ship.
        To serve an optional demand of hydrogen, methane may be split via \glsfmtlong{smr}.
    }
    \label{app:fig:esc_methaneliquid}
\end{minipage}

\begin{minipage}{\textwidth}
    \captionsetup{type=figure}
    \centering
\includegraphics[width=.7\textwidth]{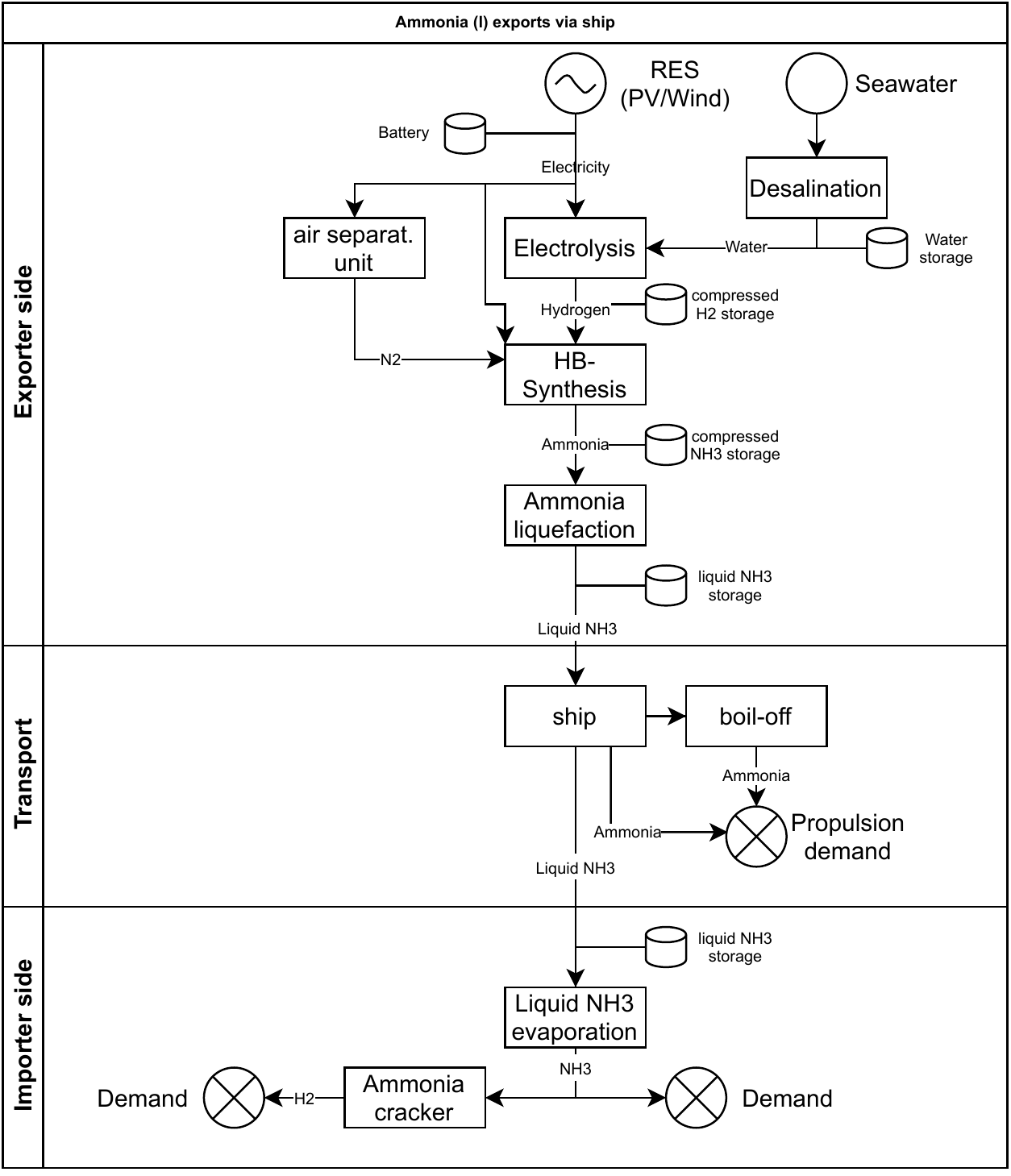}
    \caption{
        \glsfmtshort{esc} schematic for liquid ammonia imports by ship.
        The \glsfmtshort{asu} is assumed to provide nitrogen on demand without a dedicated nitrogen gas 
        feedstock storage following~\cite{banares-alcantara2015si}.
        To serve an optional demand of hydrogen, ammonia may be cracked in an ammonia cracker into hydrogen and nitrogen.
    }
    \label{app:fig:esc_ammonia}
\end{minipage}

\begin{minipage}{\textwidth}
    \captionsetup{type=figure}
    \centering
\includegraphics[width=.7\textwidth]{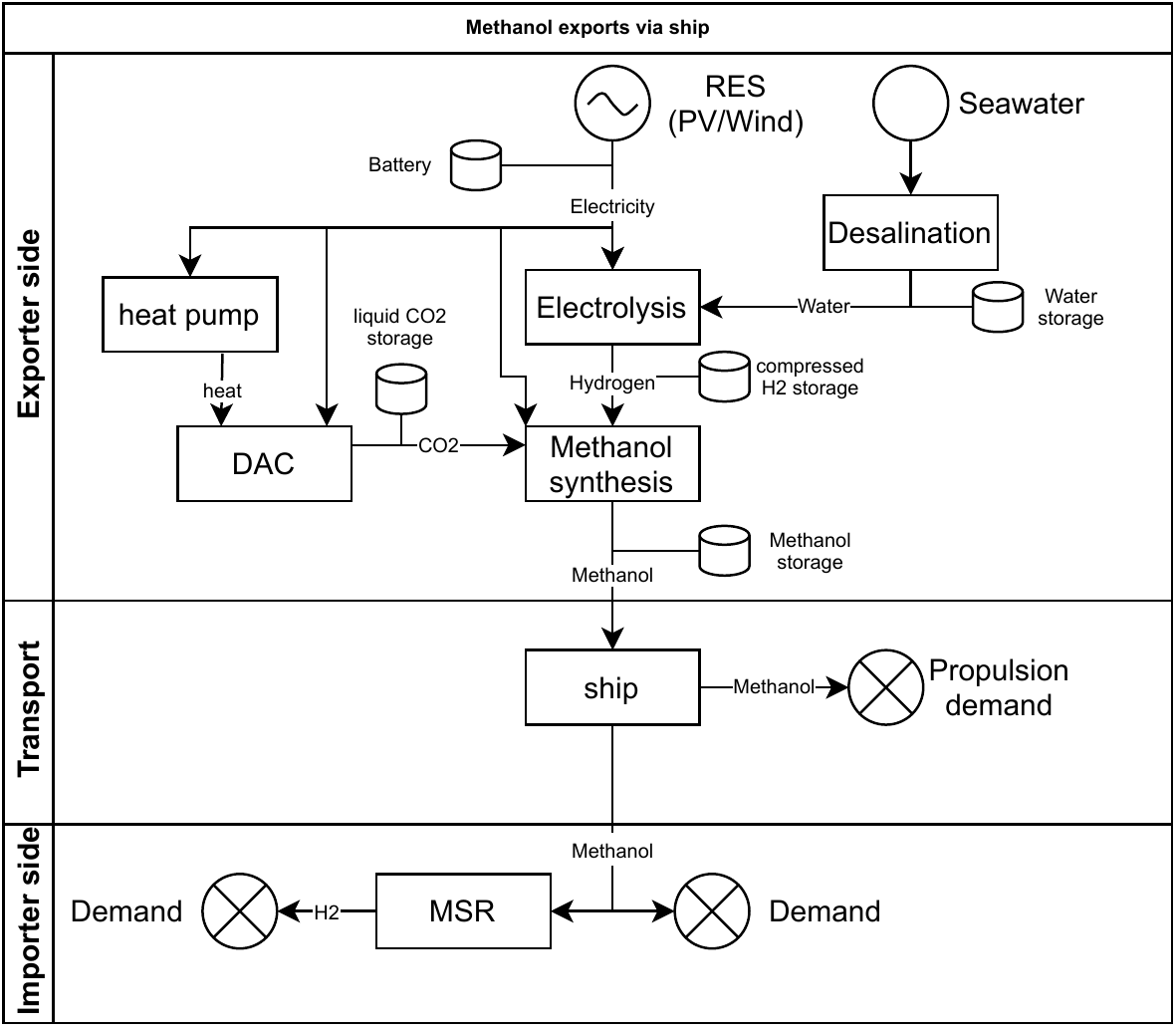}
    \caption{
        \glsfmtshort{esc} schematic for methanol imports by ship.
        To serve an optional demand of hydrogen, methanol may be split via \glsfmtlong{msr}.
    }
    \label{app:fig:esc_methanol}
\end{minipage}

\begin{minipage}{\textwidth}
    \captionsetup{type=figure}
    \centering
\includegraphics[width=.7\textwidth]{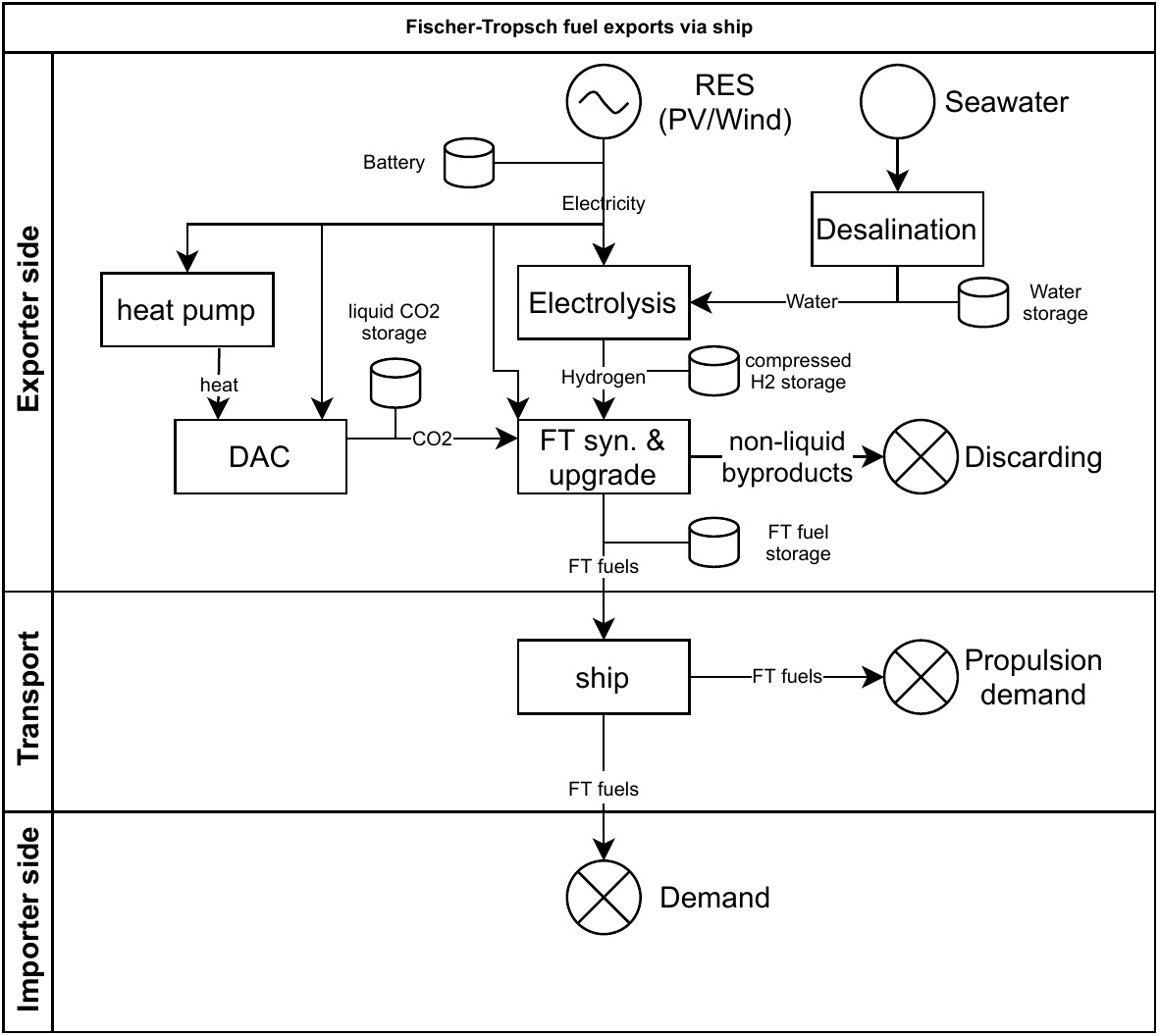}
    \caption{\glsfmtshort{esc} schematic for \glsfmtshort{ftf} imports by ship.}
    \label{fig:ftfuels}
\end{minipage}

\label{app:esc-visualisations_end}
\clearpage
\modifysubsectionname{Figs}
\subsection{Electricity generation mix and supply curves}
\label{app:res-capacity-and-generation}
\setcounter{page}{1}
\rfoot{\thepage/\pageref{app:res-capacity-and-generation_end}}

\begin{minipage}{\textwidth}
    \captionsetup{type=figure}
    \centering
\includegraphics[width=1\textwidth]{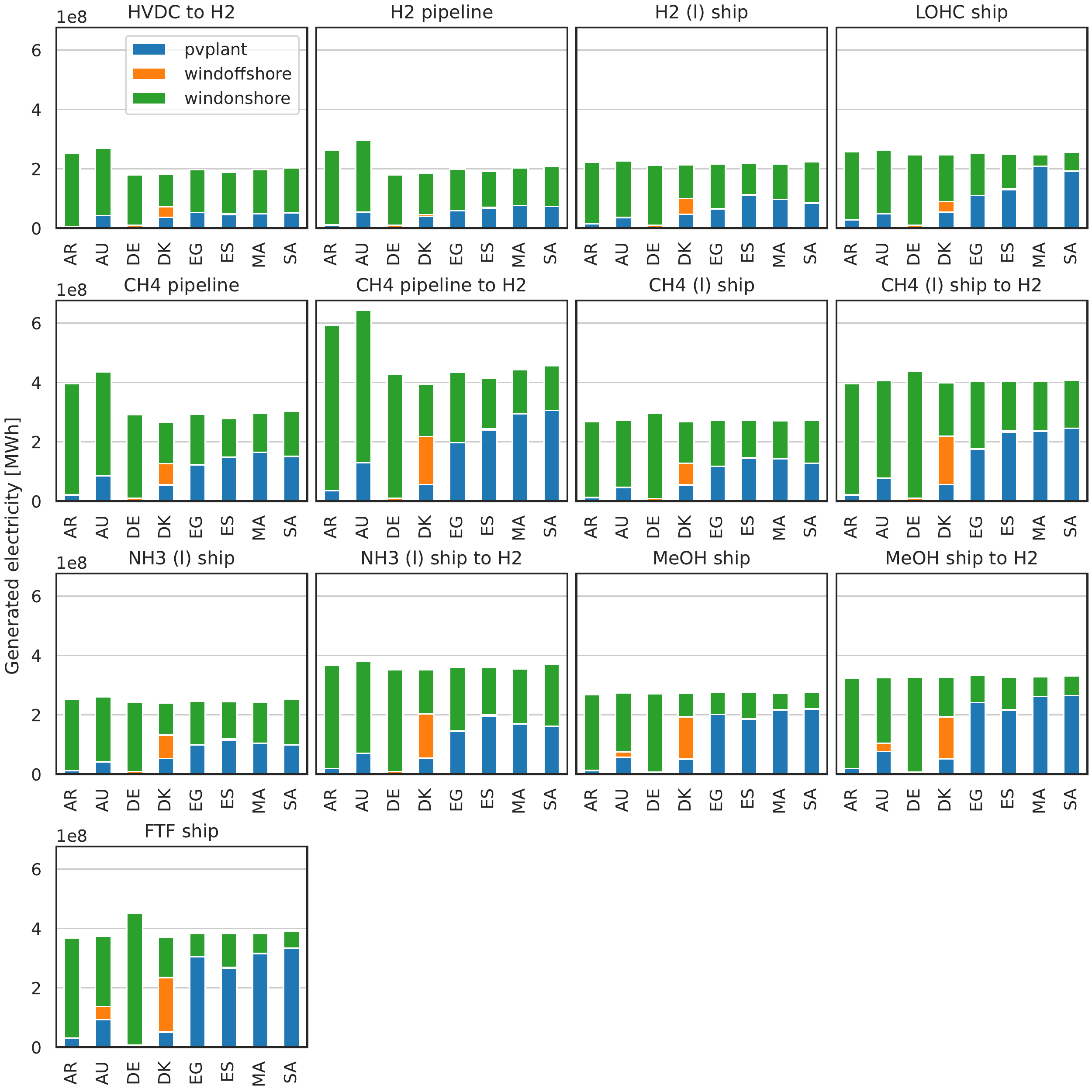}
    \caption{Electricity generation from \glsfmtshort{res} \glsfmtshort{esc} and exporting country under \SI{10}{\percentpa}  \glsfmtshort{wacc} scenario for 2030.}
\end{minipage}

\begin{minipage}{\textwidth}
    \captionsetup{type=figure}
    \centering
\includegraphics[width=1\textwidth]{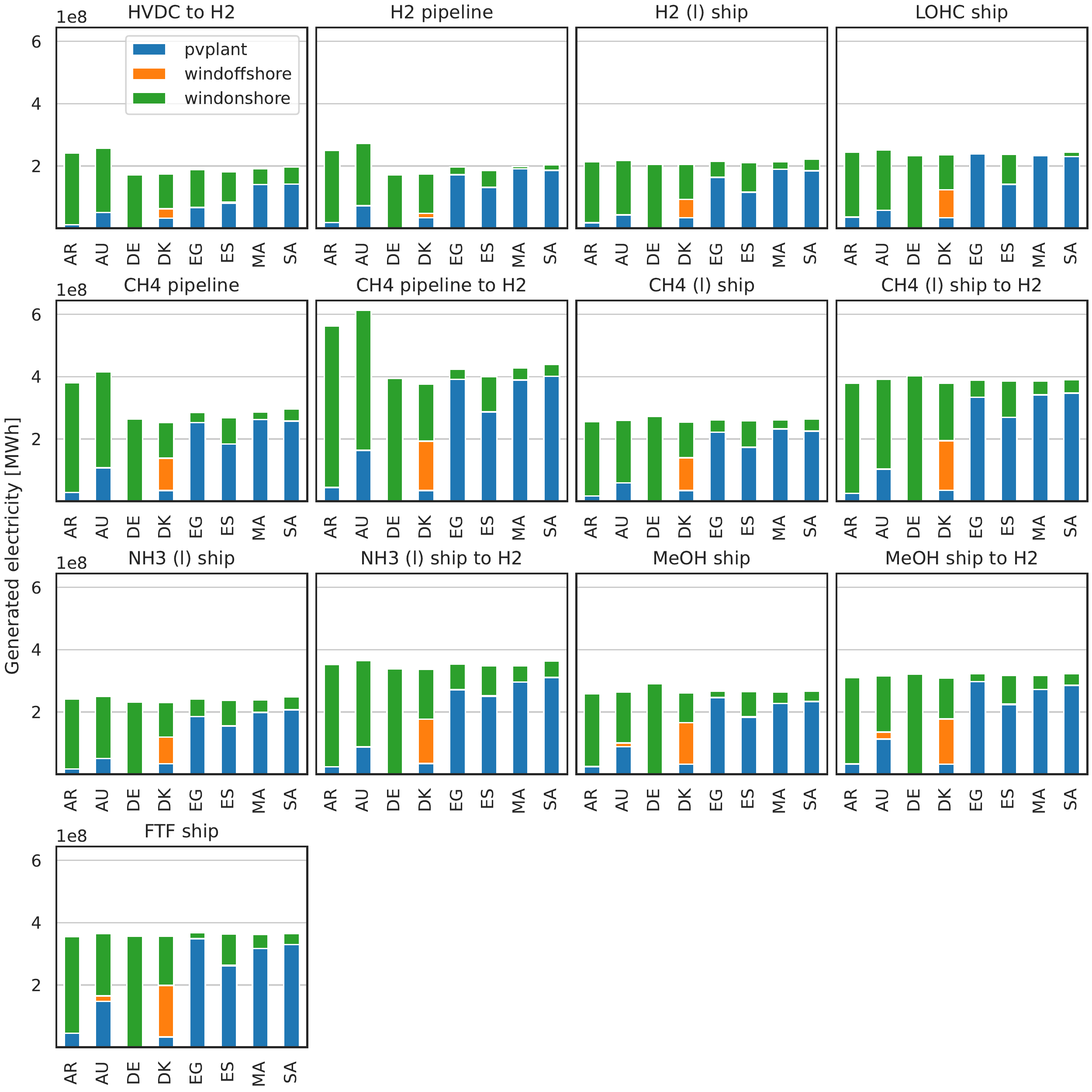}
    \caption{Electricity generation from \glsfmtshort{res} \glsfmtshort{esc} and exporting country under \SI{10}{\percentpa} \glsfmtshort{wacc} scenario for 2040.}
\end{minipage}

\begin{minipage}{\textwidth}
    \captionsetup{type=figure}
    \centering
\includegraphics[width=1\textwidth]{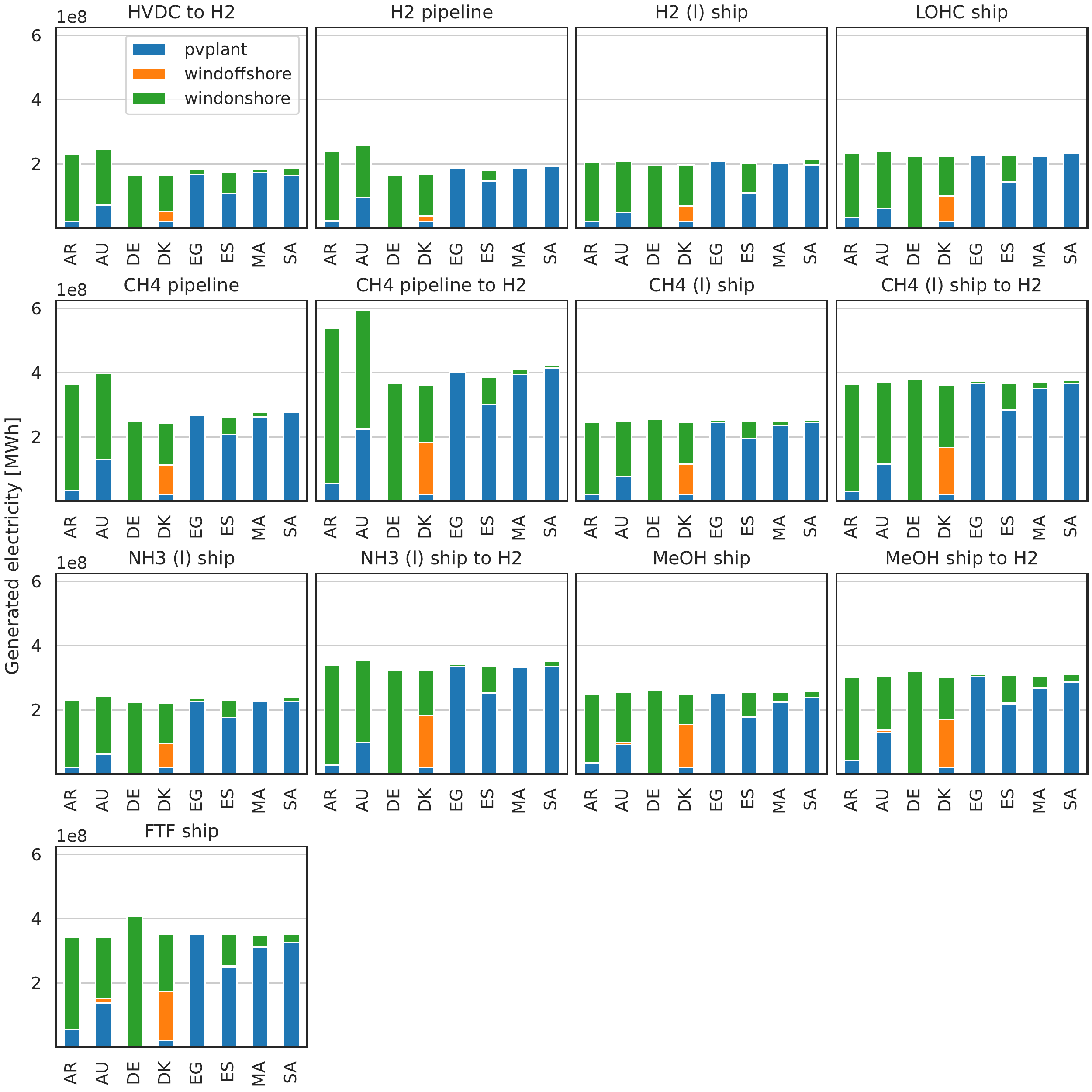}
    \caption{Electricity generation from \glsfmtshort{res} \glsfmtshort{esc} and exporting country under \SI{10}{\percentpa} \glsfmtshort{wacc} scenario for 2050.}
\end{minipage}

\begin{minipage}{\textwidth}
    \captionsetup{type=figure}
    \centering
\includegraphics[width=1\textwidth]{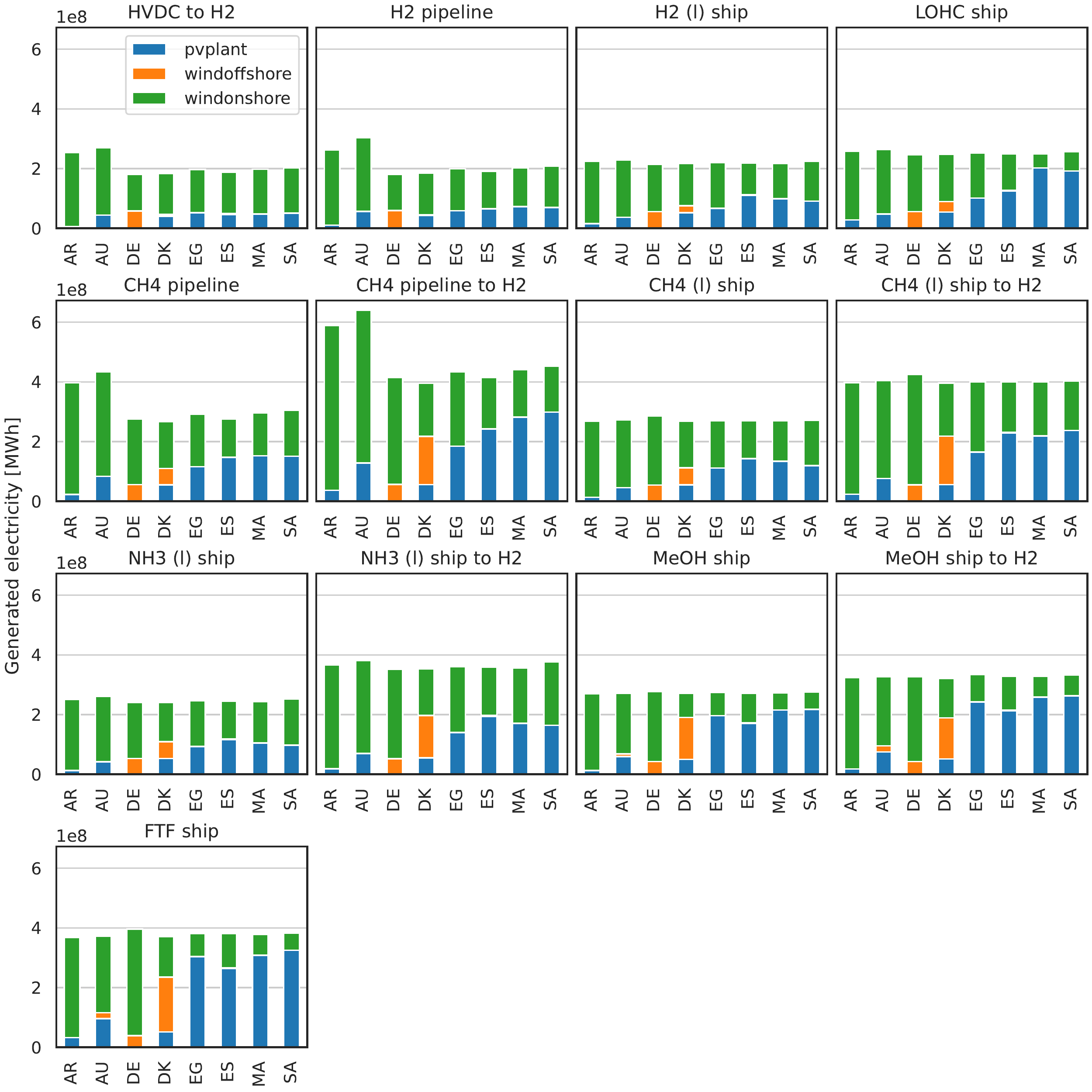}
    \caption{Electricity generation from \glsfmtshort{res} \glsfmtshort{esc} and exporting country under \SI{5}{\percentpa} \glsfmtshort{wacc} scenario for 2030.}
\end{minipage}

\begin{minipage}{\textwidth}
    \captionsetup{type=figure}
    \centering
\includegraphics[width=1\textwidth]{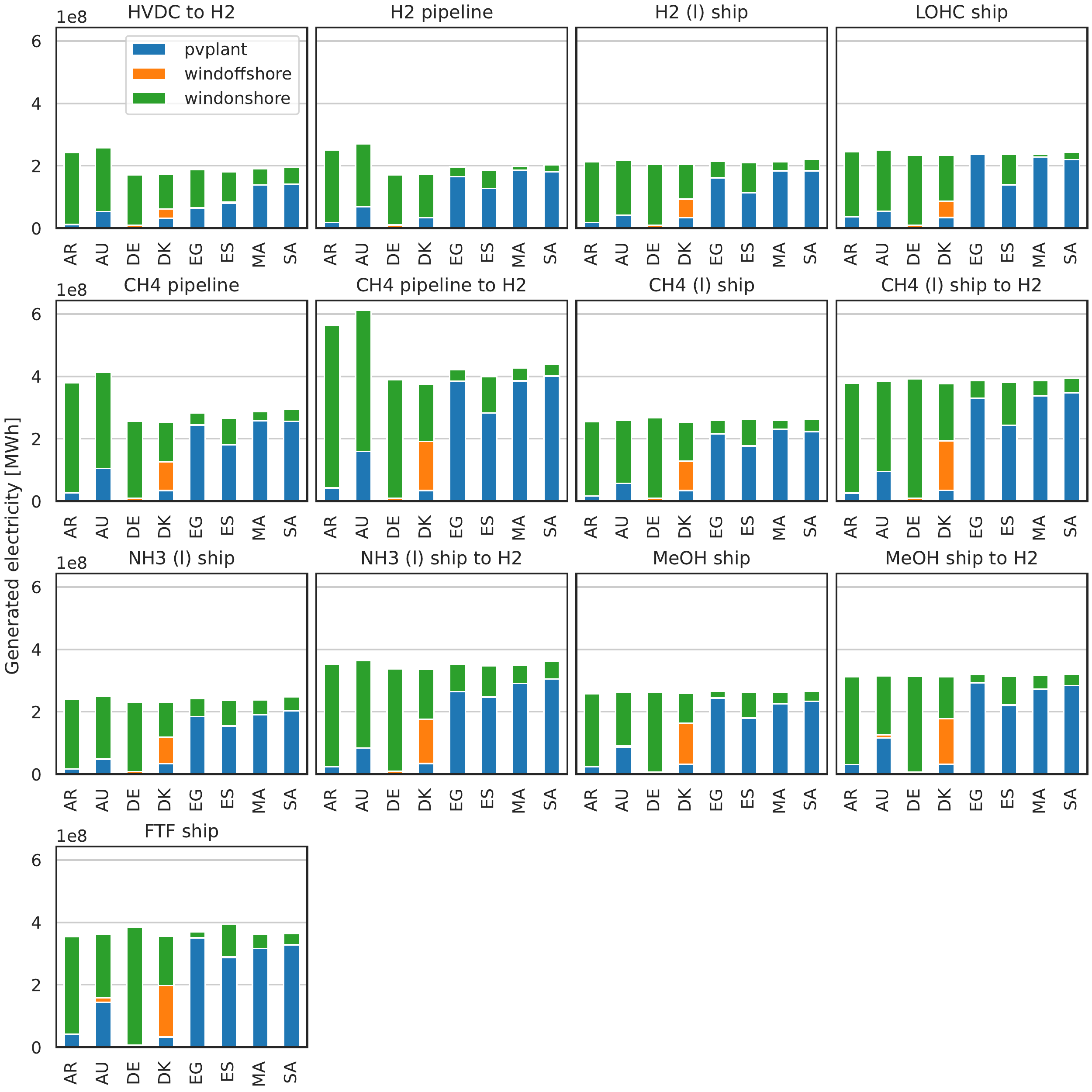}
    \caption{Electricity generation from \glsfmtshort{res} \glsfmtshort{esc} and exporting country under \SI{5}{\percentpa} \glsfmtshort{wacc} scenario for 2040.}
\end{minipage}

\begin{minipage}{\textwidth}
    \captionsetup{type=figure}
    \centering
\includegraphics[width=1\textwidth]{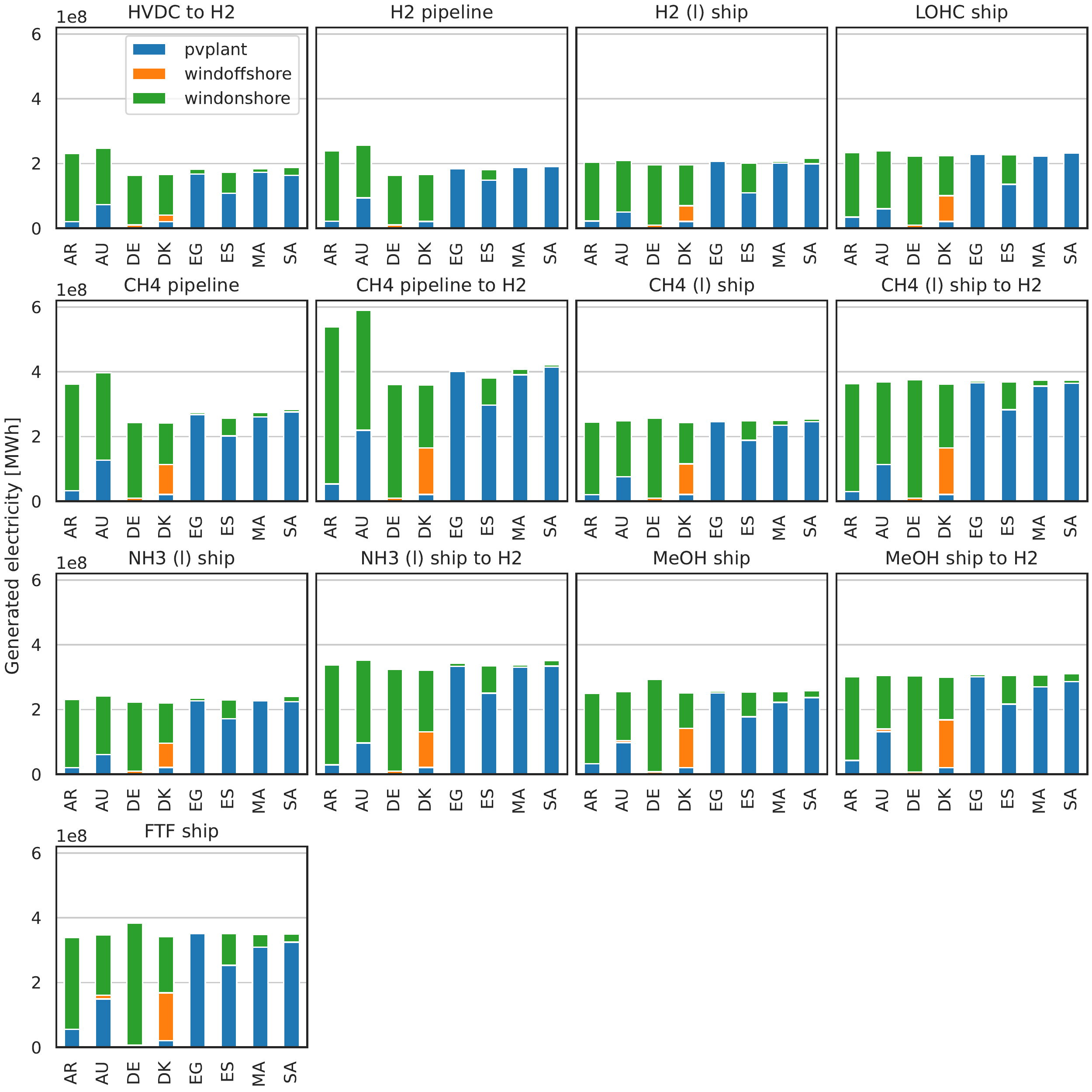}
    \caption{Electricity generation from \glsfmtshort{res} \glsfmtshort{esc} and exporting country under \SI{5}{\percentpa} \glsfmtshort{wacc} scenario for 2050.}
\end{minipage}

\begin{minipage}{\textwidth}
    \captionsetup{type=figure}
    \centering
\includegraphics[width=1\textwidth]{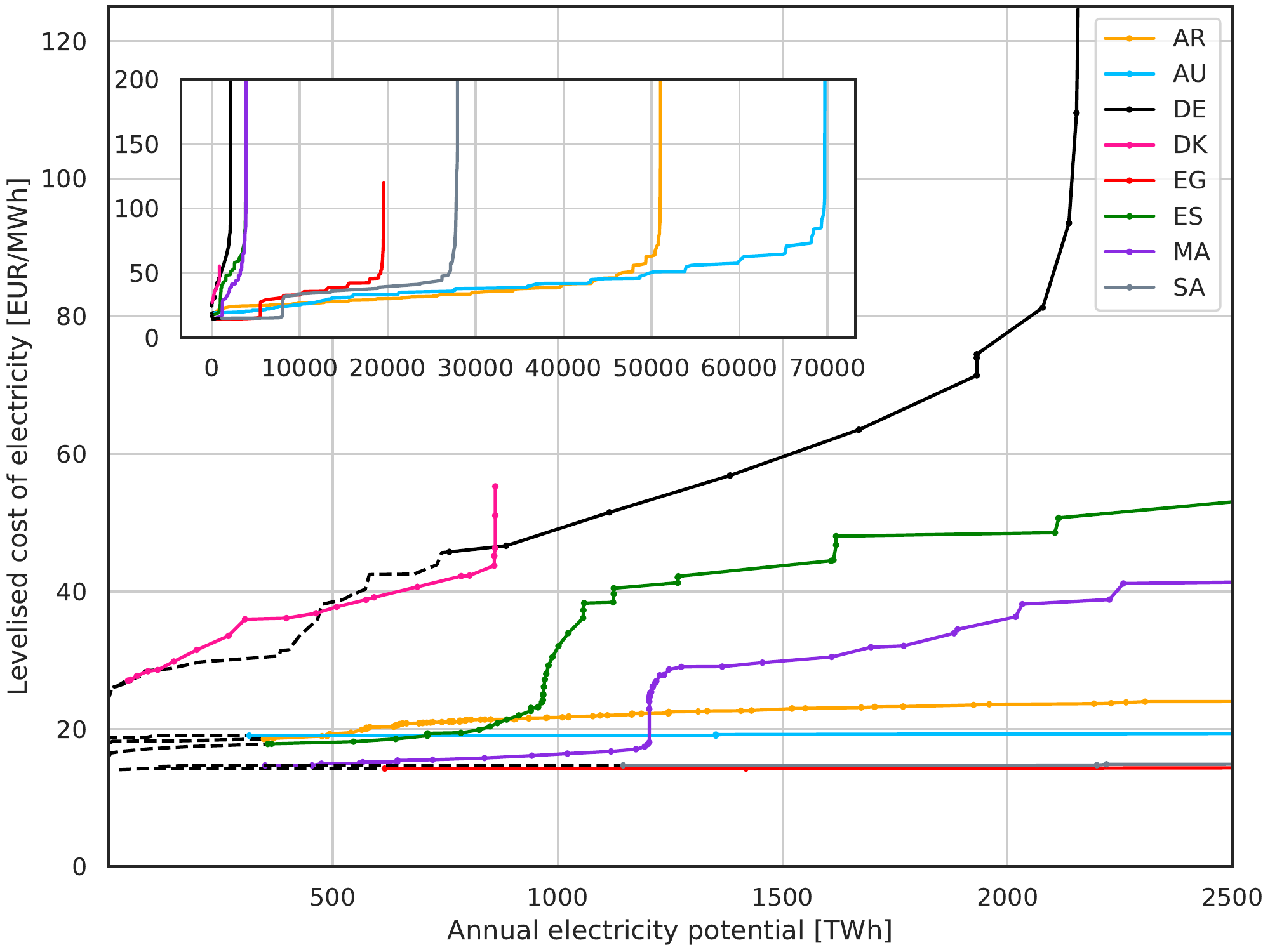}
    \caption{
        Similar figure as \autoref{fig:supply-curves} of electricity supply curves at \SI{10}{\percentpa} \glsfmtshort{wacc} for year 2040.
    }
\end{minipage}

\begin{minipage}{\textwidth}
    \captionsetup{type=figure}
    \centering
\includegraphics[width=1\textwidth]{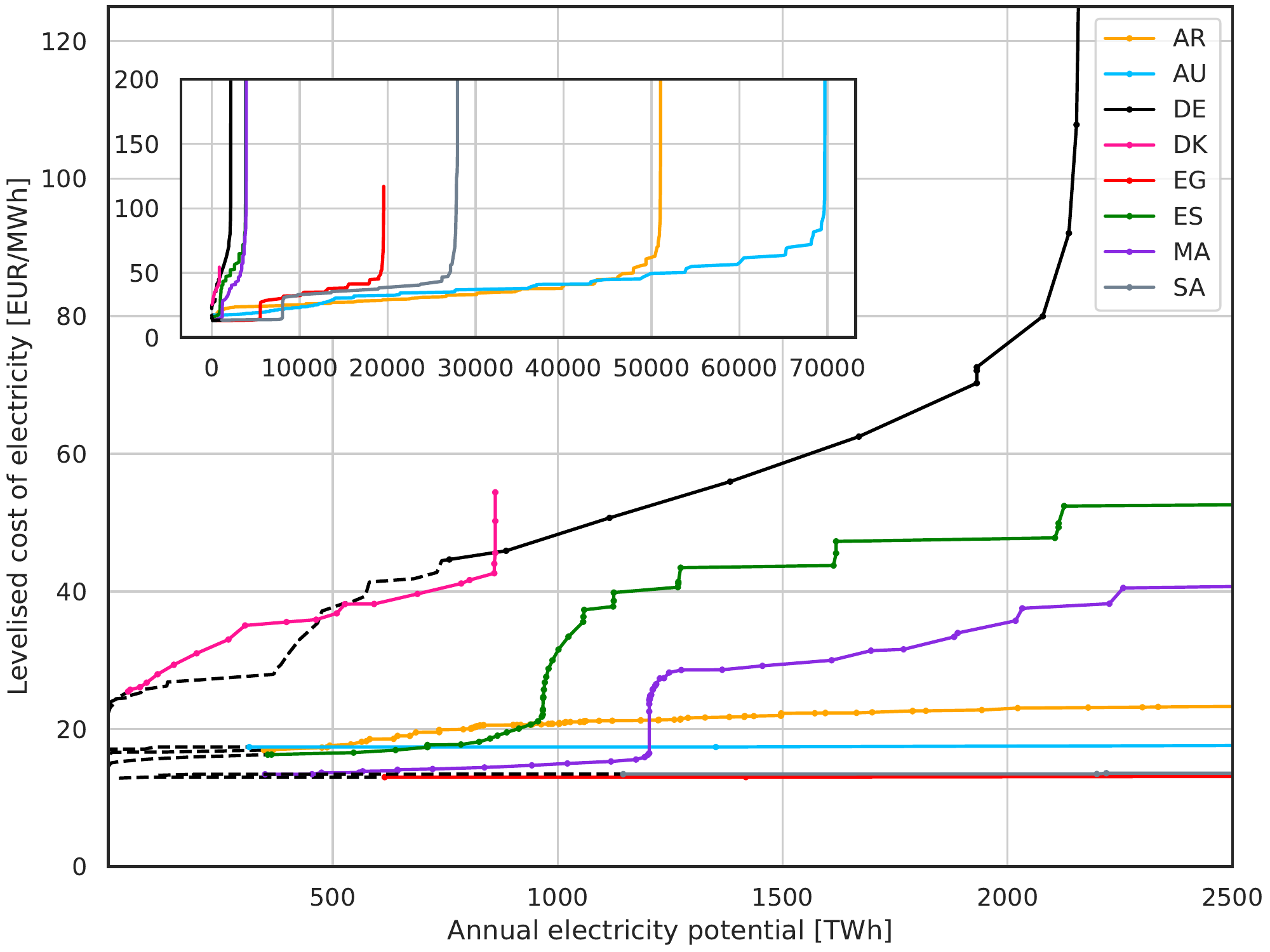}
    \caption{
        Similar figure as \autoref{fig:supply-curves} of electricity supply curves at \SI{10}{\percentpa} \glsfmtshort{wacc} for year 2050.
    }
\end{minipage}

\begin{minipage}{\textwidth}
    \captionsetup{type=figure}
    \centering
\includegraphics[width=1\textwidth]{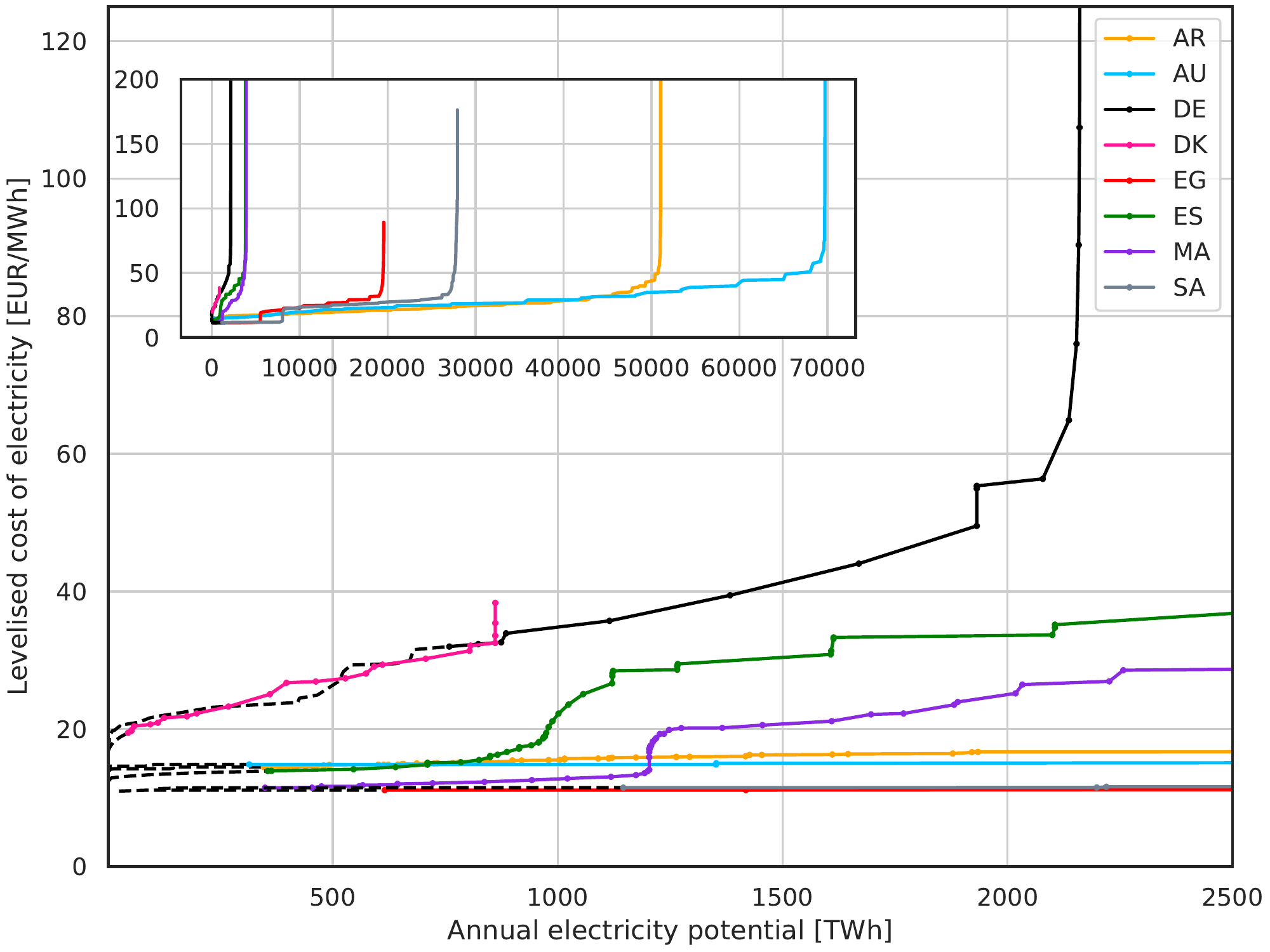}
    \caption{
        Similar figure as \autoref{fig:supply-curves} of electricity supply curves at \SI{5}{\percentpa} \glsfmtshort{wacc} for year 2030.
    }
\end{minipage}

\begin{minipage}{\textwidth}
    \captionsetup{type=figure}
    \centering
\includegraphics[width=1\textwidth]{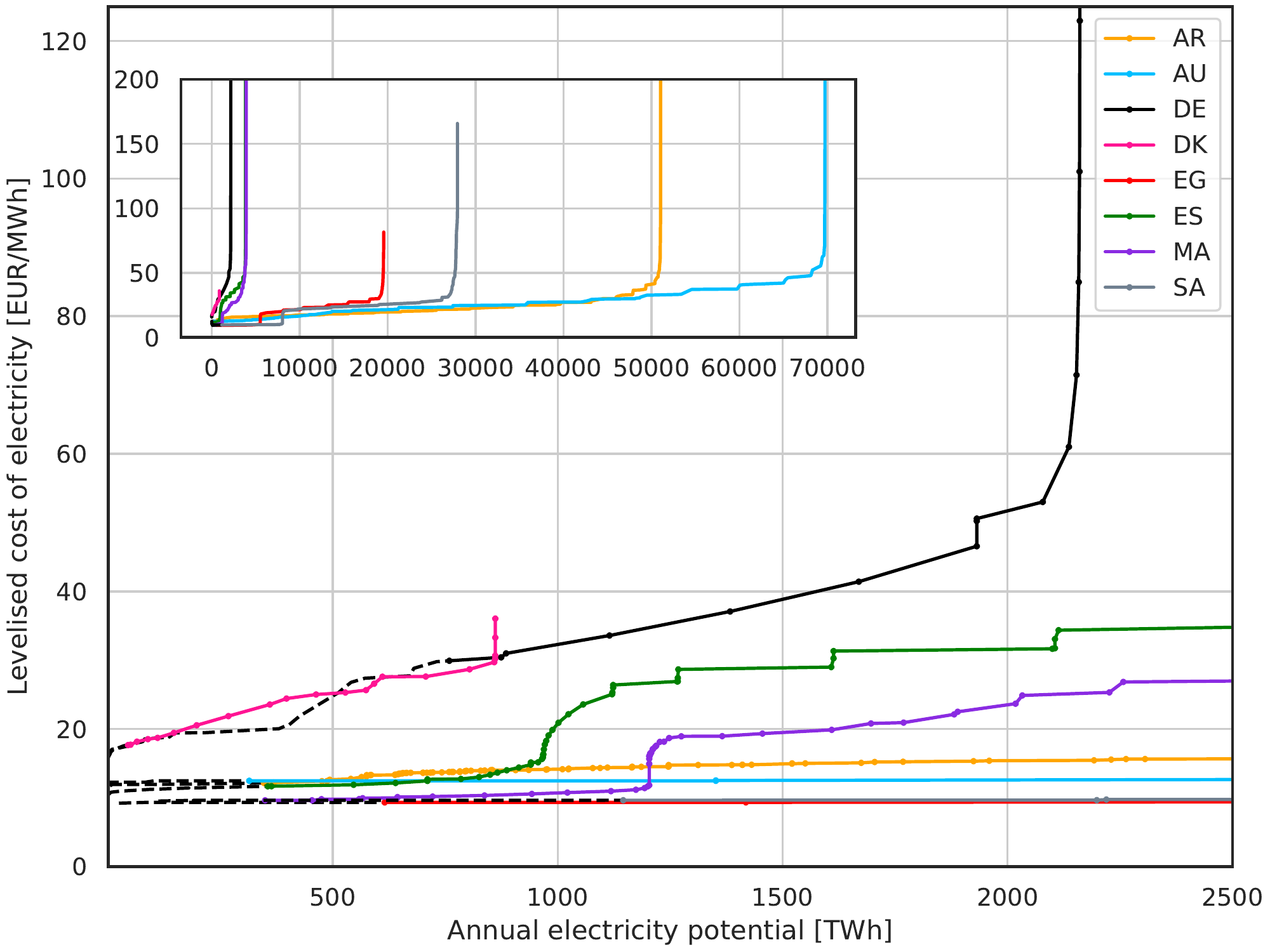}
    \caption{
        Similar figure as \autoref{fig:supply-curves} of electricity supply curves at \SI{5}{\percentpa} \glsfmtshort{wacc} for year 2040.
    }
\end{minipage}

\begin{minipage}{\textwidth}
    \captionsetup{type=figure}
    \centering
\includegraphics[width=1\textwidth]{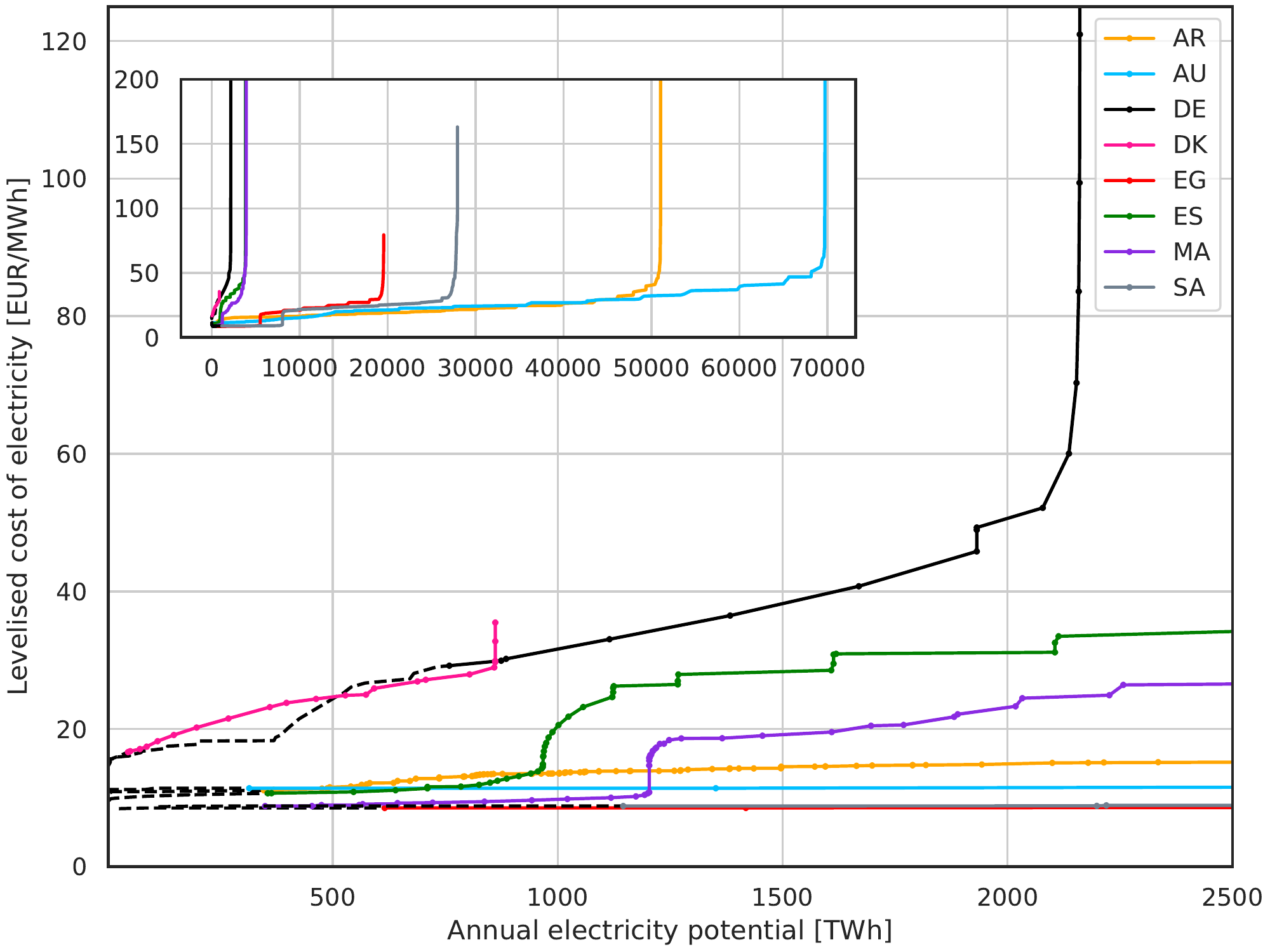}
    \caption{
        Similar figure as \autoref{fig:supply-curves} of electricity supply curves at \SI{5}{\percentpa} \glsfmtshort{wacc} for year 2050.
    }
\end{minipage}

\label{app:res-capacity-and-generation_end}
\clearpage
\modifysubsectionname{Figs}
\subsection{Cost composition for other years, lower \glsfmtshort{wacc}}
\label{app:cost-composition-other-scenarios-years}
\setcounter{page}{1}
\rfoot{\thepage/\pageref{app:cost-composition-other-scenarios-years_end}}

\begin{minipage}{\textwidth}
    \captionsetup{type=figure}
    \centering
\includegraphics[width=1\textwidth]{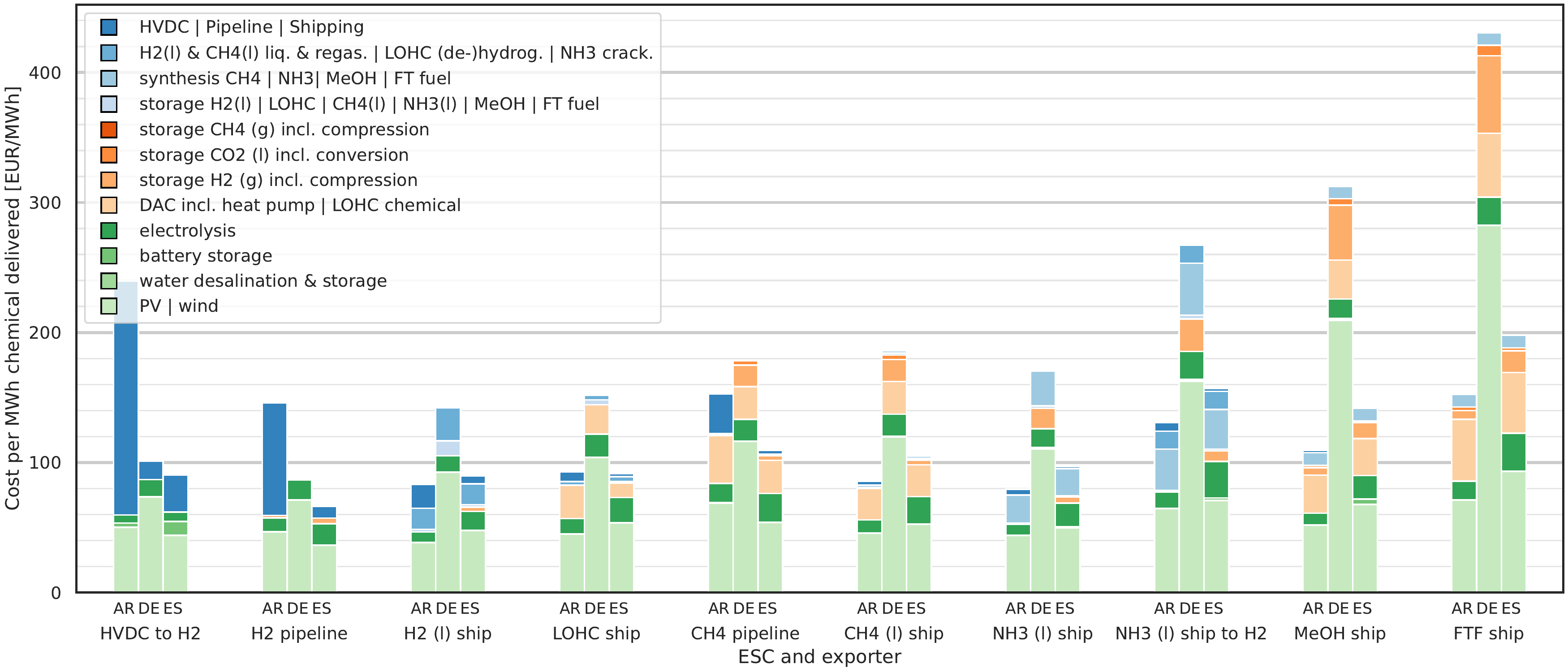}
    \caption{
        Similar figure as \autoref{fig:cost-compositionsselected-esc-exp} of cost compositions for selected \glspl{esc} at \SI{10}{\percentpa} \glsfmtshort{wacc} for year 2040.
    }
\end{minipage}

\begin{minipage}{\textwidth}
    \captionsetup{type=figure}
    \centering
\includegraphics[width=1\textwidth]{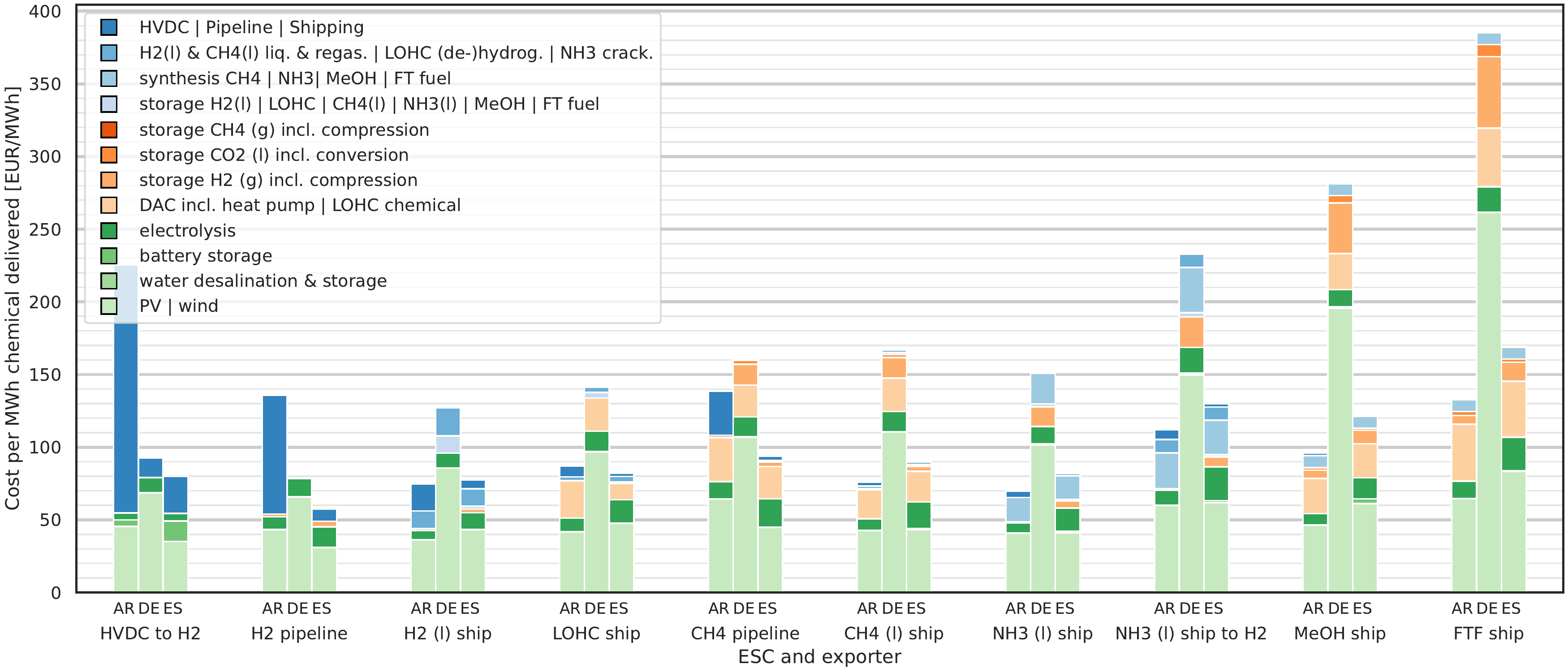}
    \caption{
        Similar figure as \autoref{fig:cost-compositionsselected-esc-exp} of cost compositions for selected \glspl{esc} at \SI{10}{\percentpa} \glsfmtshort{wacc} for year 2050.
    }
\end{minipage}

\begin{minipage}{\textwidth}
    \captionsetup{type=figure}
    \centering
\includegraphics[width=1\textwidth]{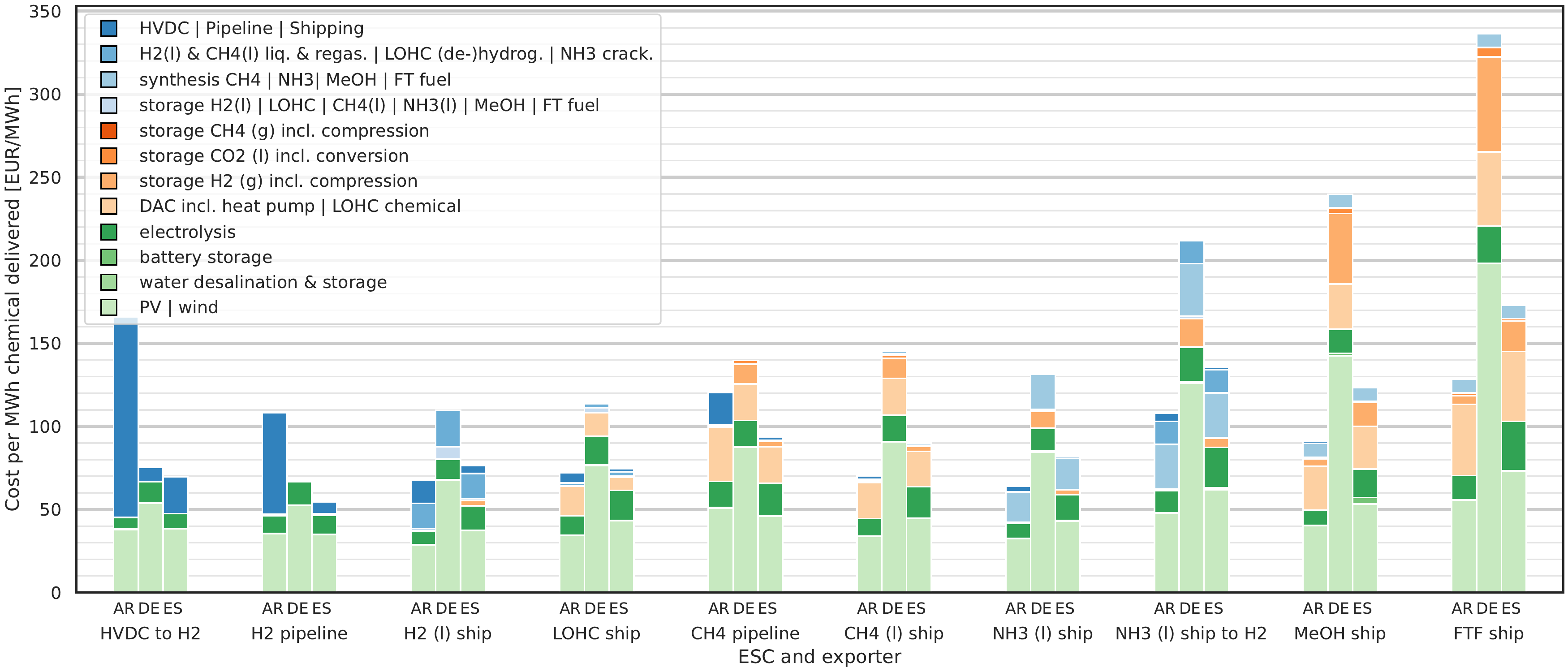}
    \caption{
        Cost compositions for selected \glspl{esc} at \SI{5}{\percentpa} \glsfmtshort{wacc} for year 2030.
    }
\end{minipage}

\begin{minipage}{\textwidth}
    \captionsetup{type=figure}
    \centering
\includegraphics[width=1\textwidth]{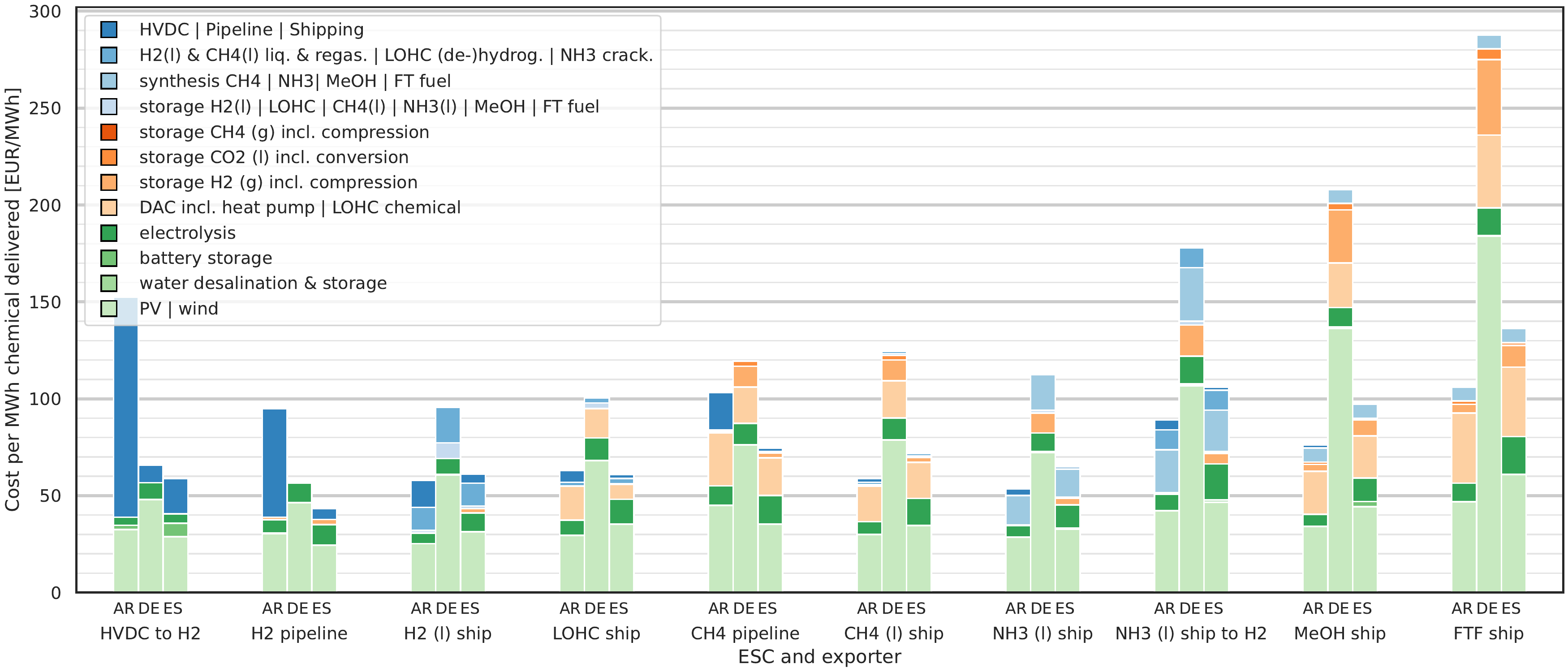}
    \caption{
        Cost compositions for selected \glspl{esc} at \SI{5}{\percentpa} \glsfmtshort{wacc} for year 2040.
    }
\end{minipage}

\begin{minipage}{\textwidth}
    \captionsetup{type=figure}
    \centering
\includegraphics[width=1\textwidth]{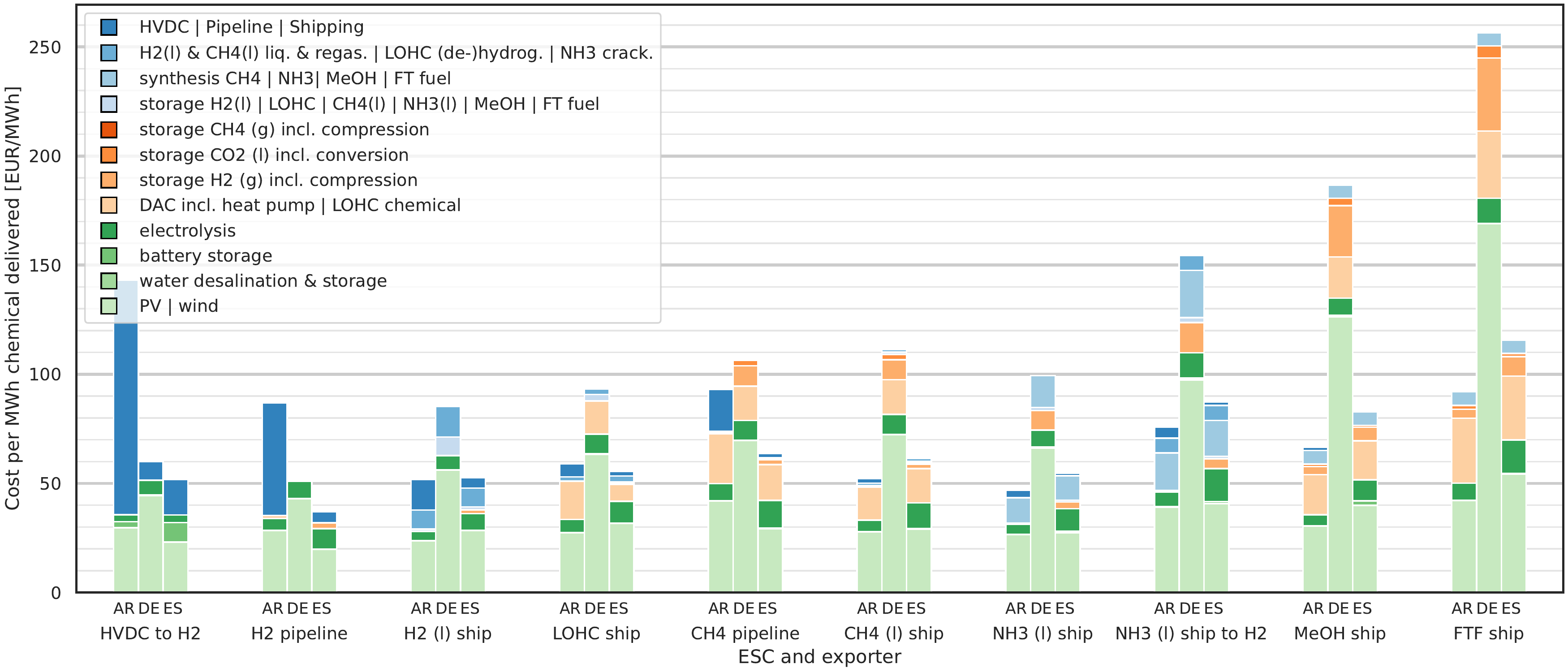}
    \caption{
        Cost compositions for selected \glspl{esc} at \SI{5}{\percentpa} \glsfmtshort{wacc} for year 2050.
    }
\end{minipage}

\label{app:cost-composition-other-scenarios-years_end}
\clearpage

\begin{landscape}
\modifysubsectionname{Table}
\subsection{Tabular results: \glsfmtlongpl{lcoe}}
\label{app:results-lcoes}
\setcounter{page}{1}
\rfoot{\thepage/\pageref{app:results-lcoes_end}}

\begin{longtable}{lSS*{8}{S[table-format=3.2, round-mode = places, round-precision = 2]}}
    \caption{
        Obtained \glsxtrfullpl{lcoe} for all \glsxtrfullpl{esc}, exporting countries and \SI{5}{\percentpa} as well as \SI{10}{\percentpa} \gls{wacc}.
    }\\
\toprule
    {ESC} & {year} & {WACC} 				& \multicolumn{8}{c}{\si{\EURpMWh}}\\
          &		   & {[\si{\percentpa}]}    & {\glsxtrshort{ar}} & {\glsxtrshort{au}} & {\glsxtrshort{de}} & {\glsxtrshort{dk}} & {\glsxtrshort{eg}} & {\glsxtrshort{es}} & {\glsxtrshort{ma}} & {\glsxtrshort{sa}} \\
    \midrule
    \endfirsthead
\toprule
    \multicolumn{11}{c}{{\tablename\ \thetable{} (continued).}} \\
    \midrule
    {ESC} & {year} & {WACC} 				& \multicolumn{8}{c}{\si{\EURpMWh}}\\
          &	       & {[\si{\percentpa}]}    & {\glsxtrshort{ar}} & {\glsxtrshort{au}} & {\glsxtrshort{de}} & {\glsxtrshort{dk}} & {\glsxtrshort{eg}} & {\glsxtrshort{es}} & {\glsxtrshort{ma}} & {\glsxtrshort{sa}} \\
        \midrule
    \endhead
\bottomrule
    \endfoot
\bottomrule
    \endlastfoot

pipeline-ch4 & 2030 & 5 & 120.35752105198084 & 151.41799706920213 & 139.76234960359312 & 95.56384733274724 & 87.3310263238632 & 93.60454081037658 & 92.4233106720082 & 94.0083082713256 \\
  & 2030 & 10 & 177.0257582441922 & 224.3715203982287 & 213.59751133196696 & 139.91746504272888 & 128.0711902401754 & 137.56931886975343 & 135.70800686164856 & 138.14170566297463 \\
  & 2040 & 5 & 103.27696379317445 & 129.78468607725853 & 119.390695698091 & 79.6033746910237 & 69.87854835195424 & 74.59092470885362 & 72.45347052216908 & 74.00972814035757 \\
  & 2040 & 10 & 152.72486935850523 & 193.1290830366381 & 178.37783306285849 & 116.16570626922156 & 102.14929334692184 & 109.46324037784068 & 106.02282820147732 & 108.53429758459863 \\
  & 2050 & 5 & 93.04483229676732 & 116.67226042523076 & 106.37076418371704 & 71.67361876151925 & 58.437203779139 & 63.63456786372226 & 60.686881424227 & 62.24361893783366 \\
  & 2050 & 10 & 138.51249519268563 & 174.69526595244008 & 159.80809647763988 & 105.18930524292132 & 85.96344318793722 & 93.74475066230347 & 89.34423615072845 & 91.85023285041262 \\
pipeline-h2 & 2030 & 5 & 108.14419718883214 & 138.70917950614196 & 66.90329856075282 & 49.6105155686853 & 54.23597949192028 & 54.70270733202344 & 58.9329022507381 & 62.52526751832517 \\
  & 2030 & 10 & 163.238665629141 & 210.11018936008807 & 103.69560777620512 & 75.06324451109967 & 82.56900417378245 & 83.15071270470315 & 89.62444567930027 & 95.199310658728 \\
\end{longtable}
 
\label{app:results-lcoes_end}
\clearpage
\modifysubsectionname{Table}
\subsection{Tabular results: \glsfmtlongpl{lcoh}}
\label{app:results-lcohs}
\setcounter{page}{1}
\rfoot{\thepage/\pageref{app:results-lcohs_end}}

\begin{longtable}{lSS*{8}{S[table-format=3.2, round-mode = places, round-precision = 2]}}
    \caption{
        Obtained \glsxtrfullpl{lcoh} for all \glsxtrfullpl{esc}, exporting countries and \SI{5}{\percentpa} as well as \SI{10}{\percentpa} \gls{wacc}.
    }\\
\toprule
    {ESC} & {year} & {WACC} 				& \multicolumn{8}{c}{\si{\EURpkgHydrogen}}\\
          &		   & {[\si{\percentpa}]}    & {\glsxtrshort{ar}} & {\glsxtrshort{au}} & {\glsxtrshort{de}} & {\glsxtrshort{dk}} & {\glsxtrshort{eg}} & {\glsxtrshort{es}} & {\glsxtrshort{ma}} & {\glsxtrshort{sa}} \\
    \midrule
    \endfirsthead
\toprule
    \multicolumn{11}{c}{{\tablename\ \thetable{} (continued).}} \\
    \midrule
    {ESC} & {year} & {WACC} 				& \multicolumn{8}{c}{\si{\EURpkgHydrogen}}\\
          &	       & {[\si{\percentpa}]}    & {\glsxtrshort{ar}} & {\glsxtrshort{au}} & {\glsxtrshort{de}} & {\glsxtrshort{dk}} & {\glsxtrshort{eg}} & {\glsxtrshort{es}} & {\glsxtrshort{ma}} & {\glsxtrshort{sa}} \\
    \midrule
    \endhead
\bottomrule
    \endfoot
\bottomrule
    \endlastfoot

hvdc & 2030 & 5 & 5.535940510079113 & 7.107059824605651 & 2.513687367390153 & 1.989754038152519 & 2.502273014777862 & 2.3208555054785367 & 2.7516672312094057 & 2.9532253969839988 \\
  & 2030 & 10 & 8.683011164837795 & 11.125728905480793 & 3.9298397166219474 & 3.026645047015176 & 3.85320158823881 & 3.558697648235839 & 4.237876641061448 & 4.549924859356461 \\
  & 2040 & 5 & 5.079541386150102 & 6.5273140117966815 & 2.1893943661271047 & 1.6627058005902018 & 2.198939911385693 & 1.963256724609485 & 2.355332937989531 & 2.561301843746767 \\
  & 2040 & 10 & 7.987206792780917 & 10.233987068675225 & 3.368902949318191 & 2.5395683834964844 & 3.393754166066027 & 3.0177207408556783 & 3.632514891287971 & 3.9533333121856797 \\
  & 2050 & 5 & 4.772804411340405 & 6.116886642490774 & 2.0024155872536955 & 1.5210534064748693 & 1.949515456416508 & 1.7275769215780628 & 2.0396971984192604 & 2.2497096778416856 \\
  & 2050 & 10 & 7.5127407372712005 & 9.599398135813095 & 3.0912163580043988 & 2.3327667047137863 & 3.0161476140367927 & 2.6649128569244414 & 3.1557833632237524 & 3.483403771963745 \\
pipeline-ch4 & 2030 & 5 & 6.188343024837631 & 7.731905443936944 & 7.41197453186251 & 5.126809268424957 & 4.527850395407097 & 4.91073711780745 & 4.789708791600259 & 4.871708427551658 \\
  & 2030 & 10 & 9.096353966016878 & 11.44895979681103 & 11.237323296780394 & 7.4646370252302585 & 6.630645735604043 & 7.210675744661293 & 7.013975782674494 & 7.142995777259211 \\
\end{longtable}

\label{app:results-lcohs_end}
\clearpage
\modifysubsectionname{Table}
\subsection{Technology assumptions}
\label{app:technology-assumptions}
\setcounter{page}{1}
\rfoot{\thepage/\pageref{app:technology-assumptions_end}}

All technology costs are given in \si{\EUR}\-2015.
Inflation adjustment was done where necessary assuming a \SI{2}{\percentpa} inflation rate.

\begin{longtable}{p{.1\linewidth}p{.07\linewidth}p{.03\linewidth}p{.075\linewidth}p{.1\linewidth}p{.55\linewidth}}
    \caption{
        Technology cost and lifetime assumptions used for 2030/2040/2050.
        A machine readable version of the input assumption can be found in the Zenodo and GitHub repositories
        listed in the data availability section.
    }\\
\toprule
    technology & parameter & year & value & unit & source \\
    \midrule
    \endfirsthead
\toprule
    \multicolumn{6}{c}{{\tablename\ \thetable{} (continued).}} \\
    \midrule
    technology & parameter & year & value & unit & source \\
    \midrule
    \endhead
\bottomrule
    \endfoot
\bottomrule
    \endlastfoot

direct air capture & \multirow{3}{*}{CAPEX} & 2030 & 6000000.0 & \multirow{3}{*}{EUR/(tCO2/h)} & \multirow{3}{\linewidth}{Danish Energy Agency, ../technology\_data\_for\_industrial\_process\_heat\_0002.xlsx} \\
             &                               & 2040 & 5000000.0 &                                                           & \\
             &                               & 2050 & 4000000.0 &                                                           & \\

 & \multirow{3}{*}{lifetime} & 2030 & 20.0 & \multirow{3}{*}{years} & \multirow{3}{\linewidth}{Danish Energy Agency, ../technology\_data\_for\_industrial\_process\_heat\_0002.xlsx} \\
 &                                  & 2040 & 20.0 &                                                         & \\
 &                                  & 2050 & 20.0 &                                                         & \\

 & \multirow{3}{*}{FOM} & 2030 & 4.95 & \multirow{3}{*}{\%/year} & \multirow{3}{\linewidth}{Danish Energy Agency, ../technology\_data\_for\_industrial\_process\_heat\_0002.xlsx} \\
 &                             & 2040 & 4.95 &                                                    & \\
\end{longtable}

\label{app:technology-assumptions_end}
\clearpage
\modifysubsectionname{Table}
\subsection{Conversion efficiencies}
\label{app:conversion-efficiencies}
\setcounter{page}{1}
\rfoot{\thepage/\pageref{app:conversion-efficiencies_end}}

\begin{longtable}{p{.15\linewidth}p{.05\linewidth}p{.1\linewidth}p{.1\linewidth}p{.1\linewidth}p{.1\linewidth}p{.35\linewidth}}
    \caption{
        Energy and feedstock efficiency assumptions used.
        A machine readable version of these assumptions can be found in the Zenodo and GitHub repositories
        listed in the data availability section.
    }\\
\toprule
    process & year & from & amount & to  & amount & details and source \\
    \midrule
    \endfirsthead
\toprule
    \multicolumn{6}{c}{{\tablename\ \thetable{} (continued).}} \\
    \midrule
    process & year & from & amount & to  & amount & details and source \\
    \midrule
    \endhead
\bottomrule
    \endfoot
\bottomrule
    \endlastfoot

Ammonia cracker & all & ammonia (g) & 3191.00 MWh(LHV) & hydrogen (g) & 2186.00 MWh(LHV) & Assuming a integrated 200t/d cracking and purification facility. Electricity demand (316 MWh per 2186 MWh(LHV) H2 output) is assumed to also be ammonia LHV input which seems a fair assumption as the facility has options for a higher degree of integration according to the report)..
 Source: ENGIE et al (2020): Ammonia to Green Hydrogen Feasibility Study (\url{https://assets.publishing.service.gov.uk/government/uploads/system/uploads/attachment\_data/file/880826/HS420\_-\_Ecuity\_-\_Ammonia\_to\_Green\_Hydrogen.pdf),} Fig. 10. \\
CCGT & all & hydrogen (g) & 1.00 MWh(LHV) & electricity & 0.58 MWh(el) & nan.
 Source: DEA technology data for electricity generation and district heating. \\
CH4 (g) fill compressor station & all & methane (g) & 1.00 p.u. & methane (g) compressed & 0.97 p.u. & Assume same value as for booster stations, need to provide initial pressure increase..
 Source: Guesstimate. \\
CH4 (g) pipeline & all & methane (g) compressed & 1.00 p.u. & methane (g) compressed submarine & 0.97 p.u. & Per 1000km for losses and boosting. Based on 0.4\%/100 miles estimate..
 Source: Asia Pacific Energy Research Centre (2000): \url{https://aperc.or.jp/file/2010/9/26/Natural\_Gas\_Infrastructure\_Development\_Northeast\_Asia\_2000.pdf} , Table 49. \\
CH4 (g) pipeline decompressor & all & methane (g) compressed & 1.00 p.u. & methane (g) & 1.00 p.u. & Arbitrary componenet for degrading gas pipeline content to regular methane (g) bus..
 Source: Guesstimate. \\
CH4 (g) submarine pipeline & all & methane (g) compressed submarine & 1.00 p.u. & methane (g) compressed & 0.97 p.u. & Per 1000km for losses and boosting. Based on 0.4\%/100 miles estimate..
 Source: Asia Pacific Energy Research Centre (2000): \url{https://aperc.or.jp/file/2010/9/26/Natural\_Gas\_Infrastructure\_Development\_Northeast\_Asia\_2000.pdf} , Table 49. \\
\end{longtable}

\label{app:conversion-efficiencies_end}
\clearpage
\modifysubsectionname{Table}
\subsection{Shipping parameters}
\label{app:shipping-assumptions}
\setcounter{page}{1}
\rfoot{\thepage/\pageref{app:shipping-assumptions_end}}

\begin{longtable}{p{.15\linewidth}p{.15\linewidth}p{.1\linewidth}S[table-format=3.2, round-mode = places]p{.35\linewidth}}
    \caption{
        Shipping parameters used.
        A machine readable version of these assumptions can be found in the Zenodo and GitHub repositories
        listed in the data availability section.
    }\\
\toprule
    ship type & parameter & unit &  {\parbox{0.15\linewidth}{\centering value}} & details and source\\
    \midrule
    \endfirsthead
\toprule
    \multicolumn{5}{c}{{\tablename\ \thetable{} (continued).}} \\
    \midrule
    ship type & parameter & unit & {value} & details and source\\
    \midrule
    \endhead
\bottomrule
    \endfoot
\bottomrule
    \endlastfoot

CH4 (l) transport ship & (un-) loading losses & \%/transfer &       0.7000 &                                                                                                                                                  Approx. 450 t for a 130 000m\textasciicircum 3 tanker based on \url{https://citeseerx.ist.psu.edu/viewdoc/download?doi=10.1.1.470.6116\&rep=rep1\&type=pdf} , pg. 59 f . \\
                       & (un-) loading time & h &      48.0000 &                                                                                                                                                                 Hurskainen 2019: \url{https://cris.vtt.fi/en/publications/liquid-organic-hydrogen-carriers-lohc-concept-evaluation-and-tech} , table 8. \\
                       & average speed & km/h &      37.0000 &                                                                                                                                                                 Hurskainen 2019: \url{https://cris.vtt.fi/en/publications/liquid-organic-hydrogen-carriers-lohc-concept-evaluation-and-tech} , table 8. \\
                       & boil-off & \%/h &       0.0050 &                                                                                                                                                                                                                                                      ca. 0.1\% per day.Lowell et al. 2013, pg. 14. \\
                       & capacity & MWh\_LHV &  809717.0000 &                                                                                                                                                                                                                            Based on CH4 LHV of 13.8888 MWh/t\_CH4 and 58300 t capacity.Calculated. \\
\end{longtable}

\label{app:shipping-assumptions_end}
\end{landscape}

\clearpage
\modifysubsectionname{Figs}
\subsection{\glsfmttext{res} eligible area masks}
\label{app:country-masks}
\setcounter{page}{1}
\rfoot{\thepage/3}

The following figures show the eligible areas considered for onshore wind and \gls{pv} \gls{res}
for the countries while determining their \gls{res} potentials.
For this study no differentiation between \enquote{plant A} and \enquote{plant B} was used.
Both eligible types of area were combined to the total eligible area for each technology.
\begin{figure}[hbt]
    \begin{subfigure}{.5\linewidth}
        \centering
\includegraphics[width=.8\textwidth]{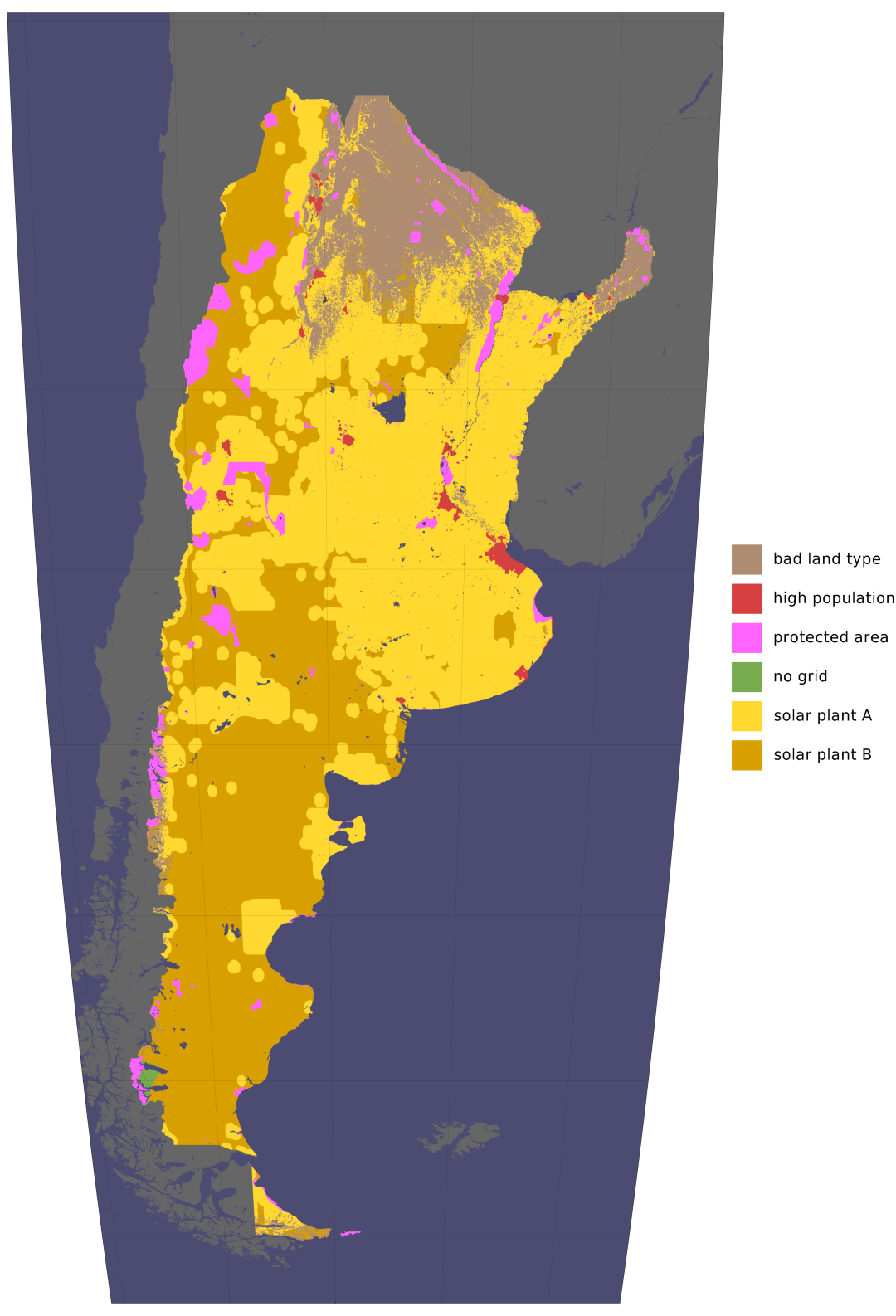}
        \caption{\gls{pv} mask for \glsxtrlong{ar}.}
    \end{subfigure}
\begin{subfigure}{.5\linewidth}
        \centering
\includegraphics[width=.8\textwidth]{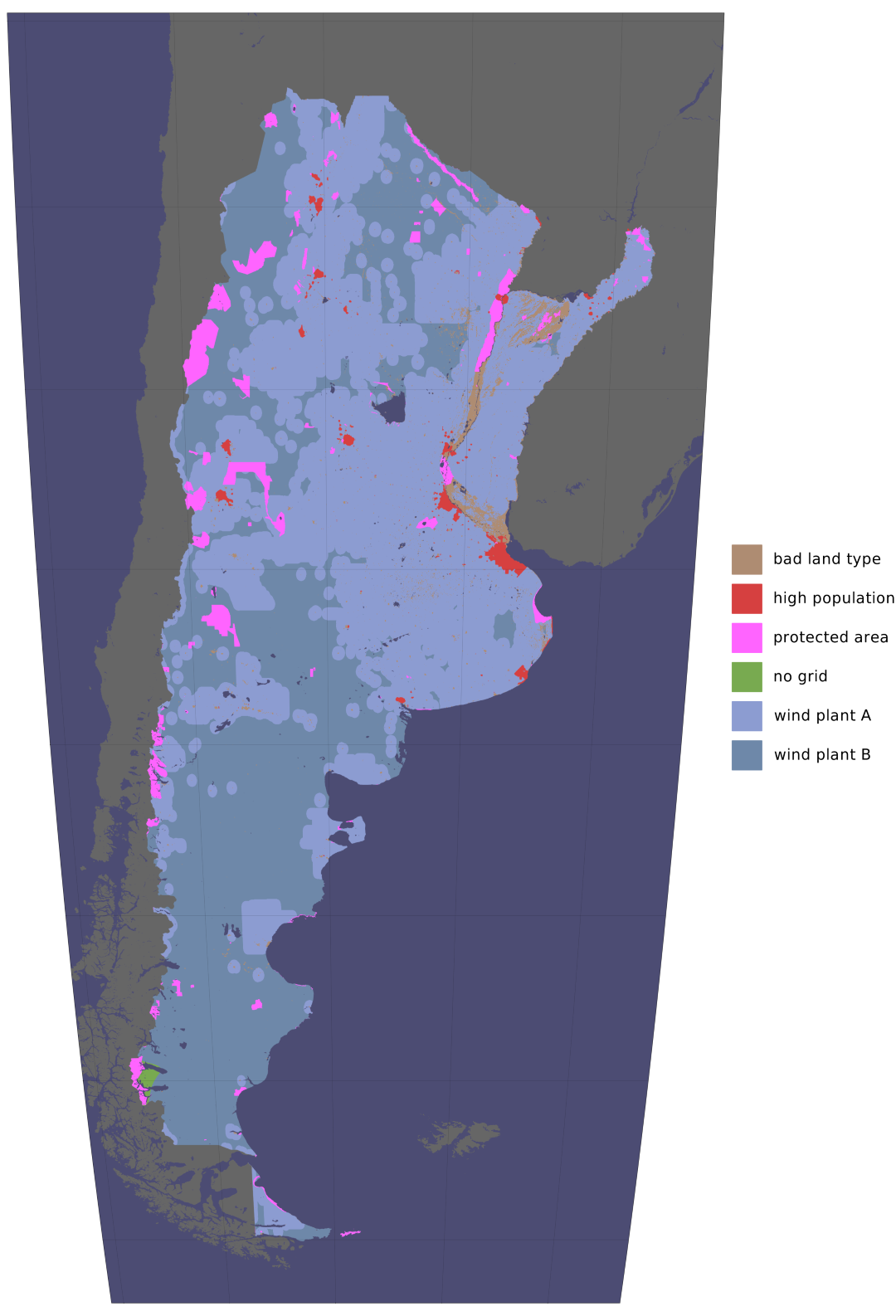}
        \caption{Onshore wind mask for \glsxtrlong{ar}.}
    \end{subfigure}

    \begin{subfigure}{.5\linewidth}
        \centering
\includegraphics[width=.8\textwidth]{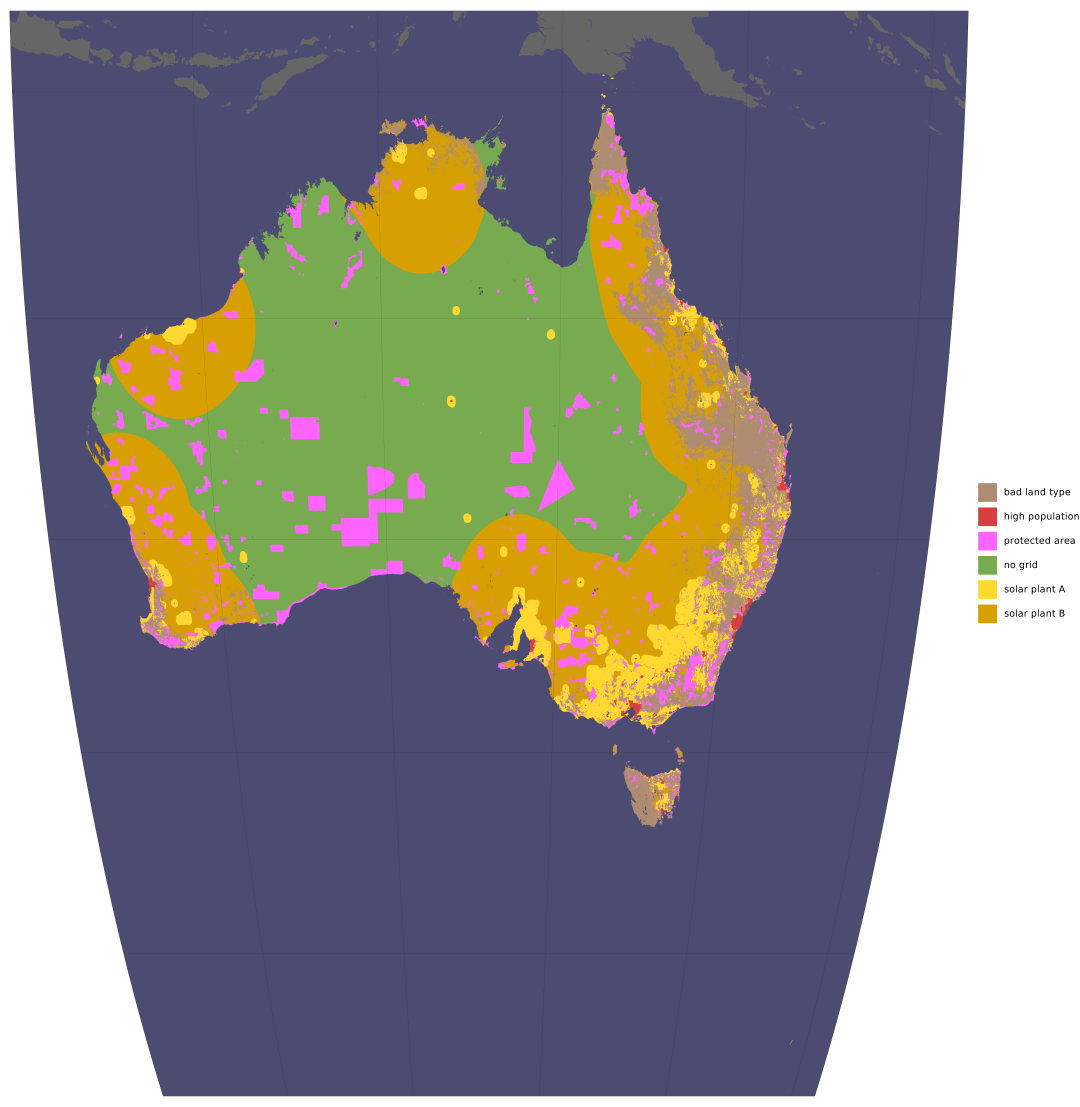}
        \caption{\gls{pv} mask for \glsxtrlong{au}.}
    \end{subfigure}
\begin{subfigure}{.5\linewidth}
        \centering
\includegraphics[width=.8\textwidth]{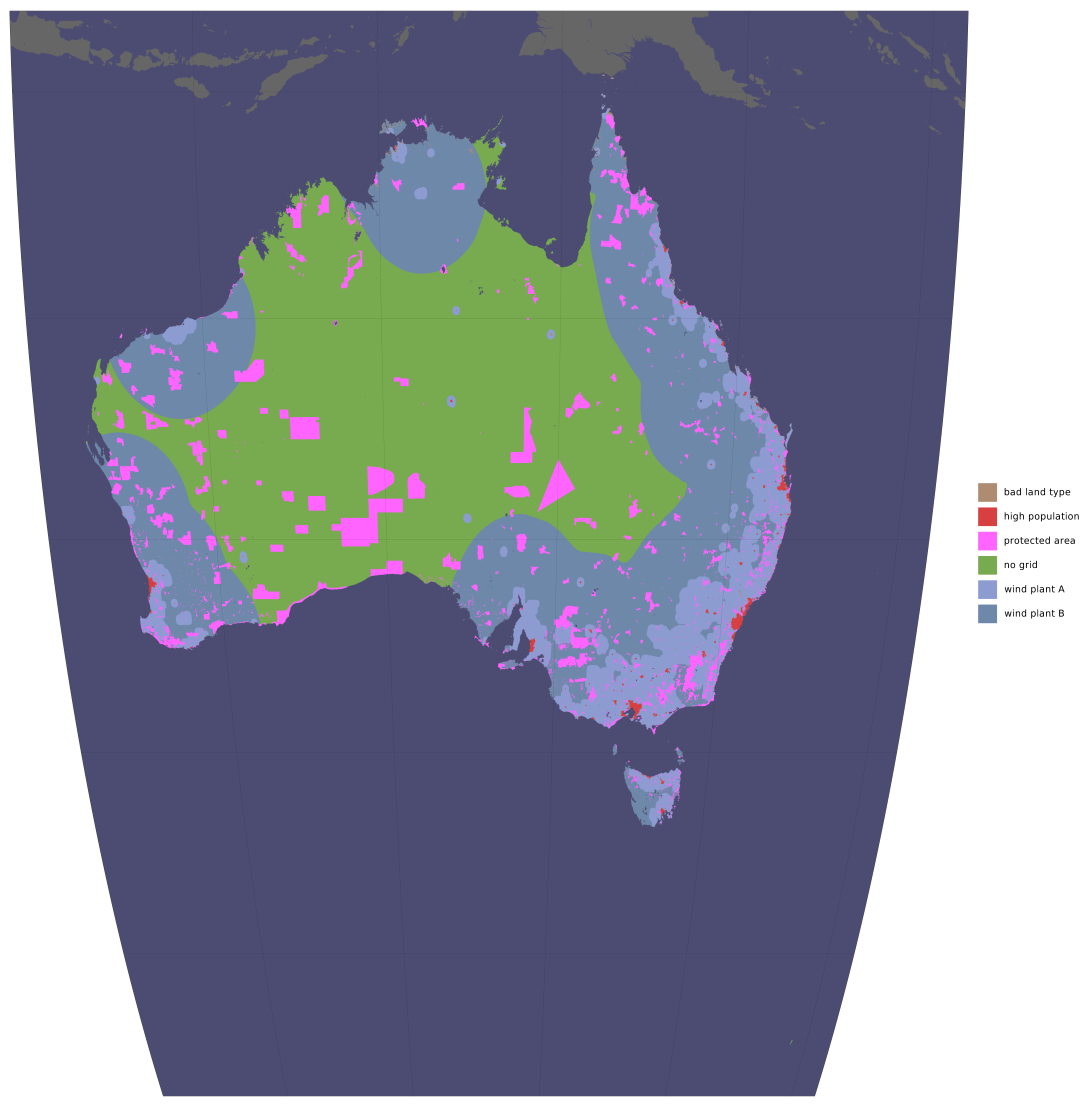}
        \caption{Onshore wind mask for \glsxtrlong{au}.}
    \end{subfigure}
\end{figure}
%\clearpage
\begin{figure}\ContinuedFloat

    \begin{subfigure}{.5\linewidth}
        \centering
\includegraphics[width=.9\textwidth]{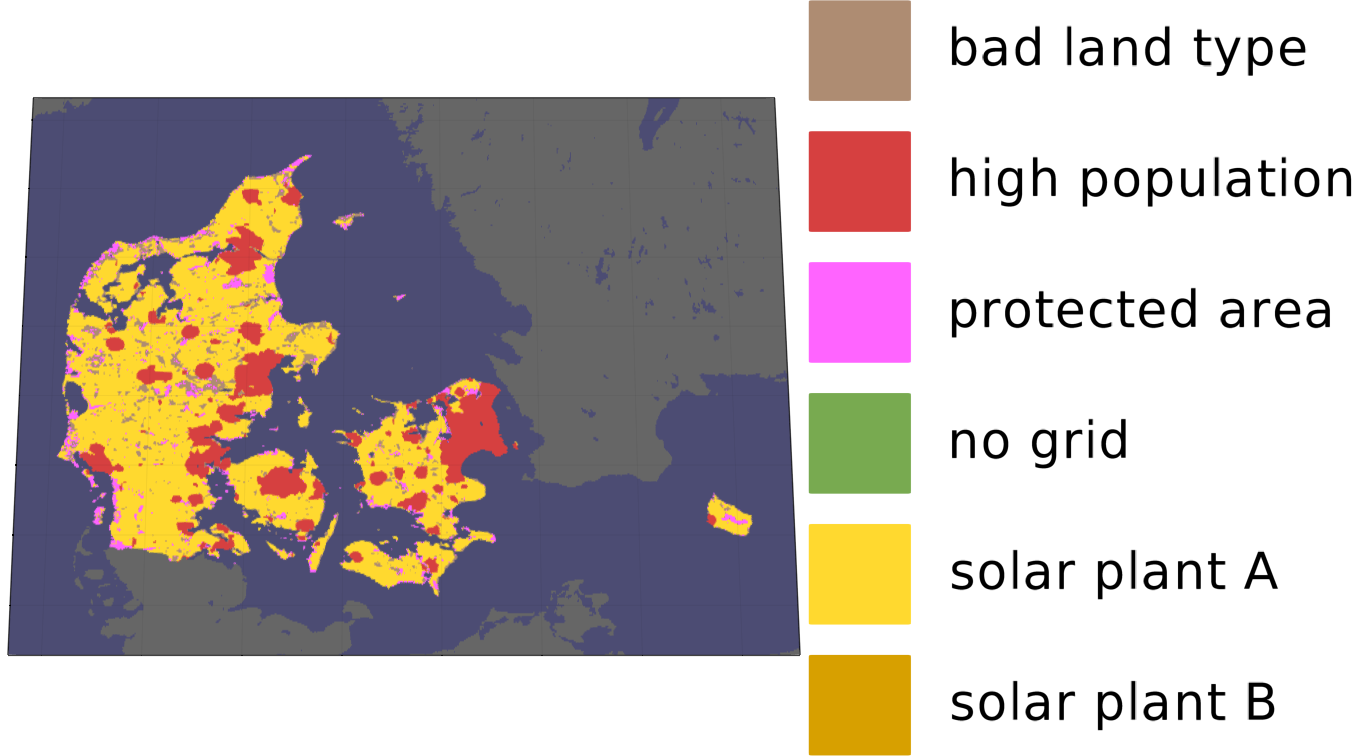}
        \caption{\gls{pv} mask for \glsxtrlong{dk}.}
    \end{subfigure}
\begin{subfigure}{.5\linewidth}
        \centering
\includegraphics[width=.9\textwidth]{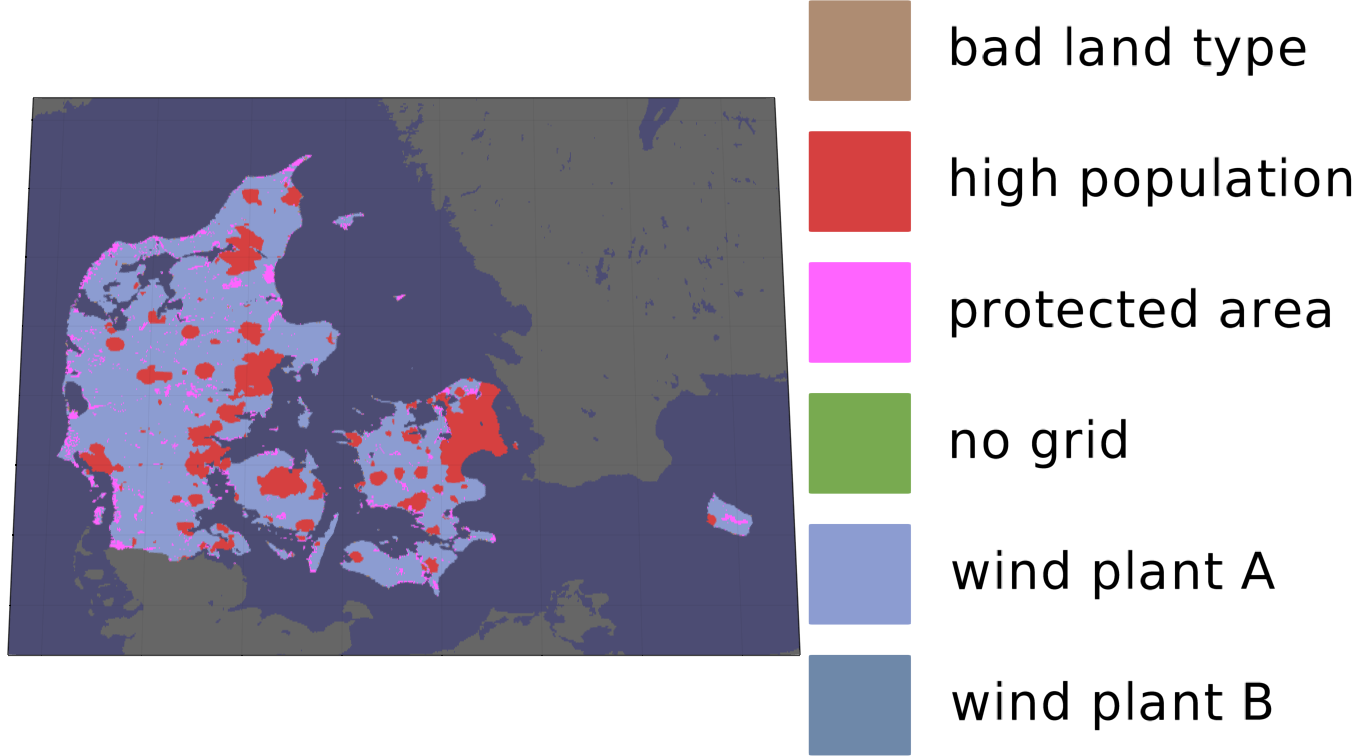}
        \caption{Onshore wind mask for \glsxtrlong{dk}.}
    \end{subfigure}

    \begin{subfigure}{.5\linewidth}
        \centering
\includegraphics[width=.9\textwidth]{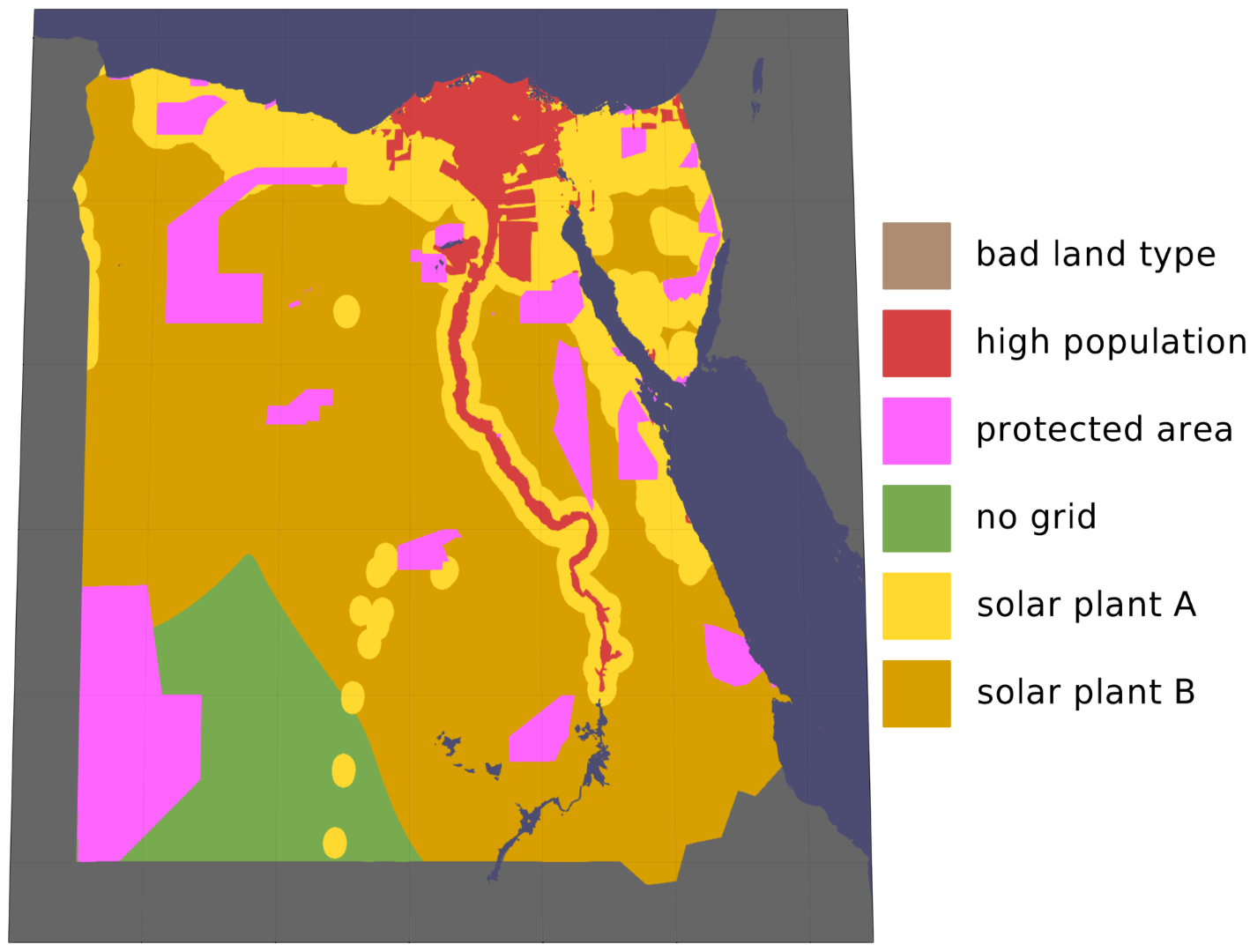}
        \caption{\gls{pv} mask for \glsxtrlong{eg}.}
    \end{subfigure}
\begin{subfigure}{.5\linewidth}
        \centering
\includegraphics[width=.9\textwidth]{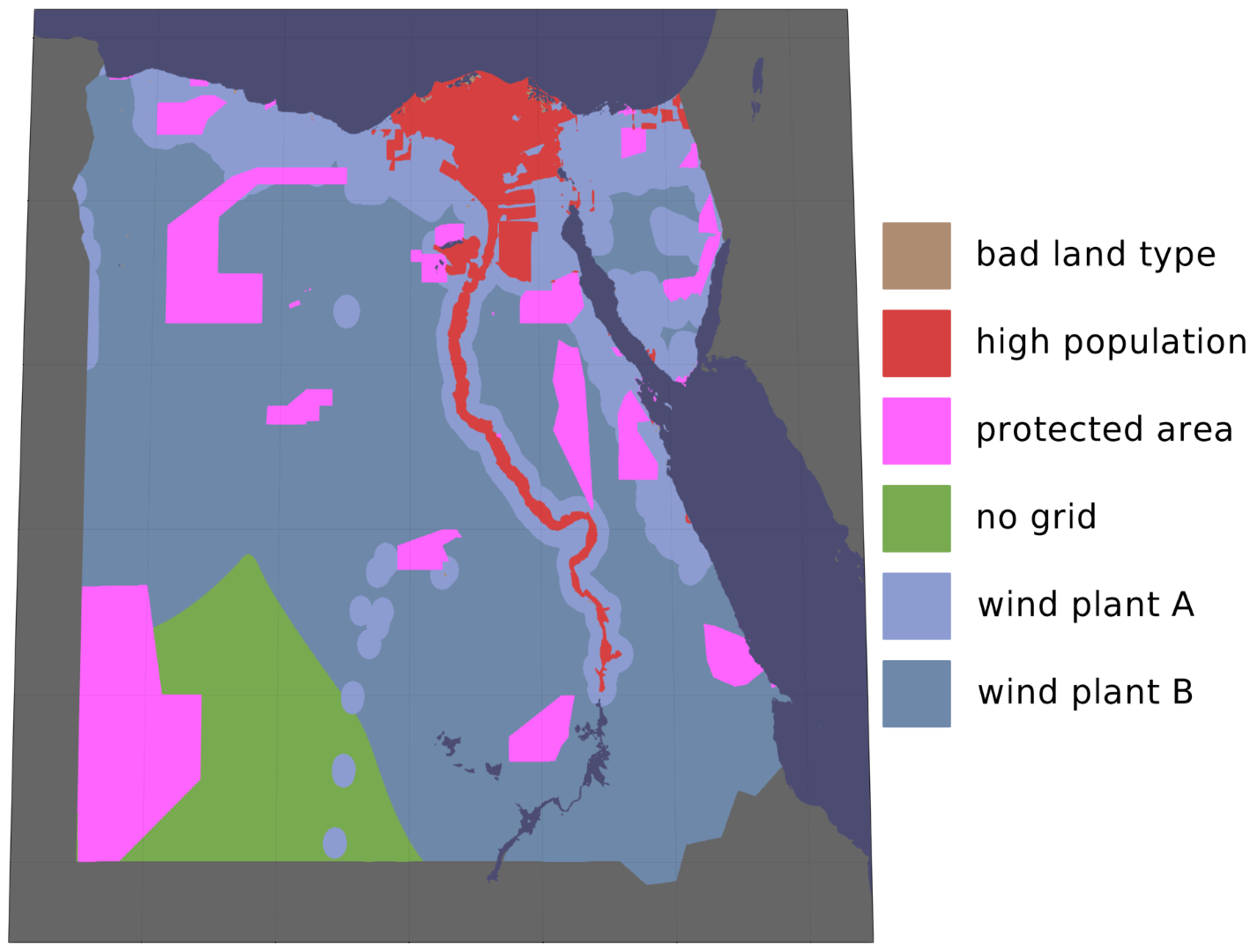}
        \caption{Onshore wind mask for \glsxtrlong{eg}.}
    \end{subfigure}

    \begin{subfigure}{.5\linewidth}
        \centering
\includegraphics[width=.9\textwidth]{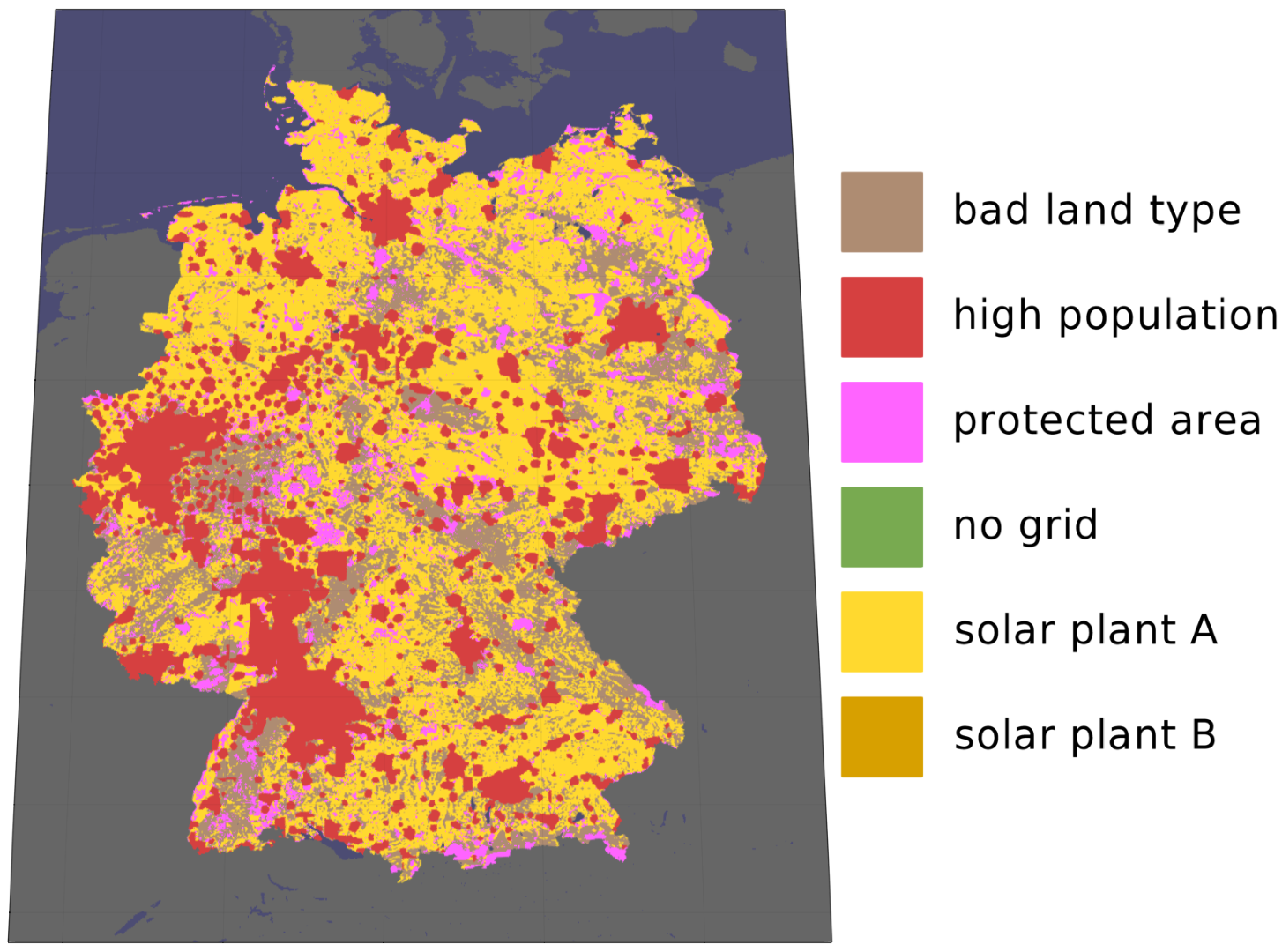}
        \caption{\gls{pv} mask for \glsxtrlong{de}.}
    \end{subfigure}
\begin{subfigure}{.5\linewidth}
        \centering
\includegraphics[width=.9\textwidth]{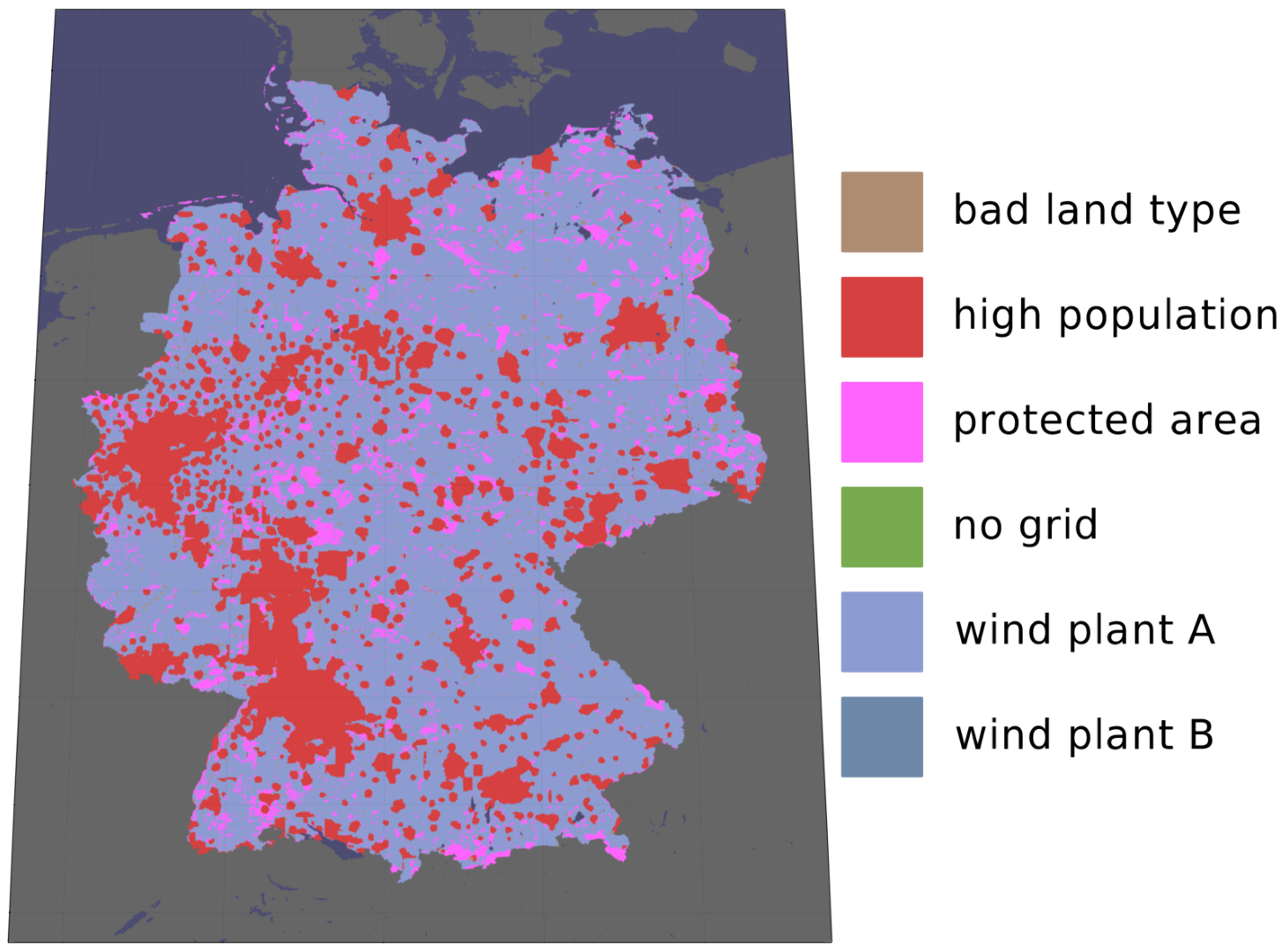}
        \caption{Onshore wind mask for \glsxtrlong{de}.}
    \end{subfigure}
\end{figure}
%\clearpage
\begin{figure}\ContinuedFloat
    
    \begin{subfigure}{.5\linewidth}
        \centering
\includegraphics[width=.9\textwidth]{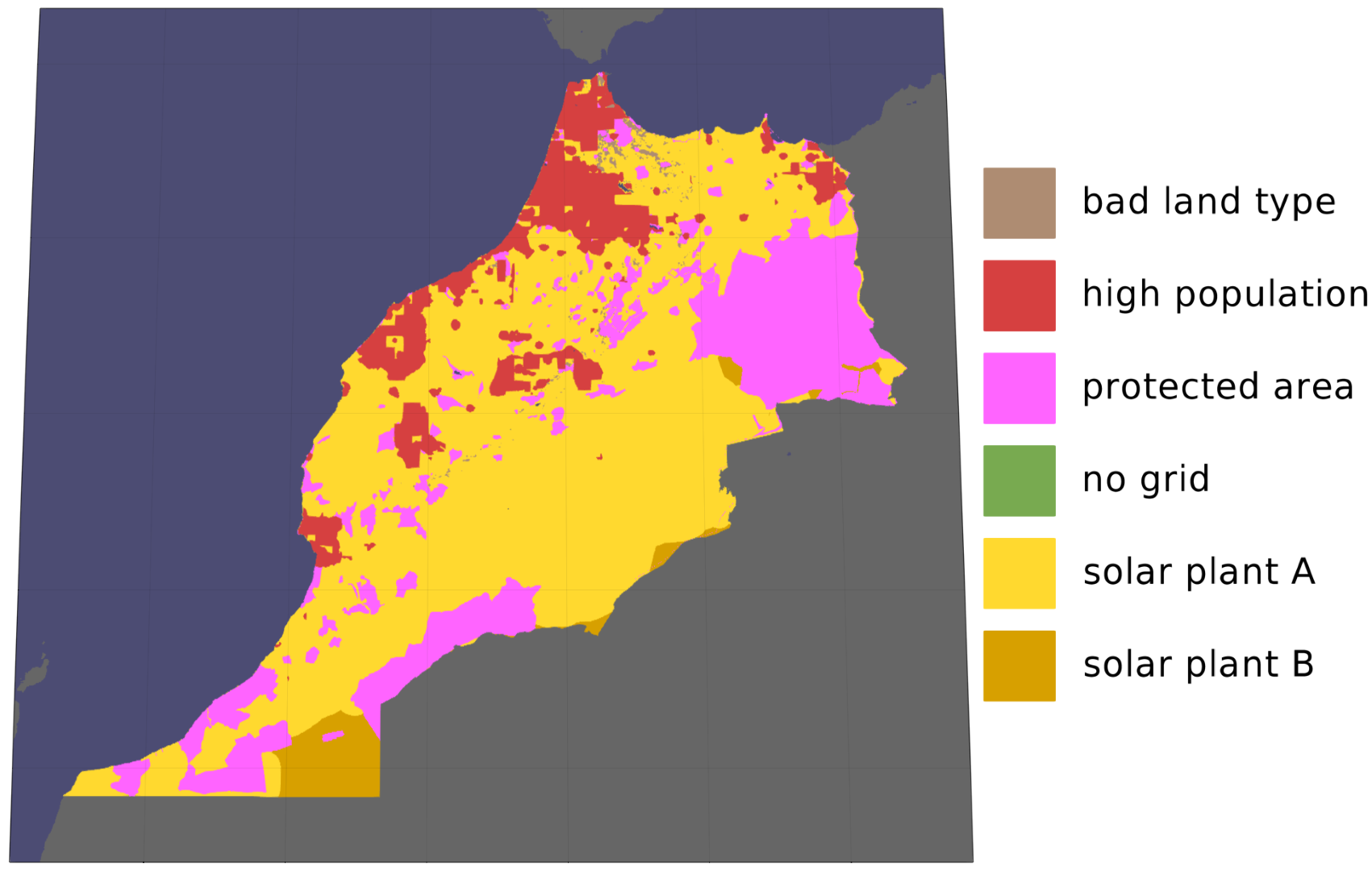}
        \caption{\gls{pv} mask for \glsxtrlong{ma}.}
    \end{subfigure}
\begin{subfigure}{.5\linewidth}
        \centering
\includegraphics[width=.9\textwidth]{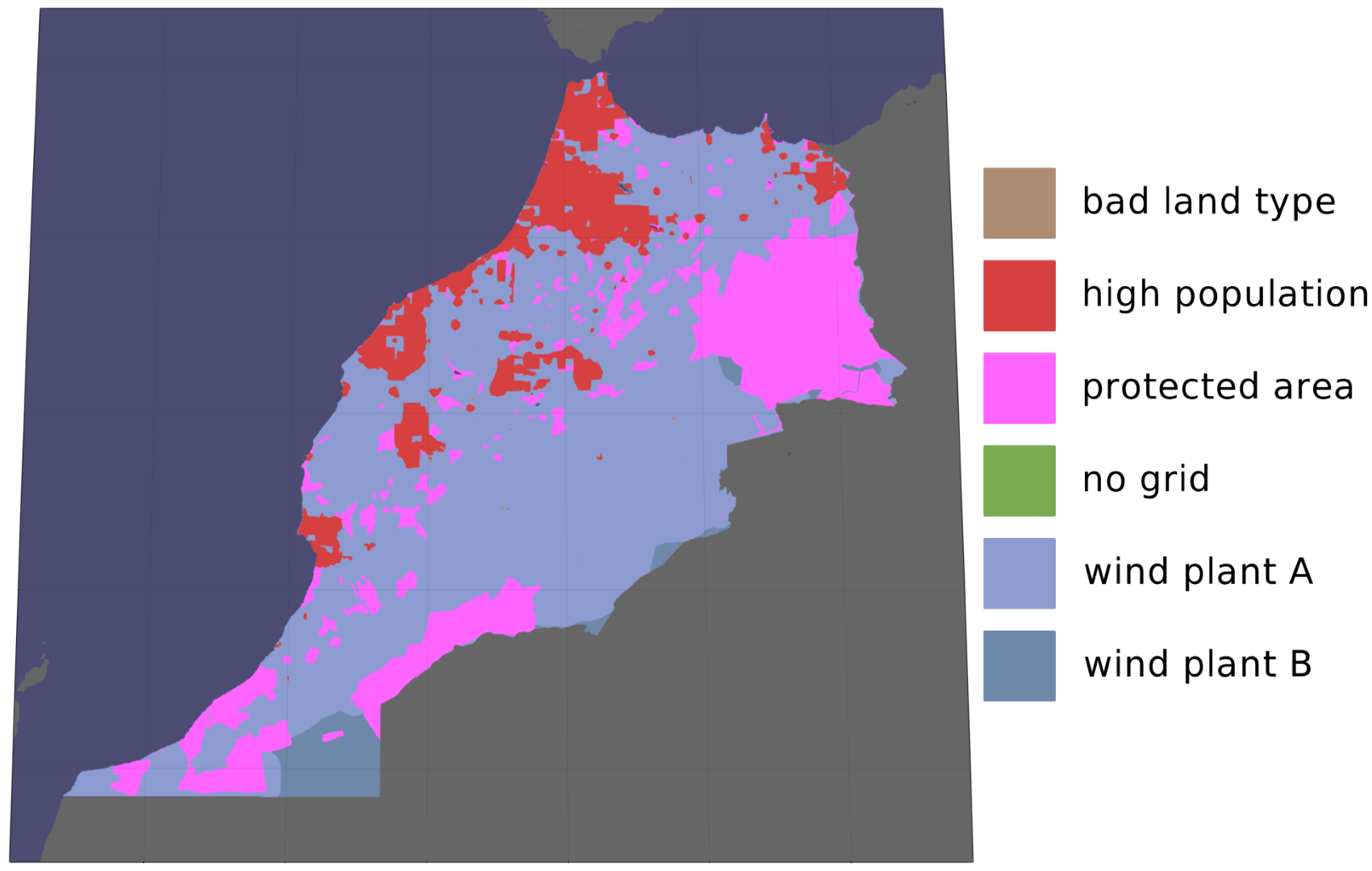}
        \caption{Onshore wind mask for \glsxtrlong{ma}.}
    \end{subfigure}

    \begin{subfigure}{.5\linewidth}
        \centering
\includegraphics[width=.9\textwidth]{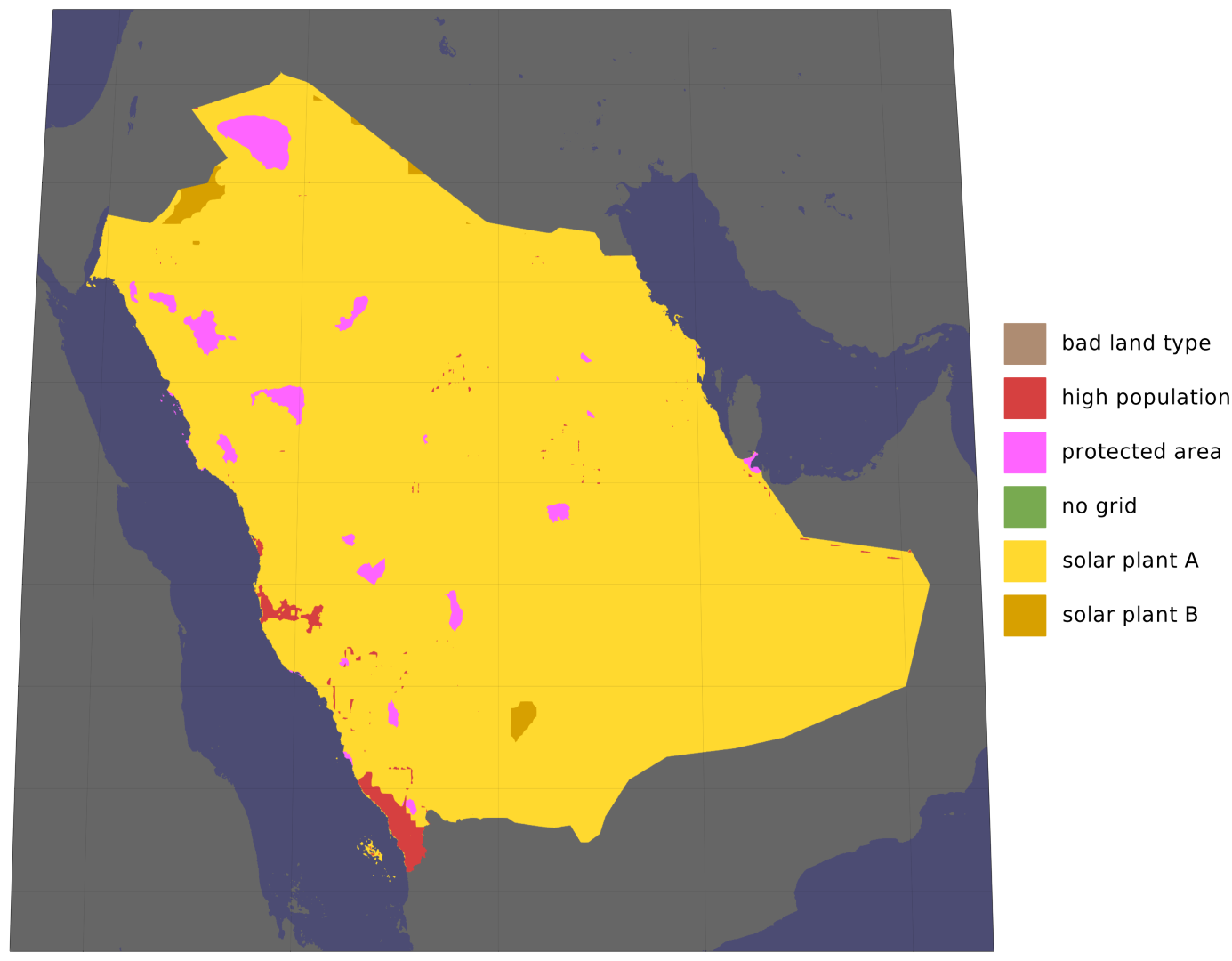}
        \caption{\gls{pv} mask for \glsxtrlong{sa}.}
    \end{subfigure}
\begin{subfigure}{.5\linewidth}
        \centering
\includegraphics[width=.9\textwidth]{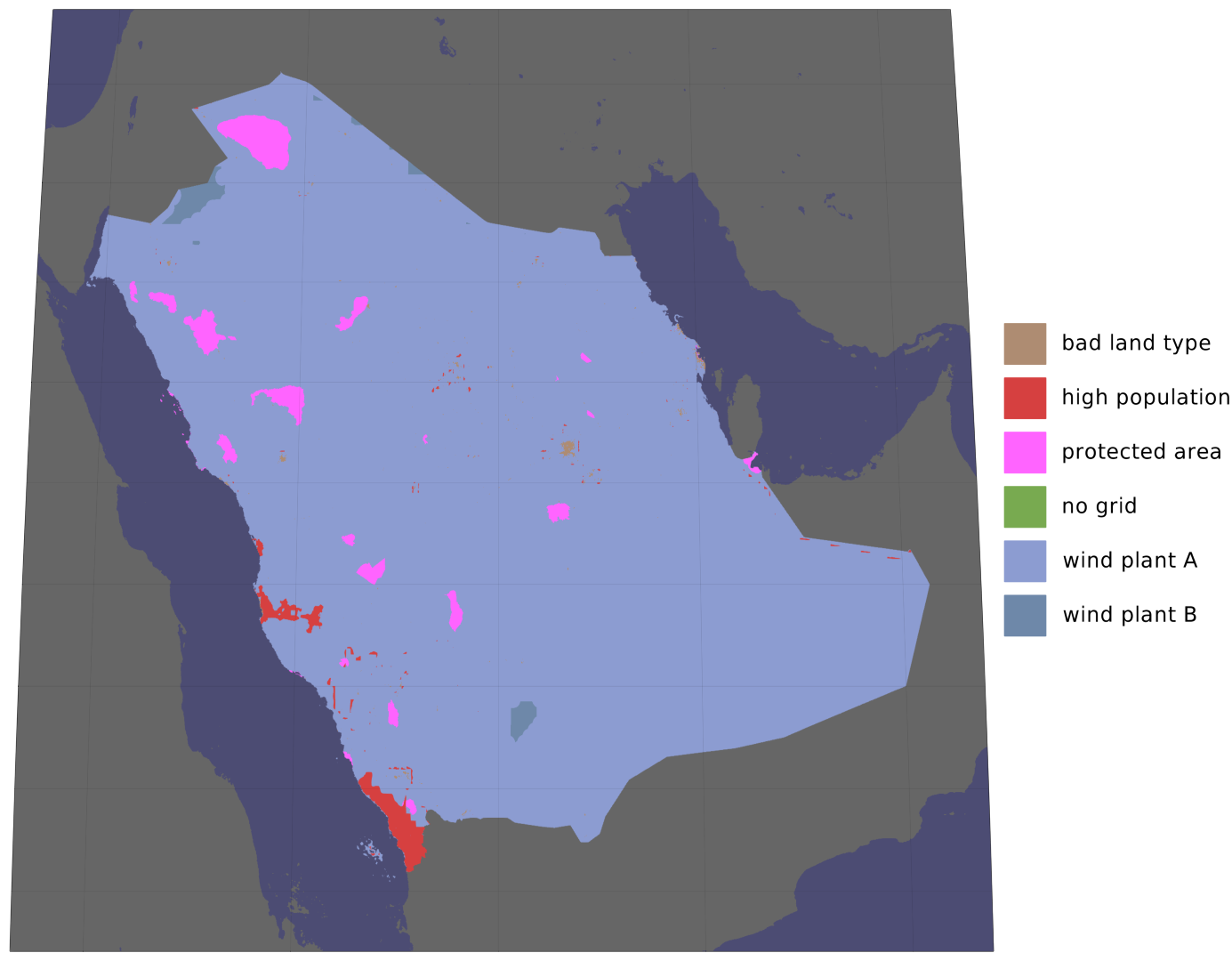}
        \caption{Onshore wind mask for \glsxtrlong{sa}.}
    \end{subfigure}
    
    \begin{subfigure}{.5\linewidth}
        \centering
\includegraphics[width=.9\textwidth]{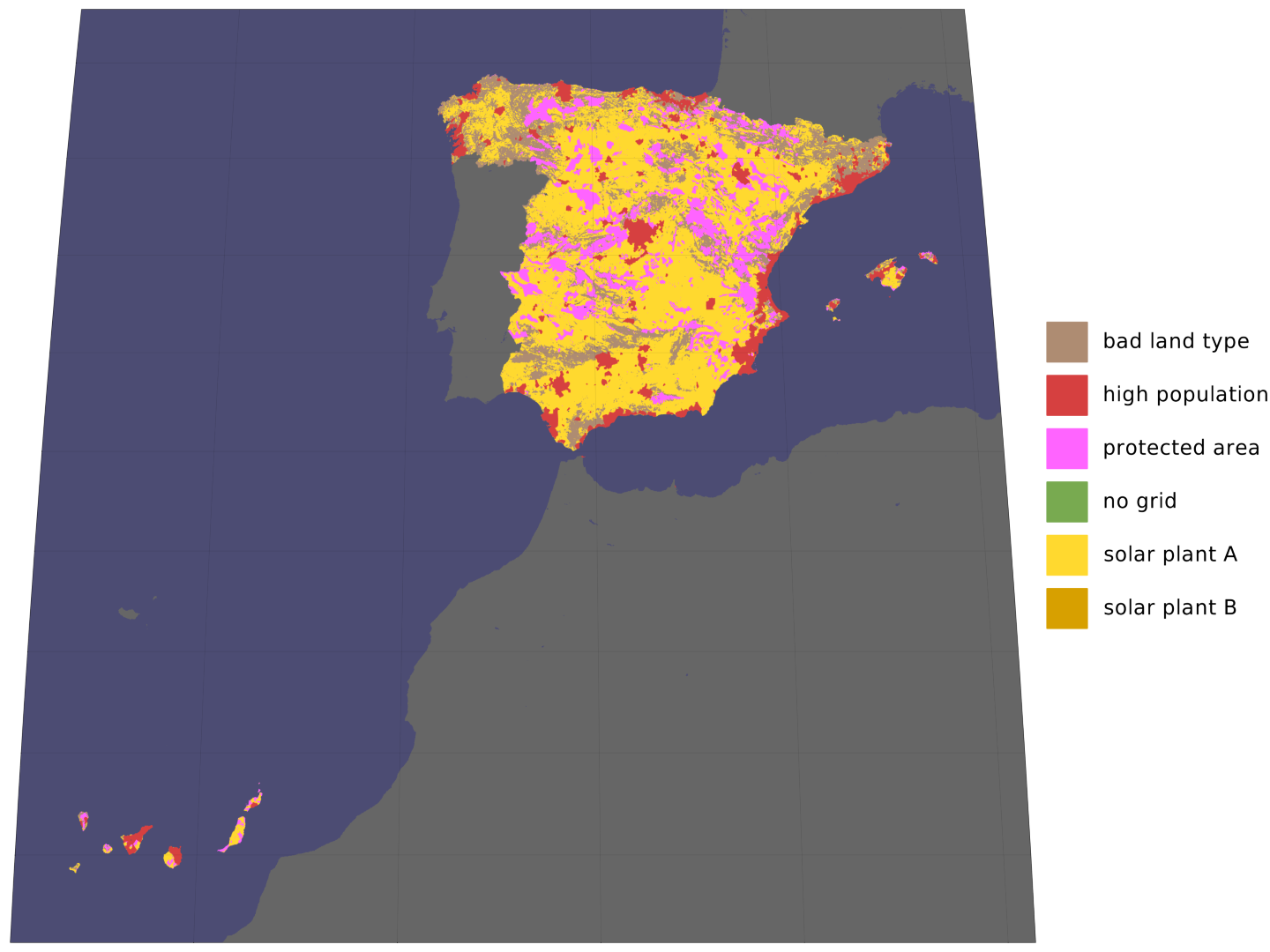}
        \caption{\gls{pv} mask for \glsxtrlong{es}.}
    \end{subfigure}
\begin{subfigure}{.5\linewidth}
        \centering
\includegraphics[width=.9\textwidth]{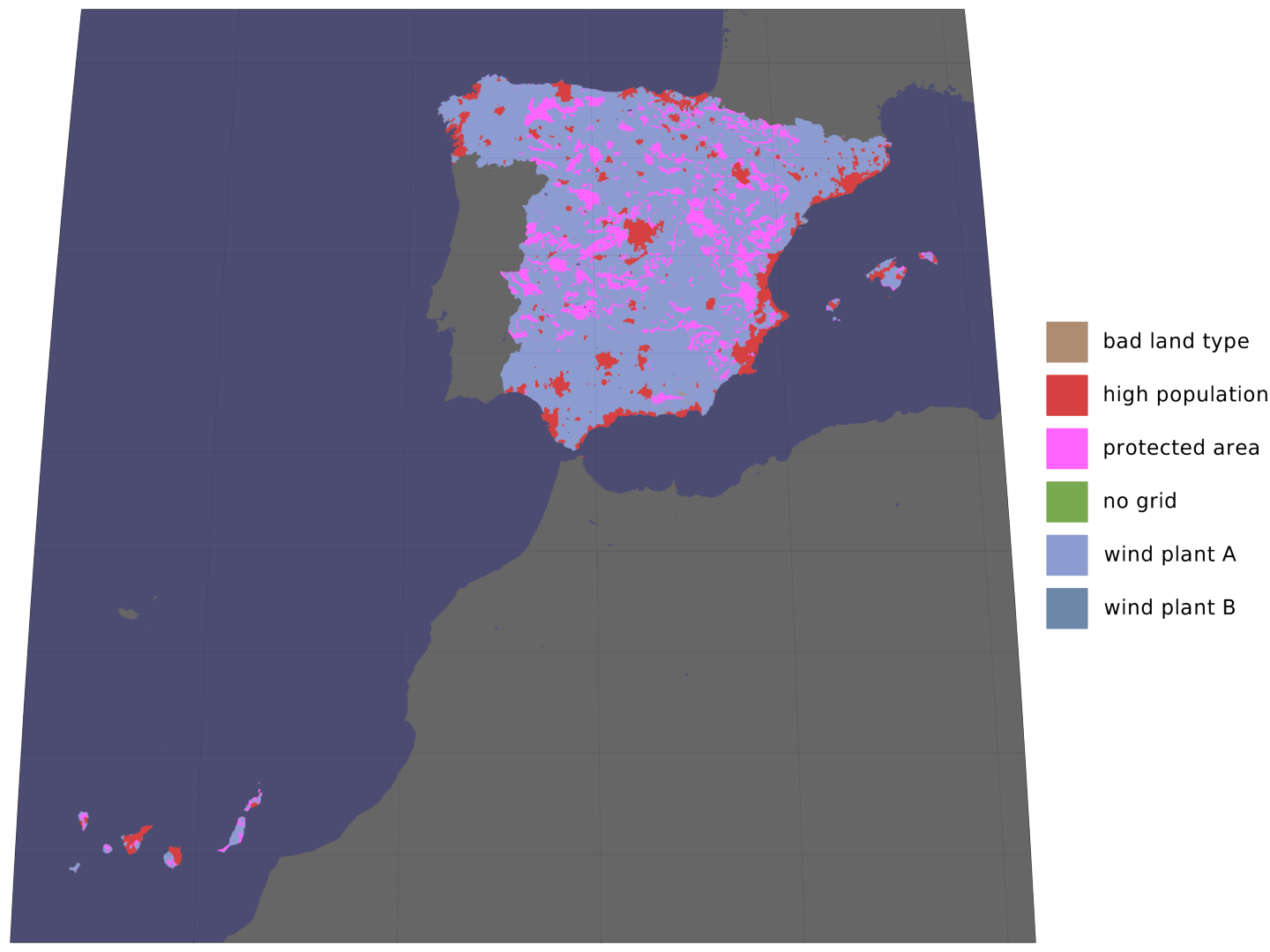}
        \caption{Onshore wind mask for \glsxtrlong{es}.}
    \end{subfigure}
    \caption{
        \gls{res} area masks for all exporting countries considered and onshore wind as well as \gls{pv}.
        No differentiation was made between \enquote{plant A} and \enquote{plant B} locations in this study.
    }
\end{figure}

\clearpage
\modifysubsectionname{Appendix}
\subsection{Data and code availability}
\label{app:data-and-code-availability}
Model results including technology and cost assumptions are available on \href{https://zenodo.org/record/7293102}{Zenodo (doi:10.5281/zenodo.7293102)}.
The model is available in \href{https://github.com/euronion/trace/releases/tag/2022.11.07}{this GitHub repository} under GPL-3.0-or-later and Creative Commons licenses.
Technology cost assumptions are also available in
\href{https://github.com/pypsa/technology-data/tree/v0.4.0}{this dedicated GitHub repository} under similar license.

\end{document}